\newcommand{\be}{\begin{equation}}
\newcommand{\ee}{\end{equation}}
\newcommand{\bea}{\begin{eqnarray}}
\newcommand{\eea}{\end{eqnarray}}
\documentclass[showpacs,preprintnumbers,amsmath,amssymb]{revtex4}
\newcommand{\sn}{{\rm sn}}
\newcommand{\ds}{{\rm ds}}
\newcommand{\cs}{{\rm cs}}
\newcommand{\ns}{{\rm ns}}
\newcommand{\dn}{{\rm dn}}
\newcommand{\cn}{{\rm cn}}
\newcommand{\sech}{{\rm sech}}

\usepackage{graphicx}
\usepackage{dcolumn}
\usepackage{bm}

\begin{document}

\preprint{APS/123-QED}

\title{Exact Static Solutions of a Generalized Discrete $\phi^4$
Model \\ Including Short-Periodic Solutions }

\author{
Avinash Khare$^1$, Sergey V. Dmitriev$^2$, and Avadh Saxena$^3$ }
\affiliation{ $^1$Institute of Physics, Bhubaneswar, Orissa
751005, India \\
$^2$Altai State Technical University, General Physics Department,
Barnaul 656038, Lenin St. 46, Russia \\
$^3$Center for Nonlinear Studies and Theoretical Division, Los
Alamos National Laboratory, Los Alamos, New Mexico 87545, USA}

\date{\today}

\begin{abstract}
For a five-parameter discrete $\phi^4$ model, we derive various
exact static solutions, including the staggered ones,
 in the form of the basic Jacobi elliptic
functions $\sn$, $\cn$, and $\dn$, and also in the form of their
hyperbolic function limits such as kink ($\tanh$) and
single-humped pulse ($\sech$) solutions. Such solutions are
admitted by the considered model in seven cases, two of which have
been discussed in the literature, and the remaining five cases are
addressed here. We also obtain $\sin$e, staggered $\sin$e as well as
a large number of short-periodic
static solutions of the generalized 5-parameter model. All the Jacobi elliptic,
hyperbolic and trigonometric function solutions (including the
staggered ones) are
translationally invariant (TI), i.e., they can be shifted along
the lattice by an arbitrary $x_0$, but among the short-periodic
solutions there are both TI and non-TI solutions.
The stability of these solutions is also investigated.
Finally, the
constructed Jacobi elliptic function solutions reveal four
new types of cubic nonlinearity with the TI property.
\end{abstract}


\maketitle

\section{Introduction and Setup} \label{Introduction}

In recent years there has been a growing interest in the analysis
of new discrete nonlinear models since they play a very important
role in many physical applications. For example, the question of
mobility of solitonic excitations in discrete media is a key issue
in many physical contexts; for example the mobility of
dislocations, a kind of topological solitons, is of importance in
the physics of plastic deformation of metals and other crystalline
bodies \cite{Nabarro}. Similar questions arise in optics for light
pulses moving in optical waveguides or in photorefractive crystal
lattices (see e.g., \cite{melvin} for a relevant recent
discussion) and in atomic physics for Bose-Einstein condensates
moving through optical lattice potentials (see e.g.,
\cite{oberthaler} for a recent review). These issues may prove
critical in aspects related to the guidance and manipulation of
coherent, nonlinear wavepackets in solid-state, atomic and optical
physics applications.

In particular, the translationally invariant (TI) discrete models
\cite{PhysicaD} have received considerable attention since they
admit static solutions that can be placed anywhere with respect to
the lattice. Such discretizations have been constructed and
investigated for the Klein-Gordon field
\cite{PhysicaD,SpeightKleinGordon,BenderTovbis,JPA2005,CKMS_PRE2005,BOP,DKY_JPA_2006Mom,
oxt1,DKYF_PRE2006,Coulomb,DKKS2007_BOP,Roy} and for the nonlinear
Schr\"odinger equation
\cite{DNLSEx,krss,pel,DKYF_JPA_2007DNLSE,DNLSE1,NewJPA}. For the
Hamiltonian TI lattices \cite{SpeightKleinGordon,CKMS_PRE2005},
this can be interpreted as the absence of the Peierls-Nabarro (PN)
potential \cite{Nabarro}. For the non-Hamiltonian lattices, the
height of the Peierls-Nabarro barrier is path-dependent but there
exists a continuous path along which the work required for a
quasi-static shift of the solution along the lattice is zero
\cite{DKYF_PRE2006}.

In general, one can state that
coherent structures in the TI models are not trapped by the
lattice and they can be accelerated by even a weak external field.
This particular property makes the TI discrete models potentially
interesting for physical applications and one such physically
meaningful model has been recently reported \cite{Coulomb}.

For some of the TI models it has been demonstrated that they
conserve momentum \cite{PhysicaD} or energy (Hamiltonian)
\cite{SpeightKleinGordon,CKMS_PRE2005} (see also
\cite{DKYF_JPA_2007DNLSE,pel}). However, we do not know a TI model
conserving both momentum and Hamiltonian and, for the Klein-Gordon
lattices with classically defined momentum, it was proved that
these two conservation laws are mutually exclusive
\cite{DKY_JPA_2006Mom}.

It may be noted that the TI discrete models can support even moving solutions,
but only for selected propagation velocities
\cite{DKYF_JPA_2007DNLSE}. In some cases, the exact static or even
moving solutions to the TI models can be expressed explicitly in
terms of the Jacobi elliptic functions (JEF). Even in the cases
when JEF solutions are impossible, static solutions to a TI
model can always be obtained iteratively from a nonlinear map
(first integral), solving at each step an algebraic equation.

While there has been no universally acceptable definition of TI
models, it is fairly easy to describe what a TI solution is. It is
a static solution which can be placed anywhere with respect to the
lattice. In particular, if there is an analytic TI solution with
an arbitrary shift $x_0$ along the chain or if one can show that
there is a corresponding Goldstone mode with zero frequency for
any $x_0$. As far as TI models are concerned, it is believed that
they should possess following properties (i) they admit static
solutions which can be placed anywhere with respect to the
lattice, which can be associated with the absence of PN barrier.
Note that in case analytical static solutions can be constructed
with an arbitrary shift along the chain, $x_0$, that would
automatically imply the absence of the PN barrier (ii) static version of
TI discrete models are integrable, i.e. the static problems are
reducible to a first-order difference equation which can be viewed
as a nonlinear map from which static solutions can be constructed
iteratively (in this study we will show that non-integrable
three-point static problems also can have particular TI solutions
derivable from a set of two lower-order finite-difference
equations, and one of those equations is a two-point one while
another is a three-point one) (iii) static solutions in TI models
possess the translational Goldstone mode with zero frequency for
any $x_0$.

A prototype class of discrete models, relevant to a variety of
applications are the so called discrete $\phi^4$ models which feature
a cubic nonlinearity. The purpose of this paper is to study in detail
several issues related to TI models. In particular we consider
a rather general discrete $\phi^4$ model with cubic nonlinearity which
is invariant under the interchange of $\phi_{n+1}$ and $\phi_{n-1}$
\begin{eqnarray} \label{DModel}
   \ddot{\phi}_n = \frac{1}{h^2} (\phi_{n+1}+\phi_{n-1}-2\phi_n)+\lambda \phi_n
   -A_1\phi_n^3
   -\frac{A_2}{2}\phi_n^2(\phi_{n+1}+\phi_{n-1}) \nonumber \\
   -\frac{A_3}{2}\phi_n(\phi_{n+1}^2+\phi_{n-1}^2)
   -A_4 \phi_n \phi_{n+1} \phi_{n-1}
   -\frac{A_5}{2}\phi_{n+1}\phi_{n-1}(\phi_{n+1}+\phi_{n-1})
   -\frac{A_6}{2}(\phi_{n+1}^3+\phi_{n-1}^3),
\end{eqnarray}
with the model parameters satisfying the constraint
\begin{eqnarray} \label{CC}
   \sum_{k=1}^{6}A_k=\lambda\,.
\end{eqnarray}
In Eq. (\ref{DModel}), $\phi_n(t)$ is the unknown function defined
on the lattice $x_n=hn$ with the lattice spacing $h>0$ and overdot
means derivative with respect to time $t$. Without the loss of
generality it is sufficient to consider $\lambda= 1$ or $\lambda=
-1$.

If model parameters $A_k$ are constant (i.e. independent of $h$),
condition of Eq. (\ref{CC}) ensures that, in the continuum limit,
Eq. (\ref{DModel}) reduces to the $\phi^4$ equation
\begin{eqnarray} \label{Contphi4}
   \phi_{tt} = \phi_{xx} + \lambda \phi (1 - \phi ^2).
\end{eqnarray}
On the other hand, if model parameters $A_k$ are functions of $h$,
then the continuum limit can be
different from Eq. (\ref{Contphi4}) even when Eq. (\ref{CC}) holds.

Static form of Eq. (\ref{Contphi4}) has the first integral
\begin{eqnarray} \label{FI}
   \phi_x^2 - \frac{\lambda }{2}\left( {1 - \phi ^2} \right)^2+ C = 0,
\end{eqnarray}
with the integration constant $C$.

So far as we are aware of, all the discrete $\phi^4$ models
discussed in the literature, under static consideration, are special
cases of the general model, Eq. (\ref{DModel}). Some of these
models are

{\em Model 1}. Only $A_1$ nonzero with other $A_k$ being equal to
zero results in the classical discretization of Eq.
(\ref{Contphi4}) that has received a great deal of attention from
the researchers in various fields. This model is not a TI one and
it will not be further discussed here.

{\em Model 2}. $A_1=A_3/2=A_4=\lambda\delta$, $A_5=A_6=2\lambda
\gamma$, $A_2=\lambda(1-4\gamma -4\delta)$ with arbitrary $\gamma$
and $\delta$. This non-Hamiltonian TI model (for arbitrary
$\gamma$ and $\delta$) conserves momentum \cite{PhysicaD}
\begin{eqnarray} \label{P1}
   P_1 =\sum_{n} \dot{\phi}_n(\phi_{n+1}-\phi_{n-1})\,.
\end{eqnarray}
Static version of
this model (with the omitted inertia term $\ddot{\phi}_n$) has the
first integral with the integration constant $C$ \cite{JPA2005},
\begin{eqnarray} \label{HbasicPhi4Transformed}
  U(\phi_{n-1},\phi_{n}) \equiv (\phi_n-\phi_{n-1})^2 + \Lambda\phi_n\phi_{n-1}
  - \Lambda\gamma \left( {\phi_n^4  + \phi_{n-1}^4 } \right) \nonumber \\
  - \Lambda\delta \phi_n\phi_{n-1}\left( {\phi_n^2
  + \phi_{n-1}^2 } \right)
  - \Lambda\left( {\frac{1}{2} -2\gamma - 2\delta } \right)
  \phi_n^2 \phi_{n-1}^2 - \frac{C\Lambda}{2} =0\,,
\end{eqnarray}
with $\Lambda$  defined by
\begin{eqnarray} \label{Lamb}
   \Lambda = \lambda h^2\,,
\end{eqnarray}
from which any static solution to Eq. (\ref{DModel}) can be
constructed iteratively, starting from any admissible value of
$\phi_0$ and solving at each step the algebraic problem. This is
so because Eq. (\ref{DModel}) is nothing but
\begin{eqnarray} \label{Representation1}
   \ddot{\phi}_n = \frac{U(\phi_{n},\phi_{n+1}) -
   U(\phi_{n-1},\phi_{n})}{\phi_{n+1} -\phi_{n-1}}\,.
\end{eqnarray}
Equation (\ref{HbasicPhi4Transformed}) is the discretized first
integral (DFI) \cite{DKYF_PRE2006}, i.e., in the continuum limit
$(h \rightarrow 0)$ it reduces to Eq. (\ref{FI}). DFI of Eq.
(\ref{HbasicPhi4Transformed}) is quartic in both $\phi_{n-1}$ and
$\phi_n$ and thus, it cannot be reduced to the integrable
nonlinear map reported in \cite{Quispel} where the corresponding
first integral is quadratic in $\phi_{n-1}$ and $\phi_n$. In this
complete form the model was first constructed in \cite{JPA2005}
and almost concurrently in \cite{BOP}, although in the latter work
any relation to the DFI was not observed. It is worth noting here that
since $\gamma$ and $\delta$ are arbitrary, one really has three
distinct models each of which (as well as their sum) conserves
momentum $P_1$ and there is a corresponding two-point map in each
case.

{\em Model 3}. Model 2 with $\gamma=\delta=0$ is the Bender-Tovbis
model \cite{BenderTovbis}. In the framework of the DFI approach
\cite{DKYF_PRE2006}, almost the entire space of static solutions
supported by this model was described and many of those solutions
were expressed in terms of the Jacobi elliptic functions (JEF)
\cite{DKYF_PRE2006}.

{\em Model 4}. Model 2 with $\delta=0$ and $\gamma=1/4$ is the Kevrekidis
model \cite{PhysicaD}.

{\em Model 5}. Discrete $\phi^4$ model
\begin{equation}
\ddot{\phi}_n  = \frac{1}{h^2}(\phi _{n-1}-2\phi_n+\phi_{n+1}) +
\frac{\lambda \left(\phi_n - \phi_n^3\right)}{1 - \Lambda
\phi_n^2/2}, \label{Saxena}
\end{equation}
discovered in \cite{CKMS_PRE2005} does not belong to Eq.
(\ref{DModel}) but its static problem coincides with that of
the Bender-Tovbis Model 3 \cite{CKMS_PRE2005,DKYF_PRE2006}. Some very
special features of this TI model are the conservation of energy
and the on-site discretization of the nonlinear term. In all other
TI $\phi^4$ models derived so far the nonlinear term is
discretized on the three neighboring nodes (i.e. lattice sites).

{\em Model 6}. With only $A_4$ nonzero and other $A_k$ being equal
to zero, one arrives at the non-Hamiltonian TI model derived by
Barashenkov, Oxtoby, and Pelinovsky \cite{BOP} and referred to as
BOP. This model conserves the momentum defined as
\cite{DKKS2007_BOP}
\begin{eqnarray} \label{P2}
   P_2 =\sum_{n} \dot{\phi}_n(\phi_{n+2}-\phi_{n-2}).
\end{eqnarray}
The first integral of the static version of this model has been
found in \cite{DKKS2007_BOP}, where an almost complete set of static
solutions supported by this model were derived and many of those
solutions were expressed in terms of JEF.

{\em Model 7}. Taking $A_1=2\lambda/9$, $A_2=A_3=\lambda/3$,
$A_4=A_5=0$, $A_6=\lambda/9$ one gets the Hamiltonian of the Speight and
Ward (SW) model \cite{SpeightKleinGordon}. For this model it is known
that, it supports TI static kinks derivable from
the two-point map
\begin{eqnarray} \label{SpeightMap}
   \phi_{n \pm 1}  =  - \frac{{\phi_n }}{2} \mp \frac{3}{{\sqrt{2} \,H}} \pm
   \frac{{\sqrt 3 }}{2}\sqrt { - \phi_n^2  \pm
\frac{{6\sqrt 2 }}{H}\phi_n
   + \frac{6}{H^2} + 4}, \quad H^2=\frac{6\Lambda}{6-\Lambda}\,,
\end{eqnarray}
where one can take either the upper or the lower sign. Note that this
map is defined in case $0 < \Lambda < 6$. To get
this map, one has to set in Eq. (\ref{FI}) $C=0$ and present it as
\begin{eqnarray} \label{SpeightFI}
   \phi_x \pm \sqrt {\frac{\lambda}{2}} (1 - \phi ^2)= 0.
\end{eqnarray}
Discretizing Eq. (\ref{SpeightFI})  as
\be \frac{\phi_n  -\phi_{n
   - 1}}{H} - \frac{1}{\sqrt{2}}
   +\frac{1}{3\sqrt{2}} (\phi_{n - 1}^2 + \phi_{n - 1}
   \phi_n + \phi_n^2)= 0\,, \ee
and solving the resulting quadratic equation we come to Eq.
(\ref{SpeightMap}). It is not known if this model supports static
TI solutions other than the kink. It is also not known if this model
has the first integral of the static problem apart from the case
of $C=0$. In the present study we will give evidence that the
answer to the second question is negative (see Sec.
\ref{FivePeriodic}) but we were able to find other TI solutions to
this model, see Eq. (\ref{zz3}) and Eq. (\ref{zz3'}).

{\em Model 8}. $A_1=4\lambda \alpha (\gamma +\beta)$,
$A_2=4\lambda [2\alpha^2 +\gamma^2 +\beta (\gamma -\alpha)]$,
$A_3=4\lambda \alpha (2\gamma +\beta)$, $A_4=4\lambda \gamma
(\alpha -\beta)$, $A_5=4\lambda \alpha (\alpha -\beta)$,
$A_6=4\lambda \alpha^2$, with two free parameters $\alpha$ and
$\beta$ with $\gamma= 1/2 -2\alpha$. This model was also offered by
Barashenkov, Oxtoby, and Pelinovsky \cite{BOP}.
This model includes as special cases the Model 3 (at $\alpha=\beta=0$),
the Model 6 (at $\alpha=0$ and $\beta=-1/2$) and also the Model 7 (at
$\alpha=\beta=1/6$).

{\em Model 9}. $A_1=0$, $A_2=2\lambda (1/2-\beta)$, $A_3=\lambda
\sigma (4+h^2)$, $A_4=2\lambda \beta +\lambda \sigma (4+h^2)$,
$A_5=-8\lambda \sigma$, $A_6=0$, with two free parameters $\beta$
and $\sigma$. This model was also proposed by Barashenkov, Oxtoby, and
Pelinovsky \cite{BOP}. Note that the coefficients $A_3$ and $A_4$
in this model are $h$-dependent and that the constraint Eq.
(\ref{CC}) for this model is satisfied only in the continuum limit
(i.e. for $h=0$).

In this paper we shall discuss four more TI models with cubic
nonlinearities.
The paper is organized as follows. In Sec. \ref{Momentum} we
discuss the subclasses of model Eq. (\ref{DModel}) that support
different conservation laws. In Sec. \ref{JEFsolutions} we report
on a number of TI static solutions to Eq. (\ref{DModel}) expressed
in closed analytical form. All seven cases, when model Eq.
(\ref{DModel}) supports the exact static JEF solutions, are
described. Two of these seven cases have been previously studied
in the literature, and for the remaining five cases, basic JEF
solutions are given here together with their hyperbolic function
limits. We also obtain a periodic sine solution for the general
5-parameter model as given by Eq. (\ref{DModel}). Section
\ref{ShortPeriodicSolutions} presents a number of short-periodic
exact static solutions. In Sec. \ref{Maps} we discuss the
two-point maps for some of the TI models that have been reported
in the literature and derive the map for the linear combination of
Models 3 (Bender-Tovbis) and 6 (BOP). Goldstone translational
modes of the TI static solutions are discussed in Sec.
\ref{Goldstone}. Numerical results that illustrate some important
properties of the TI static solutions are presented in Sec.
\ref{NumericalResults}. The discussion, conclusions, and future
challenges are described in Sec.
\ref{Discussion}. Finally in Appendix we spell out the short periodic
solutions that are admitted by many of the models discussed in this paper.

\section{Momentum and energy conservation}
\label{Momentum}

As noted above, for the discrete model of Eq. (\ref{DModel}), the
momentum operator as given by Eq. (\ref{P1}) is conserved provided
only $A_2$ is nonzero (Model 3), or if only $A_5$ and $A_6$ are nonzero with
$A_5=A_6$ (Model 4), or if only $A_1,A_3,A_4$ are nonzero with
$A_1=A_3/2=A_4$ and for a linear combination of these models which
is precisely the Model 2. It is also conserved in Model 8 (provided
$\alpha=\beta=0$) and in Model 9 (provided $\sigma=\beta=0$).
On the other hand, as was already
mentioned, the momentum defined by Eq. (\ref{P2}) is conserved for
Model 6 with only $A_4$ nonzero. It is also conserved in Model 8 (provided
$\alpha=0,\beta=-1/2$) and in Model 9 (provided $\sigma=0, \beta=1/2$).

{\em Model 10.} For
\begin{eqnarray}\label{HamCond}
   A_1=4\alpha_1 \lambda, \quad A_2=6\alpha_2 \lambda,
\quad A_3=4\alpha_3 \lambda, \quad A_6=2\alpha_2 \lambda,
\quad A_4=A_5=0, \quad {\rm with} \quad
   \alpha_1+ 2\alpha_2 +\alpha_3= \frac{1}{4}\,,
\end{eqnarray}
the model Eq. (\ref{DModel}) has the Hamiltonian
\begin{eqnarray}\label{Ham}
   H=\sum_n \left[ \frac{\dot{\phi}_n^2}{2}
   + \frac{(\phi_n -\phi_{n-1})^2}{2h^2}
   +\frac{\lambda}{4} -\frac{\lambda}{2}\phi_n^2
   +\alpha_1\phi_n^4 + \alpha_2 \phi_n\phi_{n-1}(\phi_n^2 +
   \phi_{n-1}^2)
   + \alpha_3 \phi_n^2\phi_{n-1}^2
   \right]\,,
\end{eqnarray}
and hence energy is conserved in this model.
As it can be seen, the Hamiltonian model has two free parameters.
Note that the Speight-Ward model 7 is a special case of this model for
$\alpha_1=\alpha_2=1/18$ and $\alpha_2=1/12$. Similarly, energy is also conserved
in Model 8 provided $\alpha=\beta=1/6$.

\section{Translationally invariant JEF, Hyperbolic,
and Trigonometric static solutions} \label{JEFsolutions}

We shall first discuss the JEF solutions as well as the hyperbolic
solutions which follow from the JEF solutions and later on we
shall discuss the $\sin$e solutions which exist in almost all the
models.

\subsection{JEF and Hyperbolic Solutions}

We shall now show that JEF solutions can be obtained for the
discrete model of Eq. (\ref{DModel}) in case
$A_1=A_6=0$ in the following seven cases: (i) only $A_2$
nonzero; (ii) only $A_4$ nonzero; (iii) only $A_2$ and $A_4$
nonzero; (iv) only $A_3$ and $A_5$ nonzero; (v) $A_2$, $A_3$, and
$A_5$ nonzero; (vi) $A_3$, $A_4$, and $A_5$ nonzero; (vii) $A_2$,
$A_3$, $A_4$, and $A_5$ nonzero.

The JEF solutions have already been reported in the case (i) in
\cite{DKYF_PRE2006} and in the case (ii) in \cite{DKKS2007_BOP}.
In this paper we report on the JEF solutions for the cases (iii)
to (vii). In particular, in all these cases we obtain the $\sn$
solution
\begin{eqnarray}\label{sn}
   \phi_n = SA \sn [h\beta (n+x_0),m]\,,
\end{eqnarray}
the $\cn$ solution
\begin{eqnarray}\label{cn}
   \phi_n = SA \cn [h\beta (n+x_0),m]\,,
\end{eqnarray}
and the $\dn$ solution
\begin{eqnarray}\label{dn}
   \phi_n = SA \dn [h\beta (n+x_0),m]\,,
\end{eqnarray}
where $x_0$ is an arbitrary shift, $0 \le m \le 1$ is the JEF
modulus, and $S=1$ for the non-staggered and $S=(-1)^n$ for the staggered
solutions.

We also derive the hyperbolic function limits ($m=1$) of the above
JEF solutions. In particular, while the $\sn$ solution reduces to the kink
solution of the form
\begin{eqnarray}\label{tanh}
   \phi_n = SA\tanh [h\beta (n+x_0)]\,,
\end{eqnarray}
both $\cn$ and $\dn$ solutions reduce to the single-humped pulse
solution of the form
\begin{eqnarray}\label{sech}
   \phi_n = SA \sech [h\beta (n+x_0)]\,.
\end{eqnarray}

Expressions that relate the solution parameters $A$ and $\beta$ to
the model parameters $A_k$ and, where applicable, the
relations between $A_k$, are given in what follows.

Note that if $\phi_n(t)$ is a solution to Eq. (\ref{DModel}), then
the staggered solution $(-1)^n\phi_n(t)$ is also a solution to the same
equation, i.e. Eq. (\ref{DModel}) with the coefficients $A_1,A_3$, and $A_4$
having the opposite signs and further $2-\Lambda$ is to be replaced by
$\Lambda-2$, where $\Lambda$ is as given by Eq. (\ref{Lamb}).
To make the presentation of the results as compact as possible, in most
cases, we shall
therefore not give the parameters for the staggered solutions. Only
for the case (iii) with $A_2$ and $A_4$ nonzero and for hyperbolic solutions,
parameters will be given for both the staggered and the nonstaggered solutions.

\subsection{Case (iii): only $A_2$ and $A_4$ nonzero} \label{JEFA2A4}

Since only $A_2,A_4$ are nonzero, hence Eq. (\ref{CC}) reduces to
\be A_2+A_4=\lambda\,. \ee In this case, one has both nonstaggered
$\sn$ and staggered $\sn$ solutions Eq. (\ref{sn}), with $S=1$ and
$S=(-1)^{n}$ respectively, provided the following relations are
satisfied:
\be\label{5.4}
 \frac{2m}{A^2h^2}=  A_2 \ns^2(h\beta,m)
\pm 2 A_4 \ns(h\beta,m)\ns(2h\beta,m) \,,
 \ee
 \be\label{5.5}
 \pm \frac{(2-\Lambda)m}{A^2h^2}= A_2 \cs(h\beta)\ds(h\beta)
\pm A_4 \ns^2(h\beta,m)\,,
 \ee
where $ns(x,m)=1/sn(x,m)$, $cs(x,m)=cn(x,m)/sn(x,m)$ and $ds(x,m)=
dn(x,m)/sn(x,m)$. In the above equations, the upper sign corresponds
to the nonstaggered $\sn$
solution ($S=1$), while the lower sign corresponds to the staggered case
($S=(-1)^{n}$).

In the limit of $m=1$ the nonstaggered solution reduces to the kink
solution, Eq. (\ref{tanh}), with $S=1$ and the relations (\ref{5.4}) and
(\ref{5.5}) take the simpler form
 \be\label{5.11}
 A^2=1\,,~~h^2 A_4 =\frac{2\tanh^2(h\beta)-\Lambda}{\tanh^2(h\beta)}\,,
 \quad A_2=\lambda -A_4\,.
 \ee
Note that this solution is valid for any value of $\Lambda$
including $\Lambda=2$.

On the other hand, in the limit of $m=1$, the staggered $\sn$ solution
reduces to the staggered kink
solution, Eq. (\ref{tanh}), with $S=(-1)^{n}$ and the relations (\ref{5.4}) and
(\ref{5.5}) reduce to
 \be\label{5.28}
\sech^2(h\beta)=\frac{(8-3\Lambda)h^2 A_4
+\Lambda(\Lambda-2)}{(2-h^2 A_4 )\Lambda}\,,
~~A^2=\frac{(4-\Lambda)}{(2-h^2 A_4 )\Lambda}\,.
\ee

Another set of exact JEF solutions are the nonstaggered as well as the staggered $\cn$
solutions as given by Eq. (\ref{cn}) with $S=1$ and $S=(-1)^{n}$,
respectively, provided the following relations are satisfied:
 \be\label{5.15}
 \frac{2m}{A^2h^2}= - A_2 \ds^2(h\beta,m)\mp2 A_4 \ds(h\beta,m)\ds(2h\beta,m)\,,
 \ee
 \be\label{5.16}
\pm \frac{(2-\Lambda)m}{A^2h^2}=\mp A_4 \ds^2(h\beta,m)- A_2 \cs(h\beta,m)\ns(h\beta,m)\,.
 \ee
In the above equations, the upper sign corresponds to the nonstaggered $\sn$
solution ($S=1$), while the lower sign corresponds to the staggered case
($S=(-1)^{n}$).

Yet another exact JEF solutions are the nonstaggered as well as the staggered
$\dn$ solutions as given by Eq. (\ref{dn}) with $S=1$ and $S=(-1)^{n}$
respectively provided the following relations are satisfied:
 \be\label{5.21}
 \frac{2}{A^2h^2}= - A_2 \cs^2(h\beta,m)\mp 2 A_4 \cs(h\beta,m)\cs(2h\beta,m)\,,
 \ee
 \be\label{5.22}
\pm \frac{(2-\Lambda)}{A^2h^2}=\mp A_4 \cs^2(h\beta,m)
- A_2 \ns(h\beta,m)\ds(h\beta,m)\,.
 \ee
In the above equations, the upper sign corresponds to the nonstaggered $\sn$
solution ($S=1$), while the lower sign corresponds to the staggered case
($S=(-1)^{n}$).

In the limit of $m=1$, the nonstaggered $\cn$ as well $\dn$
solutions reduce to the pulse solution Eq. (\ref{sech}) with $S=1$
and the relations (\ref{5.15}) and (\ref{5.16}) as well as
(\ref{5.21}) and (\ref{5.22}) take the simpler form
\be\label{5.18}
   \Lambda =-2[\cosh(h\beta)-1]<0\,,~~
   ~~A^2=\frac{(|\Lambda|+2)(|\Lambda|+4)}{2[(|\Lambda|+2)+h^2 A_4
   ]}\,.
\ee
Thus the pulse solution exists only if $\Lambda <0$.

On the other hand, in the limit of $m=1$ the staggered $\cn$ as
well $\dn$ solutions reduce to the staggered pulse solution Eq.
(\ref{sech}) with $S=(-1)^{n}$ and the relations (\ref{5.15}) and
(\ref{5.16}) as well as (\ref{5.21}) and (\ref{5.22}) take the
simpler form
 \be\label{5.32}
\Lambda =2[\cosh(h\beta)+1]>0\,,~~
~~A^2=\frac{(\Lambda-2)(\Lambda-4)}{2[2+h^2 A_4 -\Lambda]}\,.
\ee
Thus the staggered pulse solution (\ref{5.32}) exists only if
$\Lambda >4$.

\subsection{Case (iv): only  $A_3$  and $A_5$ nonzero}

Since only $A_3,A_5$ are nonzero, Eq. (\ref{CC}) reduces to
\be
A_3+A_5=\lambda\,.
\ee
In this case, one has the $\sn$ solution Eq. (\ref{sn}) with $S=1$ provided
the following relations are satisfied:
 \be\label{3.3aa}
  A_5 \ns(2h\beta,m)=- A_3 \ns(h\beta,m)\,,
 \ee
 \be\label{3.4aa}
 \frac{2m}{A^2h^2}= - A_3 \cs(h\beta,m)\ds(h\beta,m)
 - A_5 [\cs(2h\beta,m)\ds(2h\beta,m)-\ns^2(2h\beta,m)]\,,
 \ee
 \be\label{3.5aa}
 \frac{(2-\Lambda)m}{A^2h^2}=- A_3 \ns^2(h\beta,m)\,.
 \ee

In the limit $m=1$, this reduces to the kink solution Eq. (\ref{tanh})
with $S=1$ provided
 \be\label{3.11aa}
 A^2=1\,,~~2 A_3 =- A_5 [1+\tanh^2(h\beta)]\,,~~
\Lambda= \frac{2\tanh^2(h\beta)\sech^2(h\beta)}
{1+2\tanh^2(h\beta)-\tanh^4(h\beta)}\,.
\ee

From Eq. (\ref{3.11aa}) it follows that
$\Lambda < A_5 h^2/2$. Further, using the fact
that $\tanh^2(x)<1$, it is easily checked that the kink solution
exists in this model only if $0 < \Lambda < (2-\sqrt{2})/2$.

In case only $A_3$ and $A_5$ are nonzero, the static Eq. (\ref{DModel})
is equivalent to a highly nonlinear map which is
even more complicated than the 18 parameter map of Quispel et al.
\cite{Quispel}.  In fact, this is also true for the remaining three cases
discussed below.

The corresponding staggered solutions are easily found by changing the
signs of terms with $A_3$ and $(2-\Lambda)$. In particular, it is easily shown
that in the limit $m=1$, the staggered $\sn$ solution reduces to the staggered
kink solution (\ref{tanh}) with $S=(-1)^{n}$ provided
 \be\label{3.28aa}
2 A_3 = A_5 [1+\tanh^2(h\beta)]\,,~~ A_5
h^2=\frac{2\Lambda}{3+\tanh^2(h\beta)}\,,
~~A^2= \frac{[3+\tanh^2(h\beta)]\tanh^2(h\beta)}
{[2+3\tanh^2(h\beta)-\tanh^4(h\beta)]}\,.
\ee
One can show that this solution can only exist provided
\be
\frac{6+\sqrt{2}}{2} <
\Lambda <4\,.
\ee

Another exact JEF solution is the nonstaggered
$\dn$ solution as given by Eq. (\ref{dn}) with $S=1$
provided the following relations are satisfied:
\be\label{3.14aa}
  A_5 \cs(2h\beta,m)=- A_3 \cs(h\beta,m)\,,
 ~~\frac{2-\Lambda}{A^2h^2}= A_3 \cs^2(h\beta,m)\,,
 \ee
 \be\label{3.15aa}
 \frac{2}{A^2h^2}=  A_3 \ds(h\beta,m)\ns(h\beta,m)
 + A_5 [\ns(2h\beta,m)\ds(2h\beta,m)-\cs^2(2h\beta,m)]\,.
 \ee

Yet another exact JEF solution is the nonstaggered
$\cn$ solution as given by Eq. (\ref{cn}) with $S=1$
provided the following relations are satisfied:
 \be\label{3.20aa}
  A_5 \ds(2h\beta,m)=- A_3 \ds(h\beta,m)\,,
 ~~\frac{(2-\Lambda)m}{A^2h^2}= A_3 \ds^2(h\beta,m)\,,
 \ee
 \be\label{3.21aa}
 \frac{2m}{A^2h^2}=  A_3 \cs(h\beta,m)\ns(h\beta,m)
 + A_5 [\ns(2h\beta,m)\cs(2h\beta,m)-\ds^2(2h\beta,m)]\,.
 \ee

In the limit $m=1$, both the $\cn$ and the $\dn$ solutions go over to the pulse
solution  Eq. (\ref{sech}) with
$S=1$, and the relations (\ref{3.14aa}) and (\ref{3.15aa}) as well as
(\ref{3.20aa}) and (\ref{3.21aa}) take the simpler form
 \bea\label{3.18aa}
&&\Lambda=-2[\cosh(h\beta)-1] <0\,,~~ A_5 <0\,,~~2 A_3 =- A_5
\sech(h\beta)\,,~~
| A_5 |h^2=|\Lambda|\frac{|\Lambda|+2}{|\Lambda|+1}\,, \nonumber \\
&& A^2=\frac{(|\Lambda|+1)(|\Lambda|+2)(|\Lambda|+4)}{4}\,.
 \eea

One can also obtain the corresponding staggered $\cn$ and $\dn$ solutions
by changing the signs of $A_3$ and $(2-\Lambda)$ in
the relations (\ref{3.14aa}) and (\ref{3.15aa}) as well as
(\ref{3.20aa}) and (\ref{3.21aa}).
In the limit $m=1$ one can show that {\it no} staggered pulse
solution Eq. (\ref{sech}) with $S=(-1)^n$ exists in case
only $A_3$ and $A_5$ are nonzero.

\subsection{Case (v): only  $A_2$, $A_3$, and $A_5$ are nonzero}

Since only $A_2,A_3,A_5$ are nonzero, hence Eq. (\ref{CC}) reduces to
\be
A_2+A_3+A_5=\lambda\,.
\ee
In this case, one has the $\sn$ solution Eq. (\ref{sn}) with $S=1$ provided
the following relations are satisfied:
 \be\label{4.3}
  A_5 \ns(2h\beta,m)=- A_3 \ns(h\beta,m)\,,~~\frac{(2-\Lambda)m}{A^2h^2}=
A_2 \cs(h\beta)\ds(h\beta)- A_3 \ns^2(h\beta,m)\,,
 \ee
 \bea\label{4.4}
 &&\frac{2m}{A^2h^2}=  A_2 \ns^2(h\beta,m)
- A_3 \cs(h\beta,m)\ds(h\beta,m) \nonumber \\
&& - A_5 [\cs(2h\beta,m)\ds(2h\beta,m)-\ns^2(2h\beta,m)]\,.
 \eea

In the $m=1$ case, for the kink solution Eq. (\ref{tanh}) with
$S=1$, the relations (\ref{4.3}) to (\ref{4.4}) take the simpler
form
 \bea\label{4.11}
&& A^2=1\,,~~h^2 A_3
=\frac{[\Lambda-2\tanh^2(h\beta)][1+\tanh^2(h\beta)]}
{\tanh^2(h\beta)[3-\tanh^2(h\beta)]}\,, \nonumber \\
&& A_2
h^2=\frac{[\Lambda+2(\Lambda-1)\tanh^2(h\beta)+(2-\Lambda)\tanh^4(h\beta)]}
{\tanh^2(h\beta)[3-\tanh^2(h\beta)]}\,, ~~ A_5
h^2=-\frac{2[\Lambda-2\tanh^2(h\beta)]}
{\tanh^2(h\beta)[3-\tanh^2(h\beta)]}\,.
\eea
Note that this solution is valid for any value of $\Lambda$
including $\Lambda=2$.

The corresponding staggered $\sn$ solution is easily obtained by changing the
signs of terms with $A_3$ and $(2-\Lambda)$. In particular, it is easily shown
that in the limit $m=1$, the staggered $\sn$ solution reduces to the staggered
kink solution (\ref{tanh}) with $S=(-1)^{n}$ provided
 \bea\label{4.28}
&&h^2 A_3
=\frac{\Lambda[1+\tanh^2(h\beta)][\Lambda-2(2-\tanh^2(h\beta)]}
{[\Lambda
\tanh^4(h\beta)+(8-\Lambda)\tanh^2(h\beta)+2(\Lambda-4)]}\,,
\nonumber \\
&&h^2 A_5 =\frac{2\Lambda[\Lambda-2(2-\tanh^2(h\beta)]} {[\Lambda
\tanh^4(h\beta)+(8-\Lambda)\tanh^2(h\beta)+2(\Lambda-4)]}\,,
\nonumber \\
&&h^2 A_2 =\frac{\Lambda[(\Lambda-2)\tanh^4(h\beta)
+2(3-\Lambda)\tanh^2(h\beta)+4-\Lambda]} {[\Lambda
\tanh^4(h\beta)+(8-\Lambda)\tanh^2(h\beta)+2(\Lambda-4)]}\,,
\nonumber \\
&&A^2= \frac{(\Lambda -4)}{(2h^2 A_3 -\Lambda)}\,. \eea
Note that this solution exists over a vast range of $\Lambda$ values including
$\Lambda=2$.

For the dn solution Eq. (\ref{dn}) with $S=1$ one has
 \be\label{4.14}
  A_5 \cs(2h\beta,m)=- A_3 \cs(h\beta,m)\,,
 ~~\frac{2-\Lambda}{A^2h^2}= A_3 \cs^2(h\beta,m)- A_2
 \ds(h\beta,m)\ns(h\beta,m)\,,
 \ee
 \be\label{4.15}
 \frac{2}{A^2h^2}= - A_2 \cs^2(h\beta,m)+ A_3 \ds(h\beta,m)\ns(h\beta,m)
 + A_5 [\ns(2h\beta,m)\ds(2h\beta,m)-\cs^2(2h\beta,m)]\,.
 \ee

Parameters of the cn solution Eq. (\ref{cn}) with $S=1$ satisfy
 \be\label{4.20}
  A_5 \ds(2h\beta,m)=- A_3 \ds(h\beta,m)\,,
 ~~\frac{(2-\Lambda)m}{A^2h^2}= A_3 \ds^2(h\beta,m)
- A_2 \ns(h\beta,m)\cs(h\beta,m)\,,
 \ee
 \be\label{4.21}
 \frac{2m}{A^2h^2}= - A_2 \ds^2(h\beta,m)+ A_3 \cs(h\beta,m)\ns(h\beta,m)
 + A_5 [\ns(2h\beta,m)\cs(2h\beta,m)-\ds^2(2h\beta,m)]\,.
 \ee

In the limiting case of $m=1$, from the relations (\ref{4.14}) and
(\ref{4.15}) as well as (4.20) and (4.21), one finds the relations for the
pulse solution  Eq. (\ref{sech}) with $S=1$
 \bea\label{4.18}
&&\Lambda =-2[\cosh(h\beta)-1]<0\,,~~2 A_3 =- A_5
\sech(h\beta)\,,~~ \nonumber \\
&& A_2 h^2=h^2 A_3 [2\cosh(h\beta)-1]-2[\cosh(h\beta)-1]\,,
A^2=\frac{(|\Lambda|+2)(|\Lambda|+4)}{2[(|\Lambda|+2)-(|\Lambda|+3)h^2
A_3 ]}\,.
 \eea
Thus the pulse solution exists only if $\Lambda <0$.

One can also obtain the corresponding staggered $\dn$ and $\cn$ solutions.
In the limit $m=1$, from these solutions we obtain the
staggered pulse solution Eq. (\ref{sech}) with $S=(-1)^n$ satisfying
  \bea\label{4.37}
&&\Lambda =2[1+\cosh(h\beta)]>4\,,~~2 A_3 = A_5
\sech(h\beta)]\,,~~
 A_2 h^2=-h^2 A_3 [2\cosh(h\beta)+1]+2[1+\cosh(h\beta)]\,, \nonumber \\
&&A^2=\frac{(\Lambda-2)(\Lambda-4)}{2[(\Lambda-3)h^2 A_3
-(\Lambda-2)]}\,.
 \eea
Thus the staggered pulse solution exists only if $\Lambda >4$.

\subsection{Case (vi): only $A_3$, $A_4$, and $A_5$ are nonzero}

Since only $A_3,A_4,A_5$ are nonzero, hence Eq. (\ref{CC}) reduces to
\be
A_3+A_4+A_5=\lambda\,.
\ee
In this case, one has the $\sn$ solution Eq. (\ref{sn}) with $S=1$
satisfying
\be\label{3.3bb}
  A_5 \ns(2h\beta,m)=- A_3 \ns(h\beta,m)\,,
 ~~\frac{(2-\Lambda)m}{A^2h^2}=( A_4 - A_3 )\ns^2(h\beta,m)\,,
 \ee
 \bea\label{3.4bb}
 &&\frac{2m}{A^2h^2}= 2 A_4 \ns(h\beta,m)\ns(2h\beta,m)
- A_3 \cs(h\beta,m)\ds(h\beta,m) \nonumber \\
&& - A_5 [\cs(2h\beta,m)\ds(2h\beta,m)-\ns^2(2h\beta,m)]\,.
 \eea

In the limit $m=1$, we obtain from the $\sn$ solution the kink
solution Eq. (\ref{tanh}) with $S=1$ with the parameters satisfying
 \bea\label{3.11bb}
&& A^2=1\,,~~h^2 A_3
=\frac{[1+\tanh^2(h\beta)][(\Lambda-2)\tanh^2(h\beta)+\Lambda]}
{2\tanh^2(h\beta)}\,, \nonumber \\
&&h^2 A_4
=\frac{\Lambda+2(\Lambda-1)\tanh^2(h\beta)-(\Lambda-2)\tanh^4(h\beta)}
{2\tanh^2(h\beta)}\,,~~ h^2 A_5
=-\frac{(\Lambda-2)\tanh^2(h\beta)+\Lambda} {\tanh^2(h\beta)}\,.
\eea
Note that this solution is valid for any $\Lambda$ including
$\Lambda=2$ when conditions (\ref{3.11bb}) take particularly
simple form.

One can also obtain the corresponding staggered $\sn$ solution.
In the limit $m=1$, we obtain from the staggered $\sn$ solution the
staggered kink solution  Eq. (\ref{tanh}) with $S=(-1)^n$
provided
 \be\label{3.29}
A^2= \frac{(3\Lambda-8) \tanh^2(h\beta)+2(\Lambda-4)} {\Lambda
\tanh^2(h\beta)}\,, \ee \be\label{3.29c}
 A_3 h^2=\frac{\Lambda[1+\tanh^2(h\beta)][4-\Lambda+(2-\Lambda)\tanh^2(h\beta)]}
{2[(8-3\Lambda)\tanh^2(h\beta) +2(4-\Lambda)]}\,.
\ee
From the above equations it follows that no solution exists in case
$0< \Lambda \le 3$. In particular, note that a solution does exist
in case $\lambda<0$ provided $ A_3 <0$.

Parameters of the dn solution Eq. (\ref{dn}) with $S=1$ can be
found from
 \be\label{3.14bb}
  A_5 \cs(2h\beta,m)=- A_3 \cs(h\beta,m)\,,
 ~~\frac{2-\Lambda}{A^2h^2}=( A_3 - A_4 )\cs^2(h\beta,m)\,,
 \ee
 \be\label{3.15bb}
 \frac{2}{A^2h^2}= -2 A_4 \cs(h\beta,m)\cs(2h\beta,m)+ A_3 \ds(h\beta,m)\ns(h\beta,m)
 + A_5 [\ns(2h\beta,m)\ds(2h\beta,m)-\cs^2(2h\beta,m)]\,.
 \ee

On the other hand, the parameters of the $\cn$ solution Eq. (\ref{cn}) with
$S=1$ satisfy
 \be\label{3.20bb}
  A_5 \ds(2h\beta,m)=- A_3 \ds(h\beta,m)\,,
 ~~\frac{(2-\Lambda)m}{A^2h^2}=( A_3 - A_4 )\ds^2(h\beta,m)\,,
 \ee
 \be\label{3.21bb}
 \frac{2m}{A^2h^2}= -2 A_4 \ds(h\beta,m)\ds(2h\beta,m)+ A_3 \cs(h\beta,m)\ns(h\beta,m)
 + A_5 [\ns(2h\beta,m)\cs(2h\beta,m)-\ds^2(2h\beta,m)]\,.
 \ee

In the limit $m=1$, we obtain the parameters of the pulse solution
Eq. (\ref{sech}) with $S=1$
 \bea\label{3.18}
&&\Lambda =-2[\cosh(h\beta)-1]<0\,,~~2 A_3 =- A_5
\sech(h\beta)\,,~~
 A_4 h^2=h^2 A_3 [2\cosh(h\beta)-1]-2[\cosh(h\beta)-1]\,, \nonumber \\
&&A^2=\frac{(|\Lambda|+2)(|\Lambda|+4)}{4(1-h^2 A_3 )}\,.
 \eea
Thus the pulse solution exists only if $\Lambda <0$.

One can also obtain the corresponding staggered $\dn$ and $\cn$ solutions.
In the limit $m=1$, from these solutions we obtain the
staggered pulse solution Eq. (\ref{sech}) with $S=(-1)^n$ satisfying
  \bea\label{3.37a}
&&\Lambda =2[1+\cosh(h\beta)]>4\,,~~2 A_3 = A_5 \sech(h\beta)\,,~~
 A_4 h^2=-h^2 A_3 [2\cosh(h\beta)+1]+2[1+\cosh(h\beta)]\,, \nonumber \\
&&A^2=\frac{(\Lambda-2)(\Lambda-4)}{4(1-h^2 A_3 )}\,.
 \eea
Thus the staggered pulse solution exists only if $\Lambda >4$.

\subsection{Case (vii): nonzero $A_2$, $A_3$, $A_4$, and $A_5$ with $A_1=A_6=0$}
\label{2345}

In this case, Eq. (\ref{CC}) reduces to
\be
A_2+A_3+A_4+A_5=\lambda\,.
\ee
In this case, the sn solution Eq. (\ref{sn}) with $S=1$ is characterized by
 \be\label{3.3cc}
  A_5 \ns(2h\beta,m)=- A_3 \ns(h\beta,m)\,,
 \ee
 \bea\label{3.4cc}
 &&\frac{2m}{A^2h^2}= 2 A_4 \ns(h\beta,m)\ns(2h\beta,m)
- A_3 \cs(h\beta,m)\ds(h\beta,m) \nonumber \\
&& + A_2 \ns^2(h\beta)- A_5
[\cs(2h\beta,m)\ds(2h\beta,m)-\ns^2(2h\beta,m)]\,,
 \eea
 \be\label{3.5cc}
 \frac{(2-\Lambda)m}{A^2h^2}=( A_4 - A_3 )\ns^2(h\beta,m)
 + A_2 \cs(h\beta)\ds(h\beta)\,.
 \ee

In the limit $m=1$, this reduces to the kink solution satisfying
\bea\label{3.11}
 A^2  = 1, \quad \tanh ^2 (h\beta) = - \frac{2A_3  + A_5}{A_5}\,,
 \quad
 A_4  = \frac{2}{{h^2 }} + \frac{\Lambda A_5}{h^2 (2A_3  + A_5)}- A_3- 2A_5\,.
\eea
Note that this solution is valid for any $\Lambda$ including
$\Lambda=2$.

One can also work out the corresponding staggered
$\sn$ solution Eq. (\ref{sn}) with $S=(-1)^n$.
In the limit $m=1$, we obtain
the staggered kink solution Eq. (\ref{tanh}) with
$S=(-1)^n$ satisfying
\be\label{3.29b}
A^2= \frac{(4-\Lambda)[1+\tanh^2(h\beta)]}{4A_3h^2+(2A_2h^2-\Lambda)
[1+\tanh^2(h\beta)]}\,,
\ee
\be\label{3.29c'}
A_2h^2[2(4-\Lambda)-\Lambda
\tanh^2(h\beta)]+\frac{2A_3h^2[2(4-\Lambda)
+(8-3\Lambda)\tanh^2(h\beta)]}{[1+\tanh^2(h\beta)]} =\Lambda
[4-\Lambda+(2-\Lambda)\tanh^2(h\beta)]\,.
\ee

Another exact solution is the $\dn$ solution Eq. (\ref{dn}) with
$S=1$ provided the following relations are satisfied:
 \be\label{3.14cc}
  A_5 \cs(2h\beta,m)=- A_3 \cs(h\beta,m)\,,
 \ee
 \bea\label{3.15cc}
 &&\frac{2}{A^2h^2}= -2 A_4 \cs(h\beta,m)\cs(2h\beta,m)
+ A_3 \ds(h\beta,m)\ns(h\beta,m) \nonumber \\
&&- A_2 \cs^2(h\beta) + A_5
[\ns(2h\beta,m)\ds(2h\beta,m)-\cs^2(2h\beta,m)]\,,
 \eea
 \be\label{3.16cc}
 \frac{2-\Lambda}{A^2h^2}=(A_3-A_4)\cs^2(h\beta,m)-A_2\ns(h\beta)\ds(h\beta)\,.
 \ee

Yet another exact solution is the $\cn$ solution Eq. (\ref{dn}) with $S=1$
satisfying
 \be\label{3.20cc}
  A_5 \ds(2h\beta,m)=-A_3\ds(h\beta,m)\,,
 \ee
 \bea\label{3.21cc}
 &&\frac{2m}{A^2h^2}= -2A_4\ds(h\beta,m)\ds(2h\beta,m)
+A_3\cs(h\beta,m)\ns(h\beta,m) \nonumber \\
&&-A_2\ds^2(h\beta) + A_5
[\ns(2h\beta,m)\cs(2h\beta,m)-\ds^2(2h\beta,m)]\,,
 \eea
 \be\label{3.22cc}
 \frac{(2-\Lambda)m}{A^2h^2}=(A_3-A_4)\ds^2(h\beta,m)-A_2\ns(h\beta)\cs(h\beta)\,.
 \ee

In the limit $m=1$, we obtain the pulse solution Eq. (\ref{sech})
with $S=1$
 \bea\label{3.18cc}
&&\Lambda =-2[\cosh(h\beta)-1]<0\,,~~2A_3=- A_5 \sech(h\beta)\,,~~
(A_4+A_2)h^2=h^2A_3[2\cosh(h\beta)-1]-2[\cosh(h\beta)-1]\,, \nonumber \\
&&A^2=\frac{(|\Lambda|+2)(|\Lambda|+4)}{2(2-2h^2A_3-A_2h^2)}\,.
 \eea
Thus the pulse solution exists only if $\Lambda <0$.

One can also work out the corresponding staggered $\dn$ and $\cn$ solutions.
In the limit $m=1$, we obtain the staggered pulse solution Eq.
(\ref{sech}) with $S=(-1)^n$ with
 \bea\label{3.47}
&&\Lambda =2[1+\cosh(h\beta)]>4\,,~~2A_3= A_5 \sech(h\beta)\,,~~
(A_4+A_2)h^2=-h^2A_3[2\cosh(h\beta)+1]+2[1+\cosh(h\beta)]\,, \nonumber \\
&&A^2=\frac{(\Lambda-2)(\Lambda-4)}{2(2-2h^2A_3-A_2h^2)}\,.
 \eea
Thus the staggered pulse solution exists only if $\Lambda >4$.

Summarizing, it is worth noting that in all the five cases discussed above
(which is also true in the remaining two cases (i) and (ii)), for the kink
solution, the amplitude $A$ is always equal to 1. Further, in all the
seven cases, while the pulse solution exists only if $\Lambda<0$, the
staggered pulse solution exists only if $\Lambda>4$.

It is worth pointing out here that the trigonometric solutions for these
models do not
follow from above JEF solutions in the limit $m=0$, since both left and right
hand sides of the identities for the Jacobi elliptic functions $\sn$ and $\cn$,
which one has used in deriving these solutions, vanish identically in this
limit \cite{kls}. However, trigonometric solutions can be derived for these
models directly, as we show below.

\subsection{Trigonometric solutions}
\label{TrigonometricSolutions}

We shall now show that unlike the JEF and the hyperbolic
solutions, the static TI trigonometric solutions with an arbitrary
shift along the chain $x_0$ exist even when all the six parameters
$A_i$ are nonzero.  Ideally, the trigonometric solutions should be
the $m\rightarrow0$ limit of the JEF solutions.  However, the relevant
JEF identities we have used vanish in this limit and thus the
trigonometric solutions must be derived separately.

We look for a solution of Eq. (\ref{DModel}) of the form
\be \label{zz0}
   \phi_n=SA \sin[h\beta(n+x_0)]\,,
\ee
where $S=1$ or $(-1)^{n}$ depending on whether it is a nonstaggered or a staggered
solution, and find that it exists under the following two conditions
\be\label{zz1}
   \pm (\Lambda-2)+2\cos(h\beta)=h^2 A ^2
   \sin^2(h\beta) [\pm (A_3 - A_4) +(3A_6 -A_5 )\cos(h\beta)]\,,
\ee
\be\label{zz2}
   \pm [A_1+A_4+A_3 \cos(2h\beta)] +(A_2 +A_5 )\cos(h\beta)
   +A_6\cos(h\beta) [4\cos^2(h\beta)-3]=0\,,
\ee
where the upper (lower) sign corresponds to nonstaggered
(staggered) $\sin$e solution. As mentioned before, if $\phi_n(t)$
is a solution to Eq. (\ref{DModel}), then the staggered solution
$(-1)^{n} \phi_n(t)$ is also a solution to Eq. (\ref{DModel}) with
the coefficients of $A_1$, $A_3$ and $A_4$ having opposite signs
and with $2-\Lambda$ replaced by $\Lambda-2$, where
$\Lambda=h^2\lambda$. To make the presentation compact, in most
cases we shall therefore only give results for the nonstaggered
solutions. We shall see that these solutions exist in all the
models discussed in this paper except in the case of Model 1 (only
$A_1$ nonzero) and Model 4 (only $A_4$ nonzero).

Using the well known addition theorem for $\sin$e, it is easily shown that
the $\sin$e solution follows from the two-point quadratic map
\be
\phi_{n+1}^2+\phi_n^2 -2 \phi_{n+1} \phi_n \cos(h\beta)-A^2 \sin^2(h\beta)=0\,.
\ee

We now discuss these solutions in the several special cases.

{\bf Model 1: Only $A_1  \ne 0\,,~ (h^2A_1 =\Lambda$)}

From Eqs. (\ref{zz1}) and (\ref{zz2}) it follows that in this case
there is neither a staggered nor a nonstaggered solution of the form
of Eq. (\ref{zz0}).

{\bf Model 6: Only $A_4  \ne 0\,,~ (h^2A_4 =\Lambda$)}

From Eqs. (\ref{zz1}) and (\ref{zz2}) it follows that in this case
also there is neither a staggered nor a nonstaggered solution of the form
of Eq. (\ref{zz0}).

{\bf Model 3: Only $A_2  \ne 0\,,~ (h^2A_2 =\Lambda$)}

In this case, both staggered and nonstaggered solutions exist provided
\be
\Lambda=2\,,~~h\beta=\pm \pi /2\,.
\ee
and in this case $A$  is arbitrary.

{\bf Only $A_4 ,A_2  \ne 0\,,~ [h^2(A_4 +A_2 )=\Lambda$]}

In this case, the nonstaggered solution with $S=1$ given by Eq.
(\ref{zz0}) exists provided \be A_4 +A_2 \cos(h\beta)=0\,,~~ 2A
^2\Lambda
\cos(h\beta)\cos^2(h\beta/2)=\Lambda-2[1-\cos(h\beta)]\,. \ee

On the other hand, the staggered solution with $S=(-1)^{n}$ exists
in the same model (with only $A_2$ and $A_4$ nonzero) provided \be
A_4 +A_2 \cos(h\beta)=0\,,~~ 2A ^2\Lambda
\cos(h\beta)\sin^2(h\beta/2)=\Lambda-2[1+\cos(h\beta)]\,. \ee

{\bf Only $A_4 ,A_3 ,A_5  \ne 0\,,~ [h^2(A_4 +A_3 +A_5 )=\Lambda$]}

In this case the nonstaggered solution exists provided
\be
A_4 +A_5 \cos(h\beta)+A_3
\cos(2h\beta)=0\,,~~ 2A ^2h^2A_3
\cos^2(h\beta)\sin^2(h\beta)=\Lambda-2[1-\cos(h\beta)]\,. \ee

{\bf Only $A_2 ,A_3 ,A_5  \ne 0\,,~ [h^2(A_2 +A_3 +A_5 )=\Lambda$]}

In this case the nonstaggered solution exists provided \be (A_2 +A_5
)\cos(h\beta)+A_3 \cos(2h\beta)=0\,,~~ A ^2h^2[A_3 -A_5
\cos(h\beta)]\sin^2(h\beta)=\Lambda-2[1-\cos(h\beta)]\,. \ee

{\bf Only $A_3 ,A_5  \ne 0~ [h^2(A_3 +A_5 )=\Lambda$]}

In this case the nonstaggered solution exists provided
\be A_5 \cos(h\beta)+A_3
\cos(2h\beta)=0\,,~~ 4A
^2\Lambda\cos^3(h\beta)\cos^2(\frac{h\beta}{2})
=(\Lambda-2[1-\cos(h\beta)])[1+2\cos(h\beta)]\,. \ee

{\bf Case of $A_1 =A_6 =0, A_2 ,A_3 ,A_4 ,A_5  \ne 0\,,~
[h^2(A_4 +A_2 +A_3 +A_5 )=\Lambda$]}

In this case the nonstaggered solution exists provided \be A_4 +(A_2 +A_5
)\cos(h\beta)+A_3 \cos(2h\beta)=0\,,~~ A ^2h^2[A_3 -A_4 -A_5
\cos(h\beta)]\sin^2(h\beta)=\Lambda-2[1-\cos(h\beta)]\,. \ee

{\bf Model 2 with $\delta=0, \gamma=1/4$: Only $A_5 =A_6  \ne 0\,,~ 2h^2A_5
=\Lambda$}

In this case the nonstaggered solution exists provided \be
\cos^2(h\beta)=\frac{1}{2}\,,~~ A ^2\Lambda= 4\pm
2\sqrt{2}(\Lambda-2)\,. \ee
Thus in this case one has a solution with period 8.

{\bf Model 2 with $\gamma=0, \delta=1/4$: Only $A_1 =A_3/2 =A_4 =
\lambda/2\,\, {\bf are \,\, nonzero}\,, ~ 2h^2A_3 =\Lambda$}

In this case the nonstaggered solution exists provided
\be \label{sine4}
   \sin^2(h\beta)=1\,,~~ A ^2\Lambda=4(\Lambda-2)\,. \ee
Thus in this case one has a solution with period 4.

{\bf Model 2: Case of $A_3 =2A_1 =2A_4 , A_6 =A_5 , A_2 $ all nonzero\,,~
[$h^2(2A_3 +2A_5 +A_2 )=\Lambda$]}

In this case the nonstaggered solution exists provided
\be \cos(h\beta)[A_2 +2A_5
\cos(2h\beta)+2A_3 \cos(h\beta)]=0\,,~~ A ^2h^2[A_3 +4A_5
\cos(h\beta)]\sin^2(h\beta) =2[\Lambda-2(1-\cos(h\beta))]\,. \ee
Thus in this case one has solution with period 4.

{\bf Model 10: Hamiltonian model given by Eq. (\ref{Ham}) with $A_1
=4\alpha_1 \lambda\,,A_2 =6\alpha_2 \lambda\,, A_3 =4\alpha_3
\lambda\,,A_6 =2\alpha_2 \lambda\,,A_4 =A_5 =0$ ~
[$4(\alpha_1+2\alpha_2+\alpha_3)=1$]}

In this case the nonstaggered solution exists provided
\bea &&\label{zz4}
   2A^2\Lambda[2\alpha_2+3\alpha_2\cos(h\beta)]\sin^2(h\beta)
   =[\Lambda-2(1-\cos(h\beta))]\,, \nonumber \\
   &&\alpha_2 \cos^3(h\beta)+\alpha_3
   \cos^2(h\beta)=\alpha_2+\alpha_3-\frac{1}{8}\,.
\eea

{\bf Speight and Ward Model 7: $\alpha_1=\alpha_2
=\frac{2}{3}\alpha_3=\frac{1}{18}$}

Being a special case of the model 10, in this case Eq. (\ref{zz4}) reduces to
\be \label{zz3}
  4\cos^3(h\beta)+6\cos^2(h\beta)-1=0\,,~~
  A ^2\Lambda[1+\cos(h\beta)]\sin^2(h\beta)
  =3[\Lambda-2(1-\cos(h\beta))]\,.
\ee
The roots of the first equation in Eq. (\ref{zz3}) are
$h\beta=2\pi/3$ and $h\beta \approx 1.196$. The first root
corresponds to the three-periodic TI solution with an amplitude
that can be found from the second equation in Eq. (\ref{zz3}) as
$A ^2= 8(\Lambda -3)/ \Lambda$, so that the solution is defined
either when $\Lambda>3$ or when $\Lambda<0$.
The second root corresponds to the
TI solution with the period approximately $5.253$.

Instead, the staggered solution exists in this model provided \be
\label{zz3'}
  4\cos^3(h\beta)-6\cos^2(h\beta)+1=0\,,~~
  A ^2\Lambda[1-\cos(h\beta)]\sin^2(h\beta)
  =3[\Lambda-2(1+\cos(h\beta))]\,.
\ee
The roots of the first equation in Eq. (\ref{zz3'}) are
$h\beta=\pi/3$ and $h\beta \approx 1.944$. The first root
corresponds to the six-periodic TI solution with the amplitude
which is same as for the three-periodic nonstaggered solution, i.e.
$A ^2= 8(\Lambda -3)/ \Lambda$, so that this solution is also defined
either when $\Lambda>3$ or when $\Lambda<0$.
The second root corresponds to the
TI solution with the period approximately $3.023$.

It is worth pointing out that in the case of Model 3 (only $A_2$
nonzero), both the staggered and the nonstaggered solutions are of
period 4, i.e. $h\beta=\pi /4$. The same is also true of Model 2
in case $\delta=1/4,\gamma= 0$, i.e. $A_1=A_4=A_3 /2 =\lambda/4$.
On the other hand, in the case of Model 2 with
$\gamma=1/4,\delta=0$, i.e. $A_5=A_6=\lambda /2$, both the
staggered and the nonstaggered solutions are of period 8. Finally,
in the case of Model 7, i.e. Speight and Ward model with
$A_1=2\lambda /9,A_2=A_3=\lambda /3,A_4=A_5=0, A_6=\lambda/9$, one
of the nonstaggered solution is three-periodic while in the
staggered case, one of the solution is of period 6.

\section{Short-period solutions}
\label{ShortPeriodicSolutions}

We shall now show that apart from the JEF, hyperbolic and trigonometric, there
are also several short period and even aperiodic solutions of Eq.
(\ref{DModel}). In order to obtain these solutions, it is useful to look
at the symmetries of Eq. (\ref{DModel}). In particular,
notice that Eq. (\ref{DModel}) is invariant under $\phi_{n-1}
\rightarrow \phi_{n+1}$ and $\phi_{n+1} \rightarrow \phi_{n-1}$.
Further, Eq. (\ref{DModel}) is also invariant under
$(\phi_{n-1},\phi_{n},\phi_{n+1}) \rightarrow
(-\phi_{n-1},-\phi_{n},-\phi_{n+1})$. A consequence of these two
symmetries is that if $(\phi_{n-1},\phi_{n},\phi_{n+1})$ is a
solution to Eq. (\ref{DModel}) under certain constraints, then
$(-\phi_{n-1},-\phi_{n},-\phi_{n+1})$,
$(\phi_{n+1},\phi_{n},\phi_{n-1})$ and
$(-\phi_{n+1},-\phi_{n},-\phi_{n-1})$ are also solutions of Eq.
(\ref{DModel}) provided the same constraints are satisfied.

We list below several exact solutions to Eq. (\ref{DModel})
satisfying the constraint (\ref{CC}). We shall only
write down the exact solutions to the general model. IN Appendix we
spell ot the short-period solutions that are admitted by many of
the models discuused in this paper.

While obtaining the periodic solutions the following results have
been used which have been derived by using Eqs. (\ref{DModel}) and
(\ref{CC}).

\begin{enumerate}

\item If $\phi_{n-1}=\phi_{n}=\phi_{n+1}=a$ then it follows that
\be\label{xx2.1} a^2=1\,. \ee

\item If $\phi_{n-1}=\phi_{n}=-\phi_{n+1}=a$ then it follows that
\be\label{xx2.2} \Lambda-2=h^2a^2(A_1 +A_3 -A_4 )\,. \ee

\item If $\phi_{n-1}=-\phi_{n}=\phi_{n+1}=a$ then it follows that
\be\label{xx2.3} \Lambda-4=h^2a^2(A_1 -A_2 +A_3 +A_4 -A_5 -A_6
)\,. \ee

\item If $\phi_{n-1}=\phi_{n}=a$, and $\phi_{n+1}=0$ then it follows that
\be\label{xx2.4} 2(\Lambda-1)=h^2a^2(2A_1  +A_2 +A_3 +A_6 )\,. \ee

\item If $\phi_{n-1}=\phi_{n+1}=a$, and $\phi_{n}=0$ then it follows that
\be\label{xx2.5} 2=h^2a^2(A_5 +A_6 )\,. \ee

\item If $\phi_{n-1}=-\phi_{n}=a$, and $\phi_{n+1}=0$ then it follows that
\be\label{xx2.6} 2(\Lambda-3)=h^2a^2(2A_1  -A_2 +A_3 -A_6)\,. \ee

\item If $\phi_{n-1}=a$, and $\phi_{n}=\phi_{n+1}=0$ then it follows that
\be\label{xx2.7} 2=h^2a^2A_6 \,. \ee

\item If $\phi_{n}=a$, and $\phi_{n-1}=\phi_{n+1}=0$ then it follows that
\be\label{xx2.8} \Lambda-2=h^2a^2A_1 \,. \ee

\end{enumerate}

We now discuss the various exact solutions.

(i) {\bf Solution with period 2}: $\phi =(...,a,-a,...)$

This is an exact solution to Eq. (\ref{DModel}) satisfying the
constraint (\ref{CC}) provided \be\label{xx3.5}
\Lambda-4=h^2a^2[A_1 -A_2 +A_3 +A_4 -A_5 -A_6 ]\,. \ee In case
$\Lambda=4$, then one has a solution with $a$ being an arbitrary real
number provided $h^2(A_1 +A_3+A_4  )=h^2(A_2 +A_5 +A_6 )=2$.

(ii) {\bf second solution with period 2}: $\phi= (...,a,0,...)$

This is an exact solution to Eq. (\ref{DModel}) satisfying the
constraint (\ref{CC}) provided \be\label{xx3.13}
\Lambda-2=h^2a^2A_1 \,,~~2=(A_5 +A_6 )h^2a^2\,. \ee One can in
fact generalize this solution and show that even

$\phi=(...,a,0,~ (a,0~ p~ {\rm times}),~-a,0...)$

is an exact solution with period $2p+2~ (p \ge 2)$ provided Eq.
(\ref{xx3.13}) is satisfied.

One can also show that an {\it aperiodic solution} constructed
from the above periodic solution (with period $2p+2$) and with
``$-a,0$" added randomly after any ``0" is an exact solution to Eq.
(\ref{DModel}) provided Eq. (\ref{xx3.13}) is satisfied.

(iii) {\bf third solution with period 2}: $\phi=
(...,a,\frac{1}{a},a,\frac{1}{a},...$)

This is an exact solution to Eq. (\ref{DModel}) satisfying the
constraint (\ref{CC}) provided \be\label{xx3.14} A_1 =A_5 +A_6
=0\,,~~h^2A_2 =\Lambda-2\,,~~h^2(A_4 +A_3 )=2\,, \ee and in this
case $a$ is an arbitrary real number.

(iv) {\bf Solution with Period 3}: $\phi=(...,a,-a,0,...)$

It is easily shown that this is an exact solution to Eq.
(\ref{DModel}) satisfying the constraint (\ref{CC}) provided
\be\label{xx3.23} a^2h^2[2A_1 +A_3  -A_2 -A_6 ]=2(\Lambda-3)\,.
\ee In case $\Lambda=3$, then one has a solution with $a$ being
arbitrary real number provided $2A_1 +A_3  =A_2 +A_6 $.

(v) {\bf second solution with Period 3}: $\phi=(...,a,a,-a,...)$

It is easily shown that this is an exact solution to Eq.
(\ref{DModel}) satisfying the constraint (\ref{CC}) provided
\be\label{xx3.30} a^2h^2(A_1 +A_3 -A_4)=\Lambda-2\,,~~a^2h^2(A_1
+A_3 +A_4 -A_2 -A_5 -A_6)=\Lambda-4\,. \ee

One can in fact generalize this solution and show that even

$\phi=(...,a,-a,~ (a,-a~p~{\rm times}),~-a,...)$

is an exact solution with period $2p+1~ (p \ge 1)$ provided Eq.
(\ref{xx3.30}) is satisfied.

One can also show that an {\it aperiodic solution} with any number
of ``$a$" and ``$-a$" kept at random but with the constraint that at
most two ``$a$" or two ``$-a$" are always together, is an exact
solution to Eq. (\ref{DModel}) provided Eq. (\ref{xx3.30}) is
satisfied.

(vi) {\bf third solution with Period 3}: $\phi=(...,a,a,0,...)$

It is easily shown that this is an exact solution to Eq.
(\ref{DModel}) satisfying the constraint (\ref{CC}) provided
\be\label{xx3.27a} a^2h^2(A_2 +2A_1 +A_3  +A_6
)=2(\Lambda-1)\,,~~h^2a^2(A_5 +A_6 )=2\,. \ee

One can in fact generalize this solution and show that even

$\phi=(...,a,a,0,-a,-a,0, (a,a,0,-a,-a,0~p~{\rm
times}),~a,a,0,...)$

is an exact solution with period $6p+3~ (p \ge 1)$ provided Eq.
(\ref{xx3.27a}) is satisfied.

One can also show that an {\it aperiodic solution} can be
constructed from the above periodic solution with period $6p+3$
with ``$a,a,0$" or ``$-a,-a,0$" added at random between "0" and
``$-a$" or ``0" and ``$a$".

(vii) {\bf Solution with Period 4}: $\phi=(...,a,a,a,0,...)$

It is easily shown that this is an exact solution to Eq.
(\ref{DModel}) satisfying the constraint (\ref{CC}) provided
\be\label{xx3.33} a^2=1\,,~~h^2(A_2 +2A_1 +A_3  +A_6
)=2(\Lambda-1)\,,~~h^2(A_5 +A_6)=2\,. \ee
From here it follows that such a solution is valid provided
$A_2+A_3+2A_4+A_5=0$.

One can in fact generalize this solution and show that even

$\phi=(...,a,a,0,~ (a,a,0~p~{\rm times}),~a,...)$

is an exact solution with period $3p+1~ (p \ge 1)$ provided Eq.
(\ref{xx3.33}) is satisfied.

One can also show that an {\it aperiodic solution} with any number
of ``$a$" and ``0" but with the constraint that at least two ``$a$"
are always together and no two "0" are either nearest or
next-to-nearest neighbours is also an exact solution provided Eq.
(\ref{xx3.33}) is satisfied.

(viii) {\bf second Solution with Period 4}:
$\phi=(...,a,a,a,-a,...)$

It is easily shown that this is an exact solution to Eq.
(\ref{DModel}) satisfying the constraint (\ref{CC}) provided
\be\label{xx3.37} a^2=1\,,~~h^2(A_1 +A_3 -A_4
)=\Lambda-2\,,~~h^2(A_1 +A_3 +A_4 -A_2 -A_5 -A_6)=\Lambda-4\,. \ee
In view of the constraint (\ref{CC}) this implies that such a
solution is valid only if $A_4=0$.

One can in fact generalize this solution and show that even

$\phi=(...,a,a,a,~ (a~p~ {\rm times}),~-a,...)$

is an exact solution with period $p+1~ (p \ge 3)$ provided Eq.
(\ref{xx3.37}) is satisfied.

One can also show that an {\it aperiodic solution} with any number
of ``$a$" and ``$-a$" but with the constraint that at least once
three or more ``$a$" or ``$-a$" are together.

(ix) {\bf third solution with Period 4}:
$\phi=(...,a,b,-a,-b,...)$, where $a^2 \ne b^2$.

It is easily shown that this is an exact solution to Eq.
(\ref{DModel}) satisfying the constraint (\ref{CC}) provided
\be\label{xx3.41} (a^2+b^2)h^2A_1 =\Lambda-2\,,~~A_3 =A_1 +A_4
\,,~~a^2 \ne b^2\,,~~A_1  \ne 0\,. \ee Thus one has a one
parameter family of solutions. It is easily seen that such a solution
will exist in  model 2 (in case $\delta$ is arbitrary but nonzero,
while $\gamma$ is
arbitrary), model 8 (in case $\gamma=0,\alpha=1/4$) and Hamiltonian model
10 (in case $\alpha_1=\alpha_3$).

In the special case of $A_3 =A_4$, $A_1 =0$ and $\Lambda=2$, one,
in fact, has a two parameter family of solutions in the sense that
now both $a$ and $b$ are arbitrary real numbers.

(x) {\bf fourth solution with Period 4}:
$\phi=(...,a,a,-a,-a,...)$

It is easily shown that this is an exact solution to Eq.
(\ref{DModel}) satisfying the constraint (\ref{CC}) provided
\be\label{xx3.43} a^2h^2[A_1 +A_3 -A_4 ]=\Lambda-2\,. \ee In case
$\Lambda=2$, then one has a solution with $a$ being an arbitrary real
number provided $A_1 +A_3 =A_4 $.

(xi) {\bf fifth solution with Period 4}: $\phi=(...,a,0,-a,0,...)$

It is easily shown that this is an exact solution to Eq.
(\ref{DModel}) satisfying the constraint (\ref{CC}) provided
\be\label{xx3.48} a^2h^2A_1 =\Lambda-2\,. \ee In case $\Lambda=2$,
then one has a solution with $a$ being an arbitrary real number
provided $A_1 =0$.

(xii) {\bf sixth solution with Period 4}:
$\phi=(...,a,0,-a,a,...)$

It is easily shown that this is an exact solution to Eq.
(\ref{DModel}) satisfying the constraint (\ref{CC}) provided
\be\label{xx3.49} a^2h^2[A_1 +A_3 -A_4 ]=\Lambda-2\,,~~a^2h^2[2A_1
+A_3  +A_2 +A_6 ]=2(\Lambda-1)\,,~~ a^2h^2[2A_1 +A_3  -A_2 -A_6
]=2(\Lambda-3)\,. \ee
From here it follows that such a solution is valid provided
$A_3=2A_4$.

One can in fact generalize this solution and show that even

$\phi=(...,a,0,-a,~ (a,0,-a ~p~ {\rm times}),~a,...)$

is an exact solution with period $3p+1~ (p \ge 1)$ provided Eq.
(\ref{xx3.49}) is satisfied.

One can also show that an {\it aperiodic solution} with any number
of ``$a$" and ``$-a$" but with the constraint that at most two ``$a$"
or two ``$-a$" are always together and further ``0" are added at
random between ``$-a$" and ``$a$" or between ``$a$" and ``$-a$" but
with the proviso that no two ``0" are ever nearest or
next-to-nearest neighbours, is an exact solution to Eq.
(\ref{DModel}) satisfying the constraint (\ref{CC}) provided Eq.
(\ref{xx3.49}) is satisfied.

(xiii) {\bf seventh solution with Period 4}:
$\phi=(...,a,-a,0,a,...)$

It is easily shown that this is an exact solution to Eq.
(\ref{DModel}) satisfying the constraint (\ref{CC}) provided
\be\label{xx3.50} a^2h^2[A_1 +A_3 +A_4 -A_2 -A_5 -A_6
]=\Lambda-4\,,~~a^2h^2[2A_1 +A_3  -A_2 -A_6 ]=2(\Lambda-3)\,,~~
a^2h^2[A_6 +A_5 ]=2\,. \ee
From here it follows that such a solution is valid provided
$A_2+A_5=A_3+2A_4$.

One can in fact generalize this solution and show that even

$\phi=(...,a,-a,0,~ (a,-a,0, ~p~ {\rm times}),~-a,...)$

is an exact solution with period $3p+1~ (p \ge 1)$ provided Eq.
(\ref{xx3.50}) is satisfied.

One can also show that an {\it aperiodic solution} constructed
from the above periodic solution (with period $3p+1$) by adding at
random ``$a,-a$" between any 0 and ``$a$" is an exact solution to
Eq. (\ref{DModel}) satisfying the constraint (\ref{CC}) provided
Eq. (\ref{xx3.49}) is satisfied.

(xiv) {\bf Solution with Period 5}: $\phi=(...,a,a,a,-a,-a,...)$

It is easily shown that this is an exact solution to Eq.
(\ref{DModel}) satisfying the constraint (\ref{CC}) provided
\be\label{xx3.55} a^2=1\,,~~h^2[A_1 +A_3 -A_4 ]=\Lambda-2\,. \ee

One can in fact generalize this solution and show that even

$\phi=(...,a,a,a,~ (a ~p~ {\rm times}),~-a,-a, (-a~ q~ {\rm
times}),...)$

is an exact solution with period $p+q~ (p \ge 3, q \ge 2~or~ p \ge
2, q \ge 3)$
 provided Eq. (\ref{xx3.55}) is satisfied.

One can also show that an {\it aperiodic solution} constructed
from any numbers of ``$a$" and ``$-a$" with the constraint that two
or more of ``$a$" as well as two or more of ``$-a$" are always
together is an exact solution to Eq. (\ref{DModel}) satisfying the
constraint (\ref{CC}) provided Eq. (\ref{xx3.55}) is satisfied.

(xv) {\bf second solution with period 5}:
$\phi=(...,0,a,0,a,a,...)$

It is easily shown that this is an exact solution to Eq.
(\ref{DModel}) satisfying the constraint (\ref{CC}) provided
\be\label{xx3.74} h^2a^2[2A_1 +A_3  +A_2 +A_6
]=2(\Lambda-1)\,,~~h^2a^2(A_6 +A_5 )=2\,,~~h^2a^2A_1 =\Lambda-2\,.
\ee
From here it follows that such a solution is valid provided
$A_2+A_3=A_5$.

One can in fact generalize this solution and show that even

$\phi=(...0,a,~ (0,a ~p~ {\rm times}),~a,...)$

is an exact solution with period $2p+1~ (p \ge 2)$ provided Eq.
(\ref{xx3.74}) is satisfied.

One can also show that an {\it aperiodic solution} constructed
from the above periodic solution (with period $2p+1$) by adding at
random ``$-a,-a,0$" after any ``0" is an exact solution to Eq.
(\ref{DModel}) satisfying the constraint (\ref{CC}) provided Eq.
(\ref{xx3.74}) is satisfied.

(xvi) {\bf third solution with period 5}: $\phi
=(...,a,-a,a,-a,0,...)$

This is an exact solution to Eq. (\ref{DModel}) satisfying the
constraint (\ref{CC}) provided \be\label{xx3.75}
\Lambda-4=h^2a^2(A_4 +A_1 +A_3 -A_2 -A_5 -A_6
)\,,~~2(\Lambda-3)=h^2a^2(2A_1 +A_3  -A_2 -A_6)\,. \ee

One can in fact generalize this solution and show that even

$\phi=(...a,-a,~ (a,-a ~p~ {\rm times}),~ 0,...)$

is an exact solution with period $2p+1~ (p \ge 2)$ provided Eq.
(\ref{xx3.75}) is satisfied.

One can also show that an {\it aperiodic solution} constructed
from the above periodic solution (with period $2p+1$) by adding at
random ``0" between ``$-a$" and ``$a$" or ``$a$" and ``$-a$", is an
exact solution to Eq. (\ref{DModel}) satisfying the constraint
(\ref{CC}) provided Eq. (\ref{xx3.75}) is satisfied.

(xvii) {\bf fourth solution with period 5}: $\phi
=(...,a,0,a,0,-a,...)$

This is an exact solution to Eq. (\ref{DModel}) satisfying the
constraint (\ref{CC}) provided \be\label{xx3.76}
\Lambda-2=h^2a^2A_1 \,,~~2=h^2a^2(A_5 +A_6
)\,,~~2(\Lambda-3)=h^2a^2(2A_1 +A_3  -A_2 -A_6)\,. \ee
From here it follows that such a solution is valid provided
$A_2=A_3+A_5$.

One can in fact generalize this solution and show that even

$\phi =(...,a,0,~ (a,0~p~ {\rm times}),~-a,...)$

is an exact solution with period $2p+1~ (p \ge 2)$ provided Eq.
(\ref{xx3.76}) is satisfied.

One can also show that an {\it aperiodic solution} constructed
from the above periodic solution (with period $2p+1$) by adding
randomly ``$-a$" between ``0" and ``$a$" with a constraint that no
two ``$-a$" can be either nearest or next-to-nearest neighbors, is
an exact solution to Eq. (\ref{DModel}) satisfying the constraint
(\ref{CC}) provided Eq. (\ref{xx3.76}) is satisfied.

(xviii) {\bf fifth solution with period 5}: $\phi
=(...,a,a,-a,-a,0,...)$

This is an exact solution to Eq. (\ref{DModel}) satisfying the
constraint (\ref{CC}) provided \be\label{xx3.77}
\Lambda-2=h^2a^2(A_1 +A_3 -A_4)\,,~~2(\Lambda-1)=h^2a^2(2A_1
+A_3  +A_2 +A_6 )\,. \ee

One can in fact generalize this solution and show that even

$\phi =(...,a,a,-a,-a,~ (a,a,-a,-a~ p~{\rm times}),~0,...)$

is an exact solution with period $4p+1~ (p \ge 1)$ provided Eq.
(\ref{xx3.77}) is satisfied.

One can also show that an {\it aperiodic solution} constructed
from the above periodic solution (with period $4p+1$) by randomly
adding ``0" between ``$-a$" and ``$a$" or ``$a$" and ``$-a$" with a
constraint that no two ``0" can be either nearest or
next-to-nearest neighbors, is an exact solution to Eq.
(\ref{DModel}) satisfying the constraint (\ref{CC}) provided Eq.
(\ref{xx3.77}) is satisfied.

(xix) {\bf sixth solution with period 5}: $\phi
=(...,a,a,0,-a,0,...)$

This is an exact solution to Eq. (\ref{DModel}) satisfying the
constraint (\ref{CC}) provided \be\label{xx3.77a} \Lambda-2=h^2a^2
A_1\,,~~2(\Lambda-1)=h^2a^2(2A_1 +A_3 +A_2 +A_6 )\,. \ee

One can in fact generalize this solution and show that even

$\phi =(...,a,a,-a,-a,~ (a,a,-a,-a~ p~{\rm times}),~0,...)$

is an exact solution with period $4p+1~ (p \ge 1)$ provided Eq.
(\ref{xx3.77a}) is satisfied.

One can also show that an {\it aperiodic solution} constructed
from the above periodic solution (with period $4p+1$) by randomly
adding ``0" between ``$-a$" and ``$a$" or ``$a$" and ``$-a$" with a
constraint that no two ``0" can be either nearest or
next-to-nearest neighbors, is an exact solution to Eq.
(\ref{DModel}) satisfying the constraint (\ref{CC}) provided Eq.
(\ref{xx3.77a}) is satisfied.

(xx) {\bf seventh solution with period 5}: $\phi
=(...,a,-a,0,-a,a,...)$

This is an exact solution to Eq. (\ref{DModel}) satisfying the
constraint (\ref{CC}) provided \be\label{xx3.78}
\Lambda-2=h^2a^2(A_1 +A_3 -A_4)\,,~~2=h^2a^2(A_5 +A_6
)\,,~~2(\Lambda-3)=h^2a^2(2A_1 +A_3  -A_2 -A_6)\,. \ee
From here it follows that such a solution is valid provided
$A_2+A_3=2A_4+A_5$.

One can in fact generalize this solution and show that even

$\phi =(...,a,-a,0,~ (a,-a,0~ p~{\rm times}),~-a,a,...)$

is an exact solution with period $3p+2~ (p \ge 1)$ provided Eq.
(\ref{xx3.78}) is satisfied.

One can also show that an {\it aperiodic solution} constructed
from the above periodic solution (with period $3p+2$) by randomly
adding ``$-a,a$" between ``0" and ``$a$" is an exact solution to Eq.
(\ref{DModel}) satisfying the constraint (\ref{CC}) provided Eq.
(\ref{xx3.78}) is satisfied.

(xxi) {\bf eighth solution with period 5}: $\phi
=(...,a,0,-a,a,a,...)$

This is an exact solution to Eq. (\ref{DModel}) satisfying the
constraint (\ref{CC}) provided \be\label{xx3.79}
a^2=1\,,~~\Lambda-2=h^2(A_1 +A_3 -A_4 )\,, ~~2(\Lambda-3)=h^2(2A_1
+A_3  -A_2 -A_6 )\,, ~~2(\Lambda-1)=h^2(2A_1 +A_3  +A_2 +A_6 )\,.
\ee In view of the constraint (\ref{CC}), such a solution is thus
valid only if $A_3=2A_4=-A_5$.

One can in fact generalize this solution and show that even

$\phi =(...,a,0,-a,~ (a,0,-a~ p~{\rm times}),~a,a,...)$

is an exact solution with period $3p+2~ (p \ge 1)$ provided Eq.
(\ref{xx3.79}) is satisfied.

One can also show that an {\it aperiodic solution} constructed
from the above periodic solution (with period $3p+2$) by randomly
adding ``$a,a$" between ``$-a$" and ``$a$" is an exact solution to
Eq. (\ref{DModel}) satisfying the constraint (\ref{CC}) provided
Eq. (\ref{xx3.79}) is satisfied.

(xxii) {\bf ninth solution with period 5}: $\phi
=(...,a,a,0,a,-a,...)$

This is an exact solution to Eq. (\ref{DModel}) satisfying the
constraint (\ref{CC}) provided \bea\label{xx3.80} &&2=h^2a^2(A_5
+A_6 )\,,~~\Lambda-2=h^2a^2(A_1 +A_3 -A_4 )\,,
~~2(\Lambda-3)=h^2a^2(2A_1 +A_3  -A_2 -A_6 )\,, \nonumber \\
&&2(\Lambda-1)=h^2a^2(2A_1 +A_3  +A_2 +A_6 )\,,
~~\Lambda-4=h^2a^2(A_1 +A_3 +A_4 -A_5 -A_6 -A_2 )\,. \eea
From here it follows that such a solution is valid provided
$A_2=A_3=2A_4=A_5$.

One can in fact generalize this solution and show that even

$\phi =(...,a,a,0,~ (a,a,0~ p~{\rm times}),~a,-a,...)$

is an exact solution with period $3p+2~ (p \ge 1)$ provided Eq.
(\ref{xx3.80}) is satisfied.

One can also show that an {\it aperiodic solution} constructed
from the above periodic solution (with period $3p+2$) by randomly
adding "$-a$" between two "$a$" is an exact solution to Eq.
(\ref{DModel}) satisfying the constraint (\ref{CC}) provided Eq.
(\ref{xx3.80}) is satisfied.

(xxiii) {\bf Solution with period 6}: $\phi
=(...,a,a,0,-a,-a,0,...)$

This is an exact solution to Eq. (\ref{DModel}) satisfying the
constraint (\ref{CC}) provided \be\label{xx3.81}
2(\Lambda-1)=h^2a^2(2A_1 +A_3  +A_2 +A_6 )\,. \ee In case
$\Lambda=1$, then one has a solution with $a$ being an arbitrary real
number provided $2A_1 +A_3  +A_2 +A_6 =0$.

(xxiv) {\bf second solution with period 6}: $\phi
=(...,a,-a,0,-a,a,0,...)$

This is an exact solution to Eq. (\ref{DModel}) satisfying the
constraint (\ref{CC}) provided \be\label{xx3.82}
2(\Lambda-3)=h^2a^2(2A_1 +A_3  -A_2 -A_6 )\,,~~2=h^2a^2(A_5 +A_6
)\,. \ee

One can in fact generalize this solution and show that even

$\phi =(...,a,-a,0,~ (a,-a,0~ p~{\rm times}),~-a,a,0,...)$

is an exact solution with period $3p+3~ (p \ge 1)$ provided Eq.
(\ref{xx3.82}) is satisfied.

One can also show that an {\it aperiodic solution} constructed
from the above periodic solution (with period $3p+3$) by randomly
adding ``$-a,a,0$" between  ``0" and ``$a$" is an exact solution to
Eq. (\ref{DModel}) satisfying the constraint (\ref{CC}) provided
Eq. (\ref{xx3.82}) is satisfied.

(xxv) {\bf third solution with period 6}: $\phi
=(...,a,a,a,0,-a,-a,...)$

This is an exact solution to Eq. (\ref{DModel}) satisfying the
constraint (\ref{CC}) provided \be\label{xx3.83}
a^2=1\,,~~2(\Lambda-1)=h^2(2A_1 +A_3  +A_2 +A_6
)\,,~~\Lambda-2=h^2(A_1 +A_3 -A_4 )\,. \ee In view of the
constraint (\ref{CC}) this implies that such a solution is valid
only if $A_3=-A_5$.

One can in fact generalize this solution and show that even

$\phi =(...,a,a,a,~ (a~ p~{\rm times}),~0,-a,-a,...)$

is an exact solution with period $p+3~ (p \ge 3)$ provided Eq.
(\ref{xx3.83}) is satisfied.

One can also show that an {\it aperiodic solution} constructed
from the above periodic solution (with period $p+3$) by randomly
adding ``$0,-a,-a$" between any two ``$a$" is an exact solution to
Eq. (\ref{DModel}) satisfying the constraint (\ref{CC}) provided
Eq. (\ref{xx3.83}) is satisfied.

(xxvi) {\bf fourth solution with period 6}: $\phi
=(...,a,a,0,-a,0,a,...)$

This is an exact solution to Eq. (\ref{DModel}) satisfying the
constraint (\ref{CC}) provided \be\label{xx3.84}
a^2=1\,,~~2(\Lambda-1)=h^2(2A_1 +A_3  +A_2 +A_6
)\,,~~\Lambda-2=h^2 A_1 \,. \ee In view of the constraint
(\ref{CC}) this implies that such a solution is valid only if
$A_4=-A_5$.

One can in fact generalize this solution and show that even

$\phi =(...,a,a,0,-a,0,~ (a,a,0,-a,0~ p~{\rm times}),~a,...)$

is an exact solution with period $5p+1~ (p \ge 1)$ provided Eq.
(\ref{xx3.84}) is satisfied.

One can also show that an {\it aperiodic solution} constructed
from the above periodic solution (with period $5p+1$) by randomly
adding as many ``$a$" as one wants between any two ``$a$" is an
exact solution to Eq. (\ref{DModel}) satisfying the constraint
(\ref{CC}) provided Eq. (\ref{xx3.84}) is satisfied.

(xxvii) {\bf fifth solution with period 6}: $\phi
=(...,a,-a,a,a,-a,0,...)$

This is an exact solution to Eq. (\ref{DModel}) satisfying the
constraint (\ref{CC}) provided \be\label{xx3.85}
2(\Lambda-3)=h^2a^2(2A_1 +A_3  -A_2 -A_6 )\,, ~~\Lambda-2=h^2a^2
(A_1 +A_3 -A_4 )\,, ~~\Lambda-4=h^2a^2 (A_1 +A_3 +A_4 -A_2 -A_5
-A_6 )\,. \ee
From here it follows that such a solution is valid provided
$A_3=A_5$.

One can in fact generalize this solution and show that even

$\phi =(...,a,-a,a,a,-a,~ (a,-a,a,a,-a~ p~{\rm times}),~0,...)$

is an exact solution with period $5p+1~ (p \ge 1)$ provided Eq.
(\ref{xx3.85}) is satisfied.

One can also show that an {\it aperiodic solution} constructed
from the above periodic solution (with period $5p+1$) by randomly
adding ``$-a$" between any two ``$a$" is an exact solution to Eq.
(\ref{DModel}) satisfying the constraint (\ref{CC}) provided Eq.
(\ref{xx3.85}) is satisfied.

(xxviii) {\bf sixth solution with period 6}: $\phi
=(...,a,a,0,a,-a,0,...)$

This is an exact solution to Eq. (\ref{DModel}) satisfying the
constraint (\ref{CC}) provided \be\label{xx3.86}
2(\Lambda-1)=h^2a^2(2A_1 +A_3  +A_2 +A_6 )\,,
~~2(\Lambda-3)=h^2a^2(2A_1 +A_3  -A_2 -A_6 )\,, ~~2=h^2a^2 (A_5
+A_6 )\,. \ee
From here it follows that such a solution is valid provided
$A_2=A_5$.

One can in fact generalize this solution and show that even

$\phi =(...,a,a,0,~ (a,a,0~ p~{\rm times}),~a,-a,0,...)$

is an exact solution with period $3p+3~ (p \ge 1)$ provided Eq.
(\ref{xx3.86}) is satisfied.

One can also show that an {\it aperiodic solution} constructed
from the above periodic solution (with period $3p+3$) by randomly
adding ``$a,-a,0$" between ``0" and  ``$a$" is an exact solution to
Eq. (\ref{DModel}) satisfying the constraint (\ref{CC}) provided
Eq. (\ref{xx3.86}) is satisfied.

(xxix) {\bf seventh solution with period 6}: $\phi
=(...,a,a,-a,0,-a,a,...)$

This is an exact solution to Eq. (\ref{DModel}) satisfying the
constraint (\ref{CC}) provided \be\label{xx3.87}
a^2=1\,,~~(\Lambda-2)=h^2(A_1 +A_3 -A_4 )\,,
~~2(\Lambda-3)=h^2(2A_1 +A_3  -A_2 -A_6 )\,, ~~2=h^2 (A_5 +A_6
)\,. \ee In view of the constraint (\ref{CC}) this implies that
such a solution is valid only if $A_2=-2A_4,A_5=2A_2+A_3$.

One can in fact generalize this solution and show that even

$\phi =(...,a,a,-a,0,-a,~ (a,a,-a,0,-a~ p~{\rm times}),~a,...)$

is an exact solution with period $5p+1~ (p \ge 1)$ provided Eq.
(\ref{xx3.87}) is satisfied.

One can also show that an {\it aperiodic solution} constructed
from the above periodic solution (with period $5p+1$) by randomly
adding as many ``$a$" as one wants between two ``$a$", is an exact solution
to Eq. (\ref{DModel}) satisfying the constraint (\ref{CC})
provided Eq. (\ref{xx3.87}) is satisfied.

(xxx) {\bf eighth solution with period 6}: $\phi
=(...,a,0,a,0,a,-a,...)$

This is an exact solution to Eq. (\ref{DModel}) satisfying the
constraint (\ref{CC}) provided \bea\label{xx3.88}
&&\Lambda-2=h^2a^2 A_1 \,,~~(\Lambda-4)=h^2a^2(A_1 +A_3 +A_4 -A_5
-A_6 -A_2 )\,,
\nonumber \\
&&2(\Lambda-3)=h^2a^2(2A_1 +A_3  -A_2 -A_6 )\,, ~~2=h^2a^2 (A_5
+A_6 )\,. \eea
From here it follows that such a solution is valid provided
$A_4=A_5\,,A_2=A_3+A_4$.

One can in fact generalize this solution and show that even

$\phi =(...,a,0,a,0,~ (a,0,~ p~{\rm times}),~a,-a,...)$

is an exact solution with period $2p+2~ (p \ge 2)$ provided Eq.
(\ref{xx3.88}) is satisfied.

One can also show that an {\it aperiodic solution} constructed
from the above periodic solution (with period $2p+2$) by randomly
adding ``$-a$" between ``0" and ``$a$", is an exact solution to Eq.
(\ref{DModel}) satisfying the constraint (\ref{CC}) provided Eq.
(\ref{xx3.88}) is satisfied.

(xxxi) {\bf ninth solution with period 6}: $\phi
=(...,a,0,a,0,a,a,...)$

This is an exact solution to Eq. (\ref{DModel}) satisfying the
constraint (\ref{CC}) provided \be\label{xx3.89} \Lambda-2=h^2 A_1
\,,~~(\Lambda-2)=h^2(A_1 +A_3 -A_4 )\,, ~a^2=1\,,~~2=h^2 (A_5 +A_6
)\,. \ee In view of the constraint (\ref{CC}) this implies that
such a solution is valid only if $A_2=-2A_3=-2A_4$.

One can in fact generalize this solution and show that even

$\phi =(...,a,0,a,0,~ (a,0,~ p~{\rm times}),~a,a,...)$

is an exact solution with period $2p+2~ (p \ge 2)$ provided Eq.
(\ref{xx3.89}) is satisfied.
From here it follows that such a solution is valid provided
$A_2=-2A_3=-2A_4$.

One can also show that an {\it aperiodic solution} constructed
from the above periodic solution (with period $2p+2$) by randomly
adding ``$a,a$" between ``0" and ``$a$" or ``$a$" and ``0", is an exact
solution to Eq. (\ref{DModel}) satisfying the constraint
(\ref{CC}) provided Eq. (\ref{xx3.89}) is satisfied.

(xxxii) {\bf tenth solution with period 6}: $\phi
=(...,a,-a,a,a,0,a,...)$

This is an exact solution to Eq. (\ref{DModel}) satisfying the
constraint (\ref{CC}) provided \bea\label{xx3.90}
&&\Lambda-4=h^2a^2(A_3 +A_4 +A_1 -A_2 -A_5 -A_6
)\,,~~(\Lambda-2)=h^2a^2(A_1 +A_3 -A_4 )\,,
\nonumber \\
&&2(\Lambda-1)=h^2a^2(2A_1 +A_3  +A_2 +A_6 )\,,~~2=h^2a^2 (A_5
+A_6 )\,. \eea
From here it follows that such a solution is valid provided
$A_3+A_5=2A_2=4A_4$.

One can in fact generalize this solution and show that even

$\phi =(...,a,-a,a,~ (a,-a,a,~ p~{\rm times}),~a,0,a,...)$

is an exact solution with period $3p+3~ (p \ge 1)$ provided Eq.
(\ref{xx3.90}) is satisfied.

One can also show that an {\it aperiodic solution} constructed
from the above periodic solution (with period $3p+3$) by randomly
adding ``$-a$" between two ``$a$", is an exact solution to Eq.
(\ref{DModel}) satisfying the constraint (\ref{CC}) provided Eq.
(\ref{xx3.90}) is satisfied.

(xxxiii) {\bf eleventh solution with period 6}: $\phi
=(...,a,0,a,0,-a,-a,...)$

This is an exact solution to Eq. (\ref{DModel}) satisfying the
constraint (\ref{CC}) provided \bea\label{xx3.91}
&&2(\Lambda-3)=h^2a^2(2A_1 +A_3  +A_2 +A_6
)\,,~~(\Lambda-2)=h^2a^2(A_1 +A_3 -A_4 )\,,
\nonumber \\
&&2(\Lambda-1)=h^2a^2(2A_1 +A_3  +A_2 +A_6 )\,,~~2=h^2a^2 (A_5
+A_6 )\,,~~\Lambda-2=h^2a^2 A_1 \,. \eea
From here it follows that such a solution is valid provided
$A_2=A_5\,,=A_3=A_4=0$.

One can in fact generalize this solution and show that even

$\phi =(...,a,0,~ (a,0,~ p~{\rm times}),~-a,-a,...)$

is an exact solution with period $2p+2~ (p \ge 2)$ provided Eq.
(\ref{xx3.91}) is satisfied.

One can also show that an {\it aperiodic solution} constructed
from the above periodic solution (with period $2p+2$) by randomly
adding ``$-a$" between ``0" and  ``$a$" or ``$a$" and ``0", is an exact
solution to Eq. (\ref{DModel}) satisfying the constraint
(\ref{CC}) provided Eq. (\ref{xx3.91}) is satisfied.

(xxxiv) {\bf Solution with period 7}: $\phi
=(...,a,a,0,-a,-a,0,a,...)$

This is an exact solution to Eq. (\ref{DModel}) satisfying the
constraint (\ref{CC}) provided \be\label{xx3.92}
2(\Lambda-1)=h^2(2A_1 +A_3  +A_2 +A_6 )\,,~~a^2=1\,. \ee

One can in fact generalize this solution and show that even

$\phi =(...,a,a,0,-a,-a,0,~ (a,a,0,-a,-a,0,~ p~{\rm
times}),~a,...)$

is an exact solution with period $6p+1~ (p \ge 1)$ provided Eq.
(\ref{xx3.92}) is satisfied.

One can also show that an {\it aperiodic solution} constructed
from the above periodic solution (with period $6p+1$) by randomly
adding as many number of ``$a$" (``$-a$") as one wants between two
``$a$" (``$-a$"), is an exact solution to Eq. (\ref{DModel})
satisfying the constraint (\ref{CC}) provided Eq. (\ref{xx3.92})
is satisfied.

(xxxv) {\bf second solution with period 7}: $\phi
=(...,a,0,-a,a,0,-a,0,...)$

This is an exact solution to Eq. (\ref{DModel}) satisfying the
constraint (\ref{CC}) provided \be\label{xx3.93}
2(\Lambda-3)=h^2a^2(2A_1 +A_3  -A_2 -A_6 )\,,~~\Lambda-2=h^2a^2
A_1 \,. \ee

One can in fact generalize this solution and show that even

$\phi =(...,a,0,-a,a,o,-a,~ (a,0,-a,~ p~{\rm times}),~0,...)$

is an exact solution with period $3p+1~ (p \ge 2)$ provided Eq.
(\ref{xx3.93}) is satisfied.

One can also show that an {\it aperiodic solution} constructed
from the above periodic solution (with period $3p+1$) by randomly
adding ``0" between ``$-a$" and ``$a$" or ``$a$" and ``$-a$", is an
exact solution to Eq. (\ref{DModel}) satisfying the constraint
(\ref{CC}) provided Eq. (\ref{xx3.93}) is satisfied.

(xxxvi) {\bf third solution with period 7}: $\phi
=(...,a,a,0,-a,-a,a,-a,...)$

This is an exact solution to Eq. (\ref{DModel}) satisfying the
constraint (\ref{CC}) provided \be\label{xx3.94}
2(\Lambda-1)=h^2a^2(2A_1 +A_3  +A_2 +A_6 )\,,~~\Lambda-2=h^2a^2
A_1 \,,~~ \Lambda-4=h^2a^2(A_3 +A_4 +A_1 -A_2 -A_5 -A_6 )\,. \ee
From here it follows that such a solution is valid provided
$2A_3+A_4=A_5$.

One can in fact generalize this solution and show that even

$\phi =(...,a,a,0,-a,-a~ (a,a,0,-a,-a,~ p~{\rm times}),~a,-a,...)$

is an exact solution with period $5p+2~ (p \ge 1)$ provided Eq.
(\ref{xx3.94}) is satisfied.

One can also show that an {\it aperiodic solution} constructed
from the above periodic solution (with period $5p+2$) by randomly
adding ``$-a,a$" between ``0" and ``$-a$" or ``$a$" and ``0", is an
exact solution to Eq. (\ref{DModel}) satisfying the constraint
(\ref{CC}) provided Eq. (\ref{xx3.94}) is satisfied.

(xxxvii) {\bf fourth solution with period 7}: $\phi
=(...,a,-a,0,a,a,0,-a,...)$

This is an exact solution to Eq. (\ref{DModel}) satisfying the
constraint (\ref{CC}) provided \be\label{xx3.95}
2(\Lambda-1)=h^2a^2(2A_1 +A_3  +A_2 +A_6 )\,,~~
2(\Lambda-3)=h^2a^2(2A_1 +A_3  -A_2 -A_6 )\,,~~
\Lambda-4=h^2a^2(A_3 +A_4 +A_1 -A_2 -A_5 -A_6 )\,. \ee
From here it follows that such a solution is valid provided
$A_3+2A_4=2A_5$.

One can in fact generalize this solution and show that even

$\phi =(...,a,-a,0,~ (a,-a,0,~ p~{\rm times}),~a,a,0,-a,...)$

is an exact solution with period $3p+4~ (p \ge 1)$ provided Eq.
(\ref{xx3.95}) is satisfied.

One can also show that an {\it aperiodic solution} constructed
from the above periodic solution (with period $3p+4$) by randomly
adding ``$a,-a$" between ``0" and ``$a$" or ``$-a$" and ``0", is an
exact solution to Eq. (\ref{DModel}) satisfying the constraint
(\ref{CC}) provided Eq. (\ref{xx3.95}) is satisfied.

(xxxviii) {\bf fifth solution with period 7}: $\phi
=(...,a,a,0,a,a,0,-a,...)$

This is an exact solution to Eq. (\ref{DModel}) satisfying the
constraint (\ref{CC}) provided \bea\label{xx3.96}
&&2(\Lambda-1)=h^2a^2(2A_1 +A_3  +A_2 +A_6 )\,,~~
2(\Lambda-3)=h^2a^2(2A_1 +A_3  -A_2 -A_6 )\,,
\nonumber \\
&&\Lambda-2=h^2a^2(A_3 +A_1 -A_4 )\,,~~2=h^2a^2(A_5 +A_6 )\,. \eea
From here it follows that such a solution is valid provided
$A_3=2A_4\,,A_2=A_5$.

One can in fact generalize this solution and show that even

$\phi =(...,a,a,0,~ (a,a,0,~ p~{\rm times}),~-a,...)$

is an exact solution with period $3p+1~ (p \ge 2)$ provided Eq.
(\ref{xx3.96}) is satisfied.

One can also show that an {\it aperiodic solution} constructed
from the above periodic solution (with period $3p+1$) by randomly
adding ``$-a$" between ``0" and ``$a$" or ``$a$" and ``0", is an exact
solution to Eq. (\ref{DModel}) satisfying the constraint
(\ref{CC}) provided Eq. (\ref{xx3.96}) is satisfied.

(xxxix) {\bf sixth solution with period 7}: $\phi
=(...,a,a,-a,a,a,-a,0,...)$

This is an exact solution to Eq. (\ref{DModel}) satisfying the
constraint (\ref{CC}) provided \bea\label{xx3.97}
&&2(\Lambda-1)=h^2a^2(2A_1 +A_3  +A_2 +A_6)\,,~~
2(\Lambda-3)=h^2a^2(2A_1 +A_3  -A_2 -A_6)\,, \nonumber \\
&&\Lambda-2=h^2a^2(A_3 +A_1 -A_4)\,, ~~\Lambda-4=h^2a^2(A_3 +A_4
+A_1 -A_2 -A_5 -A_6 )\,. \eea
From here it follows that such a solution is valid provided
$A_3=2A_4=A_5$.

One can in fact generalize this solution and show that even

$\phi =(...,a,a,-a,~ (a,a,-a,~ p~{\rm times}),~0,...)$

is an exact solution with period $3p+1~ (p \ge 2)$ provided Eq.
(\ref{xx3.97}) is satisfied.

One can also show that an {\it aperiodic solution} constructed
from the above periodic solution (with period $3p+1$) by randomly
adding ``0" between ``$-a$" and ``$a$" or ``$a$" and ``$-a$" such that
no two ``0" are ever nearest or next-to-nearest neighbors, is an
exact solution to Eq. (\ref{DModel}) satisfying the constraint
(\ref{CC}) provided Eq. (\ref{xx3.97}) is satisfied.

(xxxx) {\bf seventh solution with period 7}: $\phi
=(...,a,0,-a,a,a,a,-a,...)$

This is an exact solution to Eq. (\ref{DModel}) satisfying the
constraint (\ref{CC}) provided \bea\label{xx3.98}
&&a^2=1\,,~~ 2(\Lambda-3)=h^2(2A_1 +A_3  -A_2 -A_6 )\,, \nonumber \\
&&\Lambda-2=h^2(A_3 +A_1 -A_4 )\,, ~~\Lambda-4=h^2(A_3 +A_4 +A_1
-A_2 -A_5 -A_6 )\,. \eea In view of the constraint (\ref{CC}),
such a solution is valid only if $A_4=0,A_3=A_5$.

One can in fact generalize this solution and show that even

$\phi =(...,a,0,-a,~ (a,0,-a,~ p~{\rm times}),~a,a,a,-a,...)$

is an exact solution with period $3p+4~ (p \ge 1)$ provided Eq.
(\ref{xx3.98}) is satisfied.

One can also show that an {\it aperiodic solution} constructed
from the above periodic solution (with period $3p+4$) by randomly
adding ``$-a,a$" between ``$a$" and ``0" or ``0" and ``$-a$", is an
exact solution to Eq. (\ref{DModel}) satisfying the constraint
(\ref{CC}) provided Eq. (\ref{xx3.98}) is satisfied.

(xxxxi) {\bf eighth solution with period 7}: $\phi
=(...,a,a,0,a,0,-a,-a,...)$

This is an exact solution to Eq. (\ref{DModel}) satisfying the
constraint (\ref{CC}) provided \bea\label{xx3.99}
&&\Lambda-2=h^2a^2 A_1 \,,~~ 2(\Lambda-1)=h^2a^2(2A_1 +A_3  +A_2
+A_6 )\,,
\nonumber \\
&&\Lambda-2=h^2a^2(A_3 +A_1 -A_4)\,, ~~2=h^2a^2(A_5 +A_6 )\,. \eea
From here it follows that such a solution is valid provided
$A_3=A_4\,,A_5=A_2+A_3$.

One can in fact generalize this solution and show that even

$\phi =(...,a,a,0,a,0,~ (a,a,0,a,0,~ p~{\rm times}),~-a,-a,...)$

is an exact solution with period $5p+2~ (p \ge 1)$ provided Eq.
(\ref{xx3.99}) is satisfied.

One can also show that an {\it aperiodic solution} constructed
from the above periodic solution (with period $5p+2$) by randomly
adding ``0" between any two ``$a$", is an exact solution to Eq.
(\ref{DModel}) satisfying the constraint (\ref{CC}) provided Eq.
(\ref{xx3.99}) is satisfied.

(xxxxii) {\bf ninth solution with period 7}: $\phi
=(...,a,a,-a,0,-a,0,-a,...)$

This is an exact solution to Eq. (\ref{DModel}) satisfying the
constraint (\ref{CC}) provided \bea\label{xx3.100} &&\Lambda-2=h^2
A_1 \,,~~ 2(\Lambda-3)=h^2(2A_1 +A_3  -A_2 -A_6 )\,,
\nonumber \\
&&\Lambda-2=h^2(A_3 +A_1 -A_4 )\,, ~~2=h^2(A_5 +A_6 )\,. \eea In
view of the constraint (\ref{CC}), such a solution is valid only
if $A_2=-2A_3=-2A_4\,,A_5=-3A_3$.

One can in fact generalize this solution and show that even

$\phi =(...,a,a,-a,0,-a,~ (a,a,-a,0,-a,~ p~{\rm
times}),~0,-a,...)$

is an exact solution with period $5p+2~ (p \ge 1)$ provided Eq.
(\ref{xx3.100}) is satisfied.

One can also show that an {\it aperiodic solution} constructed
from the above periodic solution (with period $5p+2$) by randomly
adding ``0" between any two ``$a$", is an exact solution to Eq.
(\ref{DModel}) satisfying the constraint (\ref{CC}) provided Eq.
(\ref{xx3.100}) is satisfied.

(xxxxiii) {\bf tenth solution with period 7}: $\phi
=(...,a,-a,a,a,a,0,a,...)$

This is an exact solution to Eq. (\ref{DModel}) satisfying the
constraint (\ref{CC}) provided \bea\label{xx3.101}
&&a^2=1\,,~~\Lambda-4=h^2 (A_1 +A_3 +A_4 -A_2 -A_5 -A_6 )\,,~~
2(\Lambda-1)=h^2(2A_1 +A_3  +A_2 +A_6 )\,, \nonumber \\
&&\Lambda-2=h^2(A_3 +A_1 -A_4 )\,, ~~2=h^2(A_5 +A_6 )\,. \eea In
view of the constraint (\ref{CC}), such a solution is valid only
if $A_4=A_2=0,A_3=-A_5$.

One can in fact generalize this solution and show that even

$\phi =(...,a,-a,a,a,a,0,~ (a,-a,a,a,a,0,~ p~{\rm times}),~a,...)$

is an exact solution with period $6p+1~ (p \ge 1)$ provided Eq.
(\ref{xx3.101}) is satisfied.

One can also show that an {\it aperiodic solution} constructed
from the above periodic solution (with period $5p+2$) by randomly
adding as many ``$a$" as one wants between two ``$a$", is an exact
solution to Eq. (\ref{DModel}) satisfying the constraint
(\ref{CC}) provided Eq. (\ref{xx3.101}) is satisfied.

(xxxxiv) {\bf eleventh solution with period 7}: $\phi
=(...,a,0,a,0,a,a,-a,...)$

This is an exact solution to Eq. (\ref{DModel}) satisfying the
constraint (\ref{CC}) provided \bea\label{xx3.102}
&&\Lambda-2=h^2a^2 A_1 \,,~~\Lambda-4=h^2a^2 (A_1 +A_3 +A_4 -A_2
-A_5 -A_6 )\,,~~
2(\Lambda-1)=h^2a^2(2A_1 +A_3  +A_2 +A_6 )\,, \nonumber \\
&&\Lambda-2=h^2a^2(A_3 +A_1 -A_4 )\,, ~~2(\Lambda-3)=h^2a^2(2A_1
+A_3  -A_2 -A_6 )\,, ~~2=h^2a^2(A_5 +A_6 )\,. \eea
From here it follows that such a solution is valid provided
$A_2=A_3=A_4=A_5=0$.

One can in fact generalize this solution and show that even

$\phi =(...,a,0,~ (a,0,~ p~{\rm times}),~a,a,-a,...)$

is an exact solution with period $2p+3~ (p \ge 2)$ provided Eq.
(\ref{xx3.102}) is satisfied.

One can also show that an {\it aperiodic solution} constructed
from the above periodic solution (with period $2p+3$) by randomly
adding ``$-a$" between ``$a$" and ``0" or ``0" and ``$a$", is an exact
solution to Eq. (\ref{DModel}) satisfying the constraint
(\ref{CC}) provided Eq. (\ref{xx3.102}) is satisfied.

(xxxxv) {\bf Solution with period 8}: $\phi
=(...,a,-a,0,a,a,a,0,-a,...)$

This is an exact solution to Eq. (\ref{DModel}) satisfying the
constraint (\ref{CC}) provided \bea\label{xx3.103} &&\Lambda-4=h^2
(A_1 +A_3 +A_4 -A_2 -A_5 -A_6 )\,,~~
2(\Lambda-1)=h^2(2A_1 +A_3  +A_2 +A_6 )\,, \nonumber \\
&&a^2=1\,, ~~2(\Lambda-3)=h^2(2A_1 +A_3  -A_2 -A_6 )\,. \eea In
view of the constraint (\ref{CC}), such a solution is valid only
if $A_5=0,A_3=-2A_4$.

One can in fact generalize this solution and show that even

$\phi =(...,a,-a,0,~ (a,-a,0,~ p~{\rm times}),~a,a,a,0,-a,...)$

is an exact solution with period $3p+5~ (p \ge 1)$ provided Eq.
(\ref{xx3.103}) is satisfied.

One can also show that an {\it aperiodic solution} constructed
from the above periodic solution (with period $3p+5$) by randomly
adding ``$a,-a$" between ``$-a$" and ``0" or ``0" and ``$a$", is an
exact solution to Eq. (\ref{DModel}) satisfying the constraint
(\ref{CC}) provided Eq. (\ref{xx3.103}) is satisfied.

(xxxxvi) {\bf second solution with period 8}: $\phi
=(...,a,0,a,0,a,a,a,-a,...)$

This is an exact solution to Eq. (\ref{DModel}) satisfying the
constraint (\ref{CC}) provided \bea\label{xx3.104} &&\Lambda-4=h^2
(A_1 +A_3 +A_4 -A_2 -A_5 -A_6 )\,,~~
\Lambda-2=h^2 A_1 \,,~~2(\Lambda-1)=h^2(2A_1 +A_3  +A_2 +A_6 )\,, \nonumber \\
&&a^2=1\,,~~2=h^2(A_5 +A_6 )\,,~~ \Lambda-2=h^2 (A_1 +A_3 -A_4
)\,,~~ ~~2(\Lambda-3)=h^2(2A_1 +A_3  -A_2 -A_6 )\,. \eea In view
of the constraint (\ref{CC}), such a solution is valid only if
$A_2=A_3=A_4=A_5=0$.

One can in fact generalize this solution and show that even

$\phi =(...,a,0,a,0,a,a,a,~ (a,0,a,0,a,a,a,~ p~{\rm
times}),~-a,...)$

is an exact solution with period $7p+1~ (p \ge 1)$ provided Eq.
(\ref{xx3.104}) is satisfied.

One can also show that an {\it aperiodic solution} constructed
from the above periodic solution (with period $7p+1$) by randomly
adding any combination of ``$-a$", ``0" and ``$a$" at any place, is
an exact solution to Eq. (\ref{DModel}) satisfying the constraint
(\ref{CC}) provided Eq. (\ref{xx3.104}) is satisfied.

(xxxxvii) {\bf Solution with period 9}: $\phi
=(...,a,a,0,-a,-a,0,a,-a,0,...)$

This is an exact solution to Eq. (\ref{DModel}) satisfying the
constraint (\ref{CC}) provided \be\label{xx3.105}
2(\Lambda-1)=h^2a^2(2A_1 +A_3  +A_2 +A_6 )\,,
~~2(\Lambda-3)=h^2a^2(2A_1 +A_3  -A_2 -A_6 )\,. \ee

One can in fact generalize this solution and show that even

$\phi =(...,a,a,0,-a,-a,0,~ (a,a,0,-a,-a,0,~ p~{\rm
times}),~a,-a,0,...)$

is an exact solution with period $6p+3~ (p \ge 1)$ provided Eq.
(\ref{xx3.105}) is satisfied.

One can also show that an {\it aperiodic solution} constructed
from the above periodic solution (with period $6p+3$) by randomly
adding ``$a,-a,0$" between ``0" and ``$-a$", is an exact solution to
Eq. (\ref{DModel}) satisfying the constraint (\ref{CC}) provided
Eq. (\ref{xx3.105}) is satisfied.

(xxxxviii) {\bf second solution with period 9}: $\phi
=(...,a,-a,0,a,0,a,a,0,-a,...)$

This is an exact solution to Eq. (\ref{DModel}) satisfying the
constraint (\ref{CC}) provided \bea\label{xx3.106}
&&\Lambda-4=h^2a^2 (A_1 +A_3 +A_4 -A_2 -A_5 -A_6 )\,,~~
\Lambda-2=a^2h^2 A_1 \,,~~2(\Lambda-1)=h^2a^2(2A_1 +A_3  +A_2 +A_6 )\,, \nonumber \\
&&2=h^2a^2(A_5 +A_6 )\,,~~ 2(\Lambda-3)=h^2a^2(2A_1 +A_3  -A_2
-A_6 )\,. \eea
From here it follows that such a solution is valid provided
$A_2=A_4=A_5\,,A_3=0$.

One can in fact generalize this solution and show that even

$\phi =(...,a,-a,0,a,0,a,a,0,~ (a,-a,0,a,0,a,a,0,~ p~{\rm
times}),~-a,...)$

is an exact solution with period $8p+1~ (p \ge 1)$ provided Eq.
(\ref{xx3.106}) is satisfied.

One can also show that an {\it aperiodic solution} constructed
from the above periodic solution (with period $8p+1$) by randomly
adding ``0" between ``$a$" and ``$-a$", is an exact solution to Eq.
(\ref{DModel}) satisfying the constraint (\ref{CC}) provided Eq.
(\ref{xx3.106}) is satisfied.

(xxxxix) {\bf Solution with period 10}: $\phi
=(...,a,a,0,-a,-a,0,a,-a,0,a,...)$

This is an exact solution to Eq. (\ref{DModel}) satisfying the
constraint (\ref{CC}) provided \be\label{xx3.107}
a^2=1\,,~~2(\Lambda-1)=h^2(2A_1 +A_3  +A_2 +A_6 )\,,
~~2(\Lambda-3)=h^2(2A_1 +A_3  -A_2 -A_6)\,. \ee In view of the
constraint (\ref{CC}), such a solution is valid only if
$A_3+2A_4+2A_5=0$.

One can in fact generalize this solution and show that even

$\phi =(...,a,a,0,-a,-a,0,a,-a,0,~ (a,a,0,-a,-a,0,a,-a,0,~ p~{\rm
times}),~a,...)$

is an exact solution with period $9p+1~ (p \ge 1)$ provided Eq.
(\ref{xx3.107}) is satisfied.

One can also show that an {\it aperiodic solution} constructed
from the above periodic solution (with period $9p+1$) by randomly
adding as many ``$a$" as one wants between two ``$a$" and by
randomly adding as many ``$-a$" between two ``$-a$" is an exact
solution to Eq. (\ref{DModel}) satisfying the constraint
(\ref{CC}) provided Eq. (\ref{xx3.107}) is satisfied.

(xxxxx) {\bf second solution with period 10}: $\phi
=(...,a,-a,0,a,0,a,a,a,0,-a,...)$

This is an exact solution to Eq. (\ref{DModel}) satisfying the
constraint (\ref{CC}) provided \bea\label{xx3.108} &&\Lambda-4=h^2
(A_1 +A_3 +A_4 -A_2 -A_5 -A_6 )\,,~~
\Lambda-2=h^2 A_1 \,,~~2(\Lambda-1)=h^2(2A_1 +A_3  +A_2 +A_6 )\,, \nonumber \\
&&2=h^2(A_5 +A_6 )\,,~~a^2=1\,,~~ 2(\Lambda-3)=h^2(2A_1 +A_3 -A_2
-A_6 )\,. \eea In view of the constraint (\ref{CC}), such a
solution is valid only if $A_2=A_3=A_4=A_5=0$.

One can in fact generalize this solution and show that even

$\phi =(...,a,-a,0,a,0,a,a,a,0,~ (a,-a,0,a,0,a,a,a,0,~ p~{\rm
times}),~-a,...)$

is an exact solution with period $9p+1~ (p \ge 1)$ provided Eq.
(\ref{xx3.108}) is satisfied.

One can also show that an {\it aperiodic solution} constructed
from the above periodic solution (with period $9p+1$) by randomly
adding as many ``$a$" as one wants between two ``$a$" and by
randomly adding ``0" between ``$-a$" and ``$a$" or ``$a$" and ``$-a$"
is an exact solution to Eq. (\ref{DModel}) satisfying the
constraint (\ref{CC}) provided Eq. (\ref{xx3.108}) is satisfied.

(xxxxxi) {\bf Solution with period 11}: $\phi
=(...,a,-a,0,a,a,a,0,-a,a,a,-a,...)$

This is an exact solution to Eq. (\ref{DModel}) satisfying the
constraint (\ref{CC}) provided \bea\label{xx3.109} &&\Lambda-4=h^2
(A_1 +A_3 +A_4 -A_2 -A_5 -A_6 )\,,~~
2(\Lambda-1)=h^2(2A_1 +A_3  +A_2 +A_6 )\,, \nonumber \\
&&a^2=1\,,~~ 2(\Lambda-3)=h^2(2A_1 +A_3  -A_2 -A_6 )\,,~~
\Lambda-2=h^2 (A_1 +A_3 -A_4 )\,. \eea In view of the constraint
(\ref{CC}), such a solution is valid only if $A_3=A_4=A_5=0$.

One can in fact generalize this solution and show that even

$\phi =(...,a,-a,0,a,a,a,0,-a,a,a,~ (a,-a,0,a,a,a,0,-a,a,a,~
p~{\rm times}),~-a,...)$

is an exact solution with period $10p+1~ (p \ge 1)$ provided Eq.
(\ref{xx3.109}) is satisfied.

One can also show that an {\it aperiodic solution} constructed
from the above periodic solution (with period $10p+1$) by randomly
adding as many ``$a$" as one wants between two ``$a$", is an exact
solution to Eq. (\ref{DModel}) satisfying the constraint
(\ref{CC}) provided Eq. (\ref{xx3.109}) is satisfied.

\section{Two-point maps}
\label{Maps}

\subsection{General maps that include integration constant}

As pointed out in Sec. \ref{Introduction}, while Model 2, for any value
of $\delta$ and $\gamma$, has the
first integral with the integration constant $C$ expressed by Eq.
(\ref{HbasicPhi4Transformed}), for the Speight and Ward Model 7,
the known two point map Eq. (\ref{SpeightMap}) does not contain
the integration constant $C$.

Remarkably, in case only $A_2$ and $A_4$ are nonzero (combination
of Models 3 and 6, i.e. an admixture of Bender-Tovbis and BOP
nonlinearities), the discrete model Eq. (\ref{DModel}) has the
following first integral with the integration constant $C$
\begin{eqnarray} \label{Two_point_map_A2_A4}
   \phi_{n + 1} = (2 - \Lambda)\frac{Z\phi_n
   \pm \sqrt{f(\phi _n)} }
   {2 - \Lambda - Y\phi _n^2}, \quad \quad 
   f(\phi_n) = \frac{Y}{{2 - \Lambda}}
   (C - X\phi _n^2  + \phi _n^4),
\end{eqnarray}
where
\begin{equation} \label{Map_A2_A4Z}
   Z = \frac{{( 2 - \Lambda )^2  - Ch^4  A_4 ^2
   }}{{2\left( {2 - \Lambda } \right) + C h^4 A_2 A_4 }}, \quad
   X= \frac{{CY^2  + \left( {2
   - \Lambda } \right)^2 \left( {1 - Z^2 } \right)}}{{\left( {2 -
   \Lambda } \right)Y}}, \quad
   Y = h^2 ( A_4  + A_2 Z).
\end{equation}
As expected, in the limit of either $A_2=0$ or $A_4=0$, the nonlinear map
reduces to the maps derived earlier
in \cite{DKYF_PRE2006,DKKS2007_BOP}. Thus
the nonlinear map defined by Eqs. (\ref{Two_point_map_A2_A4}) and
(\ref{Map_A2_A4Z}) generalizes the maps considered in
\cite{DKYF_PRE2006,DKKS2007_BOP} and it allows one to construct
static solutions to Eq. (\ref{DModel}) iteratively in a way
similar to those studies.

Remarkably, in case $\Lambda \ne 2$, this first integral can be
further factorized as
\be\label{q}
   W(\phi_n, \phi_{n+1})(2-\Lambda)(2-\Lambda-Y\phi_n^2)=0\,,
\ee
where
\be\label{qa}
   W(\phi_n, \phi_{n+1})=\phi_n^2+\phi_{n+1}^2-\frac{Y}
   {2-\Lambda}\phi_n^2\phi_{n+1}^2
   -2Z\phi_n \phi_{n+1}-\frac{CY}{2-\Lambda}\,.
\ee
Since the vanishing of the third bracket in Eq. (\ref{q}) is a
trivial possibility, for $\Lambda \neq 2$, effectively the first
integral in this case acquires rather simple form, $W(\phi_n,
\phi_{n+1})=0$, which is precisely of the Quispel form as given in
\cite{Quispel}. As expected, in the special case when $A_4=0$ and
hence $h^2A_2=\Lambda$, the first integral $W(\phi_n,
\phi_{n+1})=0$ agrees with that obtained from the map Eq.
(\ref{HbasicPhi4Transformed}) in case $\gamma=\delta=0$. We also
note that Eq. (\ref{DModel}) with only $A_2$ and $A_4$ nonzero can
be expressed in terms of Eq. (\ref{qa}) as follows
\be\label{Representation2}
   \ddot{\phi}_n=\frac{2-\Lambda}
   {2Z(\phi_{n+1} -\phi_{n-1})}\left\{ W(\phi_n, \phi_{n+1}) - W(\phi_{n-1}, \phi_n)
   +\frac{h^2A_4}{2-\Lambda}\left[ \phi_{n+1}^2 W(\phi_{n-1}, \phi_n)
   - \phi_{n-1}^2 W(\phi_n, \phi_{n+1})\right] \right\}. \ee
It is now clear that the static solutions to Eq. (\ref{DModel})
can be found from the two-point map $W(\phi_n, \phi_{n+1})=0$.
Note that for small $h$ one has $Z \approx (2- \Lambda)/2$ and the
last term in the curly bracket of Eq. (\ref{Representation2}) can
be neglected and one obtains the equation similar to Eq.
(\ref{Representation1}) of Model 2. In other words, the model with
only $A_2$ and $A_4$ nonzero can be regarded as the Model 2
modified by the $O$-term (the last term in the curly bracket),
i.e. the term which disappears in the continuum limit and vanishes
upon substituting $W(\phi_n, \phi_{n+1})=0$ \cite{DKYF_PRE2006}.

We have checked that from the two-point map $W(\phi_n,
\phi_{n+1})=0$ one can obtain the staggered as well as the
nonstaggered JEF solutions $\dn,\cn$ and $\sn$ derived in Sec. III
from the three-point static Eq. (\ref{DModel}) in case only $A_2$
and $A_4$ are nonzero. The following local identities are helpful in these
derivations \cite{kls}
\be\label{dn1}
\dn^2(x,m)\dn^2(x+a,m)+\cs^2(a)[\dn^2(x,m)+\dn^2(x+a,m)]
-2\ds(a,m)\ns(a,m)\dn(x,m)\dn(x+a,m)+1-m=0\,, \ee \be\label{cn1}
m^2\cn^2(x,m)\cn^2(x+a,m)+m\ds^2(a)[\cn^2(x,m)+\cn^2(x+a,m)]
-2m\cs(a,m)\ns(a,m)\cn(x,m)\cn(x+a,m)-m(1-m)=0\,, \ee
\be\label{sn1}
m^2\sn^2(x,m)\sn^2(x+a,m)-m\ns^2(a)[\sn^2(x,m)+\sn^2(x+a,m)]
+2m\cs(a,m)\ds(a,m)\sn(x,m)\sn(x+a,m)+m=0\,. \ee

It is worth emphasizing here that as demonstrated in
\cite{DKKS2007_BOP}, first integral can always be constructed from
the known JEF solutions. In the continuum limit, the DFI Eq.
(\ref{Two_point_map_A2_A4}) and the DFI $W(\phi_n, \phi_{n+1})=0$
as given by Eq. (\ref{qa}), reduce to Eq. (\ref{FI}), which is the first
integral of the
static continuum $\phi^4$ equation. Relation between the solutions
obtained from the map $W(\phi_n, \phi_{n+1})=0$ iteratively and
JEF solutions reported in Sec. \ref{JEFA2A4} can be established as
follows. JEF solutions include two integration constants, the
arbitrary shift $x_0$ and the modulus $m$. In the map $W(\phi_n,
\phi_{n+1})=0$, the role of $m$ is played by $C$ and the role of
$x_0$ is played by the initial value of the map $\phi_0$. Relation
between $m$ and $C$ can be found if one obtains the JEF solutions
not only from the three-point static problem of Eq. (\ref{DModel})
but also from the two-point problem $W(\phi_n, \phi_{n+1})=0$
\cite{DKYF_PRE2006,DKKS2007_BOP}.

\subsection{Two-point maps for the cases (iv) to (vii)}

What about a universal map for the cases (iv) to (vii)? Unfortunately, so far
we have not been able to find one for any of these cases. However, as
we show now, corresponding to any static JEF solution, we can always
generate a map. In particular, we now obtain maps corresponding to the
three JEF solutions $\sn,\cn$ and $\dn$ and show that they can always be
factorized and reduced to Quispel \cite{Quispel} form.

\subsubsection{{\rm sn} solution}

As shown in the last section, one of the exact JEF solution in cases (iv) to
(vii) is
 \begin{eqnarray}\label{x3.2}
 \phi_n = A \sn [h\beta (n+x_0),m]\,,
 \end{eqnarray}
with arbitrary $x_0$ and $0 \le m \le 1$.
On using the well known identity
\begin{eqnarray} \label{Identity_sn}
   {\rm sn}(u+v,m)  = \frac{{\rm sn}(u,m) \,{\rm cn}(v,m) \,{\rm dn}(v,m)
+ {\rm sn}(v,m) \,{\rm cn}(u,m) \,{\rm
   dn}(u,m)}{1 - m \,{\rm sn}^2(u,m) \, \, {\rm sn}^2(v,m)}\,,
\end{eqnarray}
it immediately follows that $\phi_{n+1}$ and $\phi_n$ are related by the map
\begin{eqnarray} \label{Identity_sn2}
   \phi_{n+1} = \frac{\phi_n\cn (h\beta,m)\dn (h\beta,m)
   + \sn (h\beta,m)\sqrt{A^2-(1+m)\phi_n^2 +m(\phi_n^4/A^2)}}
   {1 - m \,{\rm sn}^2(h\beta,m)\phi_n^2/A^2}\,.
\end{eqnarray}
This can be simplified and put in the following factorized form
\be\label{isn}
[A^2-m\sn^2(h\beta,m)\phi_n^2]W_{sn}(\phi_n,\phi_{n+1})=0\,, \ee
where \be
W_{sn}(\phi_n,\phi_{n+1})=A^2(\phi_n^2+\phi_{n+1}^2)-m\sn^2(h\beta,m)
\phi_n^2 \phi_{n+1}^2 -2A^2\cn(h\beta,m)\dn(h\beta,m)\phi_n
\phi_{n+1} -A^4\sn^2(h\beta,m)\,. \ee Since the vanishing of the
first bracket in Eq. (\ref{isn}) is a trivial possibility,
effectively the map in this case is given by
$W_{sn}(\phi_n,\phi_{n+1})=0$ which is precisely of the Quispel
form as given in \cite{Quispel}.

For any of the cases (iv) to (vii), the sn solution Eq. (\ref{x3.2}) can be
derived from this map for any initial value taken from $-A \le
\phi_0 \le A$.
In this map, $0\le m\le 1$ is the integration constant. For given
$h$, parameters of the map $\beta$ and $A$ are related to the model
parameters.

\subsubsection{{\rm dn} solution}

Similarly, on using the identity
\begin{eqnarray} \label{Identity_dn}
   {\rm dn}(u+v,m)  = \frac{{\rm dn}(u,m) \,{\rm dn}(v,m)-m{\rm sn}(v,m)
+ {\rm cn}(v,m) \,{\rm sn}(u,m) \,{\rm
   cn}(u,m)}{1 - m \,{\rm sn}^2(u,m) \, \, {\rm sn}^2(v,m)}\,,
\end{eqnarray}
 we find the map for the dn solution
\begin{eqnarray}\label{x3.2z}
 \phi_n = A \dn [h\beta (n+x_0),m]\,,
\end{eqnarray}
in the form
\begin{eqnarray} \label{Identity_dn2}
   \phi_{n+1} = \frac{\phi_n\dn (h\beta,m)
   - \sn (h\beta,m)\cn (h\beta,m)\sqrt{A^2(m-1)+(2-m)\phi_n^2 -(\phi_n^4/A^2)}}
   {1 - \,{\rm sn}^2(h\beta,m)(1-\phi_n^2/A^2)}\,.
\end{eqnarray}
This can be simplified and put in the following factorized form
\be\label{idn}
[A^2\dn^2(h\beta,m)+m\sn^2(h\beta,m)\phi_n^2]W_{dn}(\phi_n,\phi_{n+1})=0\,,
\ee where \be
W_{dn}(\phi_n,\phi_{n+1})=A^2\dn^2(h\beta,m)(\phi_n^2+\phi_{n+1}^2)
+m\sn^2(h\beta,m) \phi_n^2 \phi_{n+1}^2 -2A^2\cn(h\beta,m)\phi_n
\phi_{n+1} -A^4(1-m)\sn^2(h\beta,m)\,. \ee Since the vanishing of
the first bracket in Eq. (\ref{idn}) is a trivial possibility,
effectively the map in this case is given by
$W_{dn}(\phi_n,\phi_{n+1})=0$ which is precisely of the Quispel
form as given in \cite{Quispel}.

\subsubsection{{\rm cn} solution}

Similarly, on using the identity
\begin{eqnarray} \label{Identity_cn}
   {\rm cn}(u+v,m)  = \frac{{\rm cn}(u,m) \,{\rm cn}(v,m)-{\rm sn}(v,m)
+ {\rm dn}(v,m) \,{\rm sn}(u,m) \,{\rm
   dn}(u,m)}{1 - m \,{\rm sn}^2(u,m) \, \, {\rm sn}^2(v,m)}\,,
\end{eqnarray}
 we find the map for the cn solution
\begin{eqnarray}\label{x3.3z}
 \phi_n = A \cn [h\beta (n+x_0),m]\,,
\end{eqnarray}
in the form
\begin{eqnarray} \label{Identity_cn2}
   \phi_{n+1} = \frac{\phi_n\cn (h\beta,m)
   - \sn (h\beta,m)\dn (h\beta,m)\sqrt{A^2(1-m)+(2m-1)\phi_n^2 -m(\phi_n^4/A^2)}}
   {1 - m{\rm sn}^2(h\beta,m)(1-\phi_n^2/A^2)}\,.
\end{eqnarray}
This can be simplified and put in the following factorized form
\be\label{icn}
[A^2\cn^2(h\beta,m)+\sn^2(h\beta,m)\phi_n^2]W_{cn}(\phi_n,\phi_{n+1})=0\,,
\ee where \be
W_{cn}(\phi_n,\phi_{n+1})=A^2\cn^2(h\beta,m)(\phi_n^2+\phi_{n+1}^2)
+\sn^2(h\beta,m) \phi_n^2 \phi_{n+1}^2 -2A^2\dn(h\beta,m)\phi_n
\phi_{n+1} +A^4(1-m)\sn^2(h\beta,m)\,. \ee Since the vanishing of
the first bracket in Eq. (\ref{icn}) is a trivial possibility,
effectively the map in this case is given by
$W_{cn}(\phi_n,\phi_{n+1})=0$ which is precisely of the Quispel
form as given in \cite{Quispel}.

Summarizing, unlike the cases (i) to (iii), we do not have a universal map for
the cases (iv) to (vii). However, in these four cases, one can always obtain
a map by starting from $\sn$, $\cn$ or $\dn$ solutions and remarkably, the
maps in all three cases is effectively of the Quispel form. Obviously,
 one can also obtain the staggered as well as the nonstaggered JEF solution
from the relevant two-point map.

It is clear from these arguments that for any discrete model that admits a JEF
solution, one can easily construct the corresponding two-point map
from which that JEF solution can be iteratively generated
\cite{DKKS2007_BOP}. Further, such a map is effectively of the Quispel form.
However, the problem of finding a universal map from
which {\em any} static solution can be generated is not trivial
even when the JEF solutions are known (note that not all TI models
admit JEF solutions).

\subsection{Particular factorized static problems}
\label{FactorizedMaps}

In some cases, the static problem can be factorized and one can obtain
some of the exact solutions, such as those presented in
Sec. \ref{ShortPeriodicSolutions}, from this lower order algebraic equation.
As an illustration, we now present a few examples of such factorized static
problems.

As an example, we note that the two-point map for the Model 2 with
$\gamma=0$ and $\delta=1/4$ is given by Eq.
(\ref{HbasicPhi4Transformed})
\begin{eqnarray} \label{ForFourPeriodic}
   \frac{1}{{h^2 }}\left( {\phi _{n + 1}  - \phi _n } \right)^2  +
   \lambda \phi _n \phi _{n + 1}  - \frac{\lambda }{4}\phi _n
   \phi_{n + 1} \left( {\phi _n^2  + \phi _{n + 1}^2 } \right)
   -\frac{\lambda }{2} + C = 0\,.
\end{eqnarray}
If the integration constant is chosen as $Ch^2  =
(\Lambda-4)^2/(2\Lambda)$, then Eq. (\ref{ForFourPeriodic})
factorizes as
\begin{eqnarray} \label{MapFourPeriodic}
   - \frac{4}{\Lambda}\left( 1
   - \frac{\Lambda}{4}\phi _n \phi _{n + 1} \right)
   \left[ { \Lambda - 2  - \frac{\Lambda }{4}
   \left( {\phi _n^2  + \phi _{n + 1}^2 } \right)} \right] = 0\,.
\end{eqnarray}
The last multiplier of Eq. (\ref{MapFourPeriodic}) generates the
TI four-periodic solution $\phi_n \equiv (...,a,b,-a,-b,...)$ with
$\Lambda(a^2+b^2)=4(\Lambda-2)$. This solution can also be written
in the form of the four-periodic TI $\sin$e solution as given by
Eq. (\ref{zz0}) and Eq. (\ref{sine4}). Further, the last
multiplier also generates the staggered TI $\sin$e solution with
$h\beta=\pi /2$, although it can be noted that for the
four-periodic solution, non-staggered and staggered forms coincide
after $n$ is substituted with $-n$.

As a second example we take the Bender-Tovbis model, i.e. the
Model 2 with $\gamma=\delta=0$. In this case, the two-point map
Eq. (\ref{HbasicPhi4Transformed}) with $\Lambda=2$ and $C=1$ is
factorized as
\begin{eqnarray} \label{FactBT}
   (1-\phi_{n - 1}^2)(\phi_n^2-1) = 0\,.
\end{eqnarray}
This example explains how one can obtain aperiodic solutions from
factorized maps. Indeed, any sequence of $\pm 1$ satisfies Eq.
(\ref{FactBT}) and hence the Bender-Tovbis model at $\Lambda=2$
and $C=1$. To obtain this aperiodic solution one can use either of the
multipliers of Eq. (\ref{FactBT}). This is possible because the
solutions are derived from the two-point rather than the three-point
map.

In the above examples the short-periodic solutions were obtained
from the factorized two-point map. More examples of this sort can
be found in \cite{DKKS2007_BOP} where short-periodic solutions are
derived from the two-point map Eq. (\ref{Two_point_map_A2_A4})
factorized for $\Lambda=2$ for the cases $A_2=\lambda$ and
$A_4=\lambda$.

As another example, the Speight and Ward Model 7 can be written in
the form \cite{DKYF_PRE2006}
\begin{eqnarray} \label{ForKink}
   \ddot{\phi}_n =- v(\phi_{n-1},\phi_n)
   \frac{\partial}{\partial \phi_n} v(\phi_{n-1},\phi_n)
   - v(\phi_{n},\phi_{n+1})
   \frac{\partial}{\partial \phi_n} v(\phi_{n},\phi_{n+1})\,,
\end{eqnarray}
where
\begin{eqnarray} \label{MapForKink}
   v(\phi_{n-1},\phi_n) =\frac{\phi_n  - \phi_{n - 1}}{H}
- \frac{1}{\sqrt 2} +
   \frac{(\phi_{n - 1}^2
+ \phi_{n - 1} \phi_n + \phi_n^2)}{3\sqrt 2}\,,
   \quad H^2=\frac{6\Lambda}{6-\Lambda}\,,
\end{eqnarray}
from which it is clear that the equation $v(\phi_{n-1},\phi_n)=0$
generates static kink and inverted kink solutions of their model.

The following two examples are interesting because they give the
short-periodic solutions, not from a two-point map, but, from a set
of {\em two} finite-difference equations.

We note that the Speight and Ward Model 7 can be written in the
form
\begin{eqnarray} \label{SWModel}
   \ddot{\phi}_n = \frac{1}{h^2} (\phi_{n+1}+\phi_{n-1}-2\phi_n)+\lambda
   \phi_n \nonumber \\
   -\frac{\lambda}{18}(\phi_{n-1} +2\phi_n)(\phi_{n-1}^2
   +\phi_{n-1}\phi_n +\phi_n^2)
   -\frac{\lambda}{18}(\phi_{n+1} +2\phi_n)(\phi_n^2 +\phi_n
   \phi_{n+1}+\phi_{n+1}^2).
\end{eqnarray}
If the following two-point equation holds
\begin{eqnarray} \label{MapForThreePeriodic}
   \phi_{n-1}^2 +\phi_{n-1}\phi_n +\phi_n^2 = 6\frac{\Lambda-3}{\Lambda},
\end{eqnarray}
then the static version of Eq. (\ref{SWModel}) reduces to
\begin{eqnarray} \label{ForThreePeriodic}
   (\Lambda-6)(\phi_{n-1} +\phi_n +\phi_{n+1})=0\,.
\end{eqnarray}
Two-point map Eq. (\ref{MapForThreePeriodic}) generates the exact
three-periodic $\sin$ as well as the six-periodic staggered $\sin$ solutions
 to the Speight and Ward model [see Eqs.
(\ref{zz0}), (\ref{zz3}) and (\ref{zz3'})] and for these solutions
Eq. (\ref{ForThreePeriodic}) is also satisfied. Further, Eqs.
(\ref{MapForThreePeriodic}) and (\ref{ForThreePeriodic}) for the
case of $\Lambda=6$ also generate the short periodic solutions
$\phi_n \equiv (...,a,0,..)$, $(...,a,-a,0,...)$,
$(...,a,-a,a,-a,0,...)$, $(...,a,0,a,0,-a,...)$,
$(...,a,-a,0,-a,a,0,...)$, $(a,0,a,0,a,-a)$, and
$(a,0,-a,a,0,-a,0)$.

Similarly, the equation of motion for the Hamiltonian Model 10 can
be written in the form
\begin{eqnarray} \label{HModel}
   \ddot{\phi}_n = \frac{1}{h^2} (\phi_{n+1}+\phi_{n-1}-2\phi_n)+\lambda
   \phi_n \nonumber \\
   -\lambda(\alpha_2 \phi_{n-1} +2\alpha_1 \phi_n)(\phi_{n-1}^2
   +\frac{\alpha_2}{\alpha_1}\phi_{n-1}\phi_n +\phi_n^2)
   -\lambda(\alpha_2 \phi_{n+1} +2\alpha_1 \phi_n)(\phi_{n+1}^2
   +\frac{\alpha_2}{\alpha_1}\phi_{n+1}\phi_n +\phi_n^2)\,.
\end{eqnarray}
In case the following two-point equation holds
\be\label{MapForH}
   \phi_{n-1}^2 +\frac{\alpha_2}{\alpha_1}\phi_{n-1}\phi_n +\phi_n^2
= H\,,
\ee
then the static version of Eq. (\ref{HModel}) reduces to
\be\label{Hpmodel} (\phi_{n-1}+\phi_{n+1}) (1-\Lambda H \alpha_2)
+(\Lambda -2 -4\alpha_1 \Lambda H) \phi_n = 0\,,~~
H=\frac{\Lambda-2-\alpha_2/\alpha_1}
{\Lambda(4\alpha_1-\alpha_2^2/\alpha_1)}\,, \ee
provided \be \alpha_3=\alpha_1+\frac{\alpha_2^2}{2\alpha_1}\,. \ee
In the special case of $\alpha_1=\alpha_2=1/18$, this reduces to
the SW model. The two-point map Eq. (\ref{MapForH}) generates the
exact $\sin$e as well as the staggered $\sin$e solutions
 to the Hamiltonian Model 10 with $\cos(h\beta)
=\mp \alpha_2/(2\alpha_1)$, and for these solutions Eq.
(\ref{Hpmodel}) is also satisfied.

Once again, it is interesting that for the factorization Eq.
(\ref{MapForThreePeriodic}) and Eq. (\ref{ForThreePeriodic}), and
also for the factorization Eq. (\ref{MapForH}) and Eq.
(\ref{Hpmodel}), one has to satisfy two lower-order
finite-difference equations {\em simultaneously}, and one of those
equations is a two-point one while another is a three-point one.

All factorized problems discussed in this section do not contain
the integration constant and thus they generate only particular
solutions. Some of them are TI solutions, for example, the
three-periodic solution derivable from Eq.
(\ref{MapForThreePeriodic}) and Eq. (\ref{ForThreePeriodic}),
while others are not, for example, arbitrary sequence of $\pm 1$,
derivable from Eq. (\ref{FactBT}).

Solutions constructed in this section from factorized problems do
not survive in the continuum limit because the factorized problems do not
reduce to Eq. (\ref{DModel}) or Eq. (\ref{FI}) in the continuum
limit.

\section{Goldstone modes}
\label{Goldstone}

\subsection{Goldstone mode of a TI static solution}
\label{GoldstoneTI}

Let $\phi_n^0$ be a static solution to Eq. (\ref{DModel}). To
study the dynamics in the vicinity of this solution we substitute
the ansatz $\phi_n(t) =\phi_n^0 +\varepsilon_n(t)$ into Eq.
(\ref{DModel}), and obtain the following linearized equation
\begin{eqnarray}\label{Linearization}
  \ddot \varepsilon _n = K_{n,n-1} \varepsilon _{n - 1}
  + K_{n,n} \varepsilon _n + K_{n,n+1} \varepsilon _{n + 1}\,,
\end{eqnarray}
with
\begin{eqnarray}\label{LinearCoeff}
  K_{n,n-1}  = \frac{1}{h^2}
  - \frac{A_2}{2}(\phi _n^0)^2
  - A_3 \phi _{n - 1}^0 \phi _n^0  - A_4 \phi _n^0 \phi _{n + 1}^0
  - \frac{A_5}{2}\phi _{n + 1}^0 (2\phi _{n - 1}^0  + \phi _{n + 1}^0 )
  - \frac{3A_6}{2}(\phi _{n - 1}^0)^2,  \nonumber \\
  K_{n,n}  = \lambda  - \frac{2}{h^2 } - 3A_1(\phi _n^0)^2
  - A_2\phi _n^0 ( \phi _{n - 1}^0  + \phi _{n + 1}^0 )
  - \frac{A_3}{2}\left[ {(\phi _{n - 1}^0)^2
  + (\phi _{n + 1}^0)^2 } \right]
  - A_4 \phi _{n - 1}^0 \phi _{n + 1}^0, \nonumber  \\
  K_{n,n+1}  = \frac{1}{h^2} - \frac{A_2}{2}(\phi _n^0)^2
  - A_3 \phi _n^0 \phi _{n + 1}^0  - A_4 \phi _{n - 1}^0 \phi _n^0
  - \frac{A_5}{2}\phi _{n - 1}^0 (\phi _{n - 1}^0  + 2\phi _{n + 1}^0)
  - \frac{3A_6}{2}(\phi _{n + 1}^0)^2 \,.
\end{eqnarray}
Looking for solutions of Eq. (\ref{Linearization}) of the form
$\varepsilon_n(t) =U_n \exp(\pm i \omega t)$ we come to the
eigen-value problem
\begin{eqnarray}\label{EigProblem}
   \left[ K \right]{\bf{U}} = - \omega ^2 {\bf{U}}\,,
\end{eqnarray}
where vector $\bf{U}$ contains $U_n$ and the nonzero coefficients
of matrix $\left[ K \right]$ are given by Eq. (\ref{LinearCoeff}).

If $\phi_n^0$ is a TI static solution then it can be shifted along
the chain by an arbitrary $x_0$, $\phi_n^0=\phi(n+x_0)$. The
eigenvector corresponding to the zero-frequency translational
Goldstone mode, $\bf{U}_{\rm G}$, has components $\phi^\prime_n$,
where prime means derivative of $\phi$ with respect to its
argument. To confirm that, we substitute $\bf{U}_{\rm G}=\{
\phi^\prime_{\it n} \}$ into Eq. (\ref{EigProblem}) with
$\omega^2=0$ and obtain
\begin{eqnarray}\label{Check}
  K_{n,n-1} \phi_{n- 1}^\prime + K_{n,n} \phi_{n}^\prime
  + K_{n,n+1} \phi_{n + 1}^\prime =0\,.
\end{eqnarray}
The last expression is an identity because it coincides with the
derivative of static version of Eq. (\ref{DModel}) with respect to
$x_0$, and such a derivation is possible for the TI static solution,
which is an equilibrium solution for any $x_0$. We thus have proved that
any TI static solution has the zero-frequency translational mode
$\bf{U}_{\rm G}=\{ \phi^\prime_{\it n} \}$. Particularly, for any
static JEF, hyperbolic or trigonometric function solutions given in Sec.
\ref{JEFsolutions}, one can easily find the corresponding TI mode
as it is proportional to the derivative of the solution with
respect to its argument.

Looking for solutions of Eq. (\ref{Linearization}) and Eq.
(\ref{LinearCoeff}) with $\phi_n^0=1$ of the form of
small-amplitude phonons, $\varepsilon_n(t) \sim \exp(ikn \pm i
\omega t)$, where $k$ denotes wavenumber and $\omega$ is frequency, one
obtains the spectrum of the vacuum for the discrete model of Eq.
(\ref{DModel}),
\begin{eqnarray}\label{VacuumSpectrum}
   \omega^2 =2\lambda + 2\left[ \frac{2}{h^2} -A_2-2A_3-2A_4-3A_5-3A_6 \right]
   \sin^2\left( \frac{k}{2} \right) \,.
\end{eqnarray}

\subsection{Goldstone modes of some short-period static solutions}
\label{GoldstoneShort}

If a static solution does not possess the zero-frequency
translational Goldstone mode, then this solution is not a TI one.
The opposite, in general is not true, i.e., a particular
static solution may have the Goldstone mode only at certain
positions with respect to the lattice $x_0$, but a TI solution must
have such a mode at any $x_0$. It is interesting to check whether the
short-periodic solutions derived in Sec.
\ref{ShortPeriodicSolutions} can have the Goldstone mode.


{\em Four periodic solution of the form $...,a,b,-a,-b,...$}, as
found in Sec. \ref{ShortPeriodicSolutions}, exists in case $A_1 \neq
0$ under the constraint $A_1-A_3+A_4=0$ and has $a^2  = (\Lambda -
2)/(A_1 h^2 )- b^2$. This one-parameter solution is a TI solution
because it can be expressed in the form
$\phi_n=A\cos[\pi(n+x_0)/2]$ with arbitrary shift $x_0$ and
$A^2=(\Lambda -2)/(A_1 h^2)$. Being a TI solution, it possesses
the Goldstone mode at any $x_0$, as it was demonstrated in Sec.
\ref{GoldstoneTI}. Note that Model 2, Model 8, Model 10 and Speight and Ward
Model 7 have $A_1 \neq 0$ but the constraint $A_1-A_3+A_4=0$ is
not satisfied for the SW model 7, while it is
satisfied for Model 2 (at arbitrary $\gamma$ and arbitrary
nonzero $\delta$), Model 8 (at $\gamma=0,\alpha=1/4$) and Model 10
(at $\alpha_1=\alpha_3$). Thus, while Models 2,8 and 10
support this TI four-periodic solution, Model 7 does not.

It may be noted here that in case $\Lambda=2$ and $A_1=0,A_3=A_4$, then
the TI solution is a two-parameter solution which can be expressed in
the form $\phi_n=A\cos[\pi(n+x_0)/2]$ with an arbitrary shift $x_0$ and
an arbitrary amplitude $A$. This is satisfied by Model 9 at $\beta=0$ as
well as by models where only $A_3,A_4,A_5$ or where only $A_2,A_3,A_4,A_5$
are nonzero.

{\em Four periodic solution of the form $...,a,0, - a,0,...$}, as
found in Sec. \ref{ShortPeriodicSolutions}, exists for $A_1 \neq
0$ and has $a^2  = \left( {\Lambda  - 2} \right)/\left( {A_1 h^2 }
\right)$. Inserting this solution into Eq. (\ref{EigProblem}) one
finds
\begin{eqnarray}\label{Eig1}
\left[ {\begin{array}{*{20}c}
   \alpha  & \beta  & 0 & \beta   \\
   \delta  & \gamma  & \delta  & 0  \\
   0 & \beta  & \alpha  & \beta   \\
   \delta  & 0 & \delta  & \gamma   \\
\end{array}} \right]\left\{ {\begin{array}{*{20}c}
   {U_0 }  \\
   {U_1 }  \\
   {U_2 }  \\
   {U_3 }  \\
\end{array}} \right\} =  - \omega ^2 \left\{ {\begin{array}{*{20}c}
   {U_0 }  \\
   {U_1 }  \\
   {U_2 }  \\
   {U_3 }  \\
\end{array}} \right\},
\end{eqnarray}
where
\begin{eqnarray}\label{Coeff1}
 \alpha  =  - 2\frac{{\Lambda  - 2}}{{h^2 }}, \quad
 \beta  = \frac{1}{{h^2 }} - A_2 \frac{{\Lambda  - 2}}{{2A_1 h^2 }}, \nonumber \\
 \gamma  = (A_1  - A_3  + A_4)\frac{\Lambda  - 2}{A_1 h^2}, \quad
 \delta = \frac{1}{{h^2 }} + (A_5 - 3A_6 ) \frac{\Lambda - 2}{2A_1 h^2}.
\end{eqnarray}
Characteristic equation of Eq. (\ref{Eig1}) is
\begin{eqnarray}\label{Charact1}
  \left( {\alpha  + \omega ^2 } \right)\left( {\gamma  + \omega ^2 }
  \right)\left[ {\left( {\alpha  + \omega ^2 } \right)\left( {\gamma
  + \omega ^2 } \right) - 4\beta \delta } \right] = 0,
\end{eqnarray}
and zero-frequency modes are possible when any of the following
three conditions is satisfied
\begin{eqnarray}\label{Cond1}
  \alpha  = 0, \quad \gamma  =
  0, \quad \alpha \gamma  - 4\beta \delta  = 0.
\end{eqnarray}
The considered four-periodic solution can also be expressed as
$\phi_n = A\cos (\pi n/2)$ with $A^2=(\Lambda-2)/(A_1h^2)$.
The expected Goldstone mode is
proportional to the derivative of this function with respect to
its argument, and thus, should have the form $U_0=U_2=0$,
$U_1=-U_3=k$ with arbitrary $k \neq 0$. Substituting this into Eq.
(\ref{Eig1}) with $\omega^2=0$ one finds that the Goldstone mode
corresponds to $\gamma=0$. Taking into account Eq. (\ref{Coeff1}),
we conclude that the considered four-periodic solution possesses the
Goldstone mode under the condition
\begin{eqnarray}\label{FinalCond1}
 A_1  - A_3  + A_4=0\,.
\end{eqnarray}
Particularly, the Speight and Ward Model 7 supports the considered
four-periodic solution but it does not satisfy the condition Eq.
(\ref{FinalCond1}) so that the solution is {\em not} a TI one. On
the other hand, Model 2 supports the considered four-periodic
solution and it satisfies the condition of Eq. (\ref{FinalCond1}) so that
the solution possesses the Goldstone mode.

It may be noted here that the solution (...,a,0,-a,0,...) also
exists in case $\Lambda=2$ and $A_1=0$, and in that case $a$ is
any real number. In this case the solution can be expressed in the
form $\phi_n=a\cos[\pi n/2]$. Further, such a solution is {\em
always} a TI one since according to Eq. (\ref{Cond1}), $\gamma$ is
proportional to $\Lambda-2$ and hence is equal to 0. Note that
such a solution exists in almost all the models at $\Lambda=2$.
For example it exists in Model 2 at $\delta=0$, Model 3, Model 4,
Model 6, Model 8 (at $\alpha=0$ or/and $\gamma=-\beta=1/2$), Model
9 and Model 10 (at $\alpha_1=0$) as well as the five models
discussed in Sec. III.

{\em Four periodic solution of the form $...,a,a, - a,-a,...$}, as
found in Sec. \ref{ShortPeriodicSolutions}, exists for $A_1+ A_3 -
A_4 \ne 0$ and has $h^2 a^2 = (\Lambda - 2)/(A_1 + A_3 - A_4)$.
The solution can also be presented as $\phi_n = \sqrt{2}\,A\cos(\pi
n/2 + \pi /4)$ where $A^2=(\Lambda-2)/[h^2(A_1+A_3-A_4)]$, so that
the Goldstone mode is $U_0=-U_1=-U_2=U_3=k$
with an arbitrary $k \neq 0$. Performing the calculations similar to
the previous case one finds that the considered four-periodic
solution possesses the Goldstone mode under the same condition of
Eq. (\ref{FinalCond1}). Again, this solution has the Goldstone
mode in Model 2 but not in the Speight and Ward Model 7, so
that in the latter model the solution is {\em not} a TI one.

It may be noted here that the solution (...,a,a,-a,-a,...) also
exists in case $\Lambda=2$ and $A_1+A_3=A_4$, and in this case $a$ is
any real number. In this case the solution can be expressed in the form
$\phi_n=\sqrt{2}a\cos[\pi n/2+\pi /4]$.
Further, such a solution is {\em always} a
TI one since according to Eq. (\ref{FinalCond1}), $\gamma$ is proportional
to $\Lambda-2$ and hence is equal to 0.
It is easily checked that such a solution at $\Lambda=2$ exists in
Model 2 (at $\delta=0$ and arbitrary $\gamma$), Model 8
(at $8\alpha^2=2\alpha+\beta$), Model 9 (at $\beta=0$) and Model 10
(at $\alpha_1=-\alpha_3$).

{\em Three periodic solution of the form $...,a,0, - a,...$}, as
found in Sec. \ref{ShortPeriodicSolutions}, exists for $2A_1  -
A_2  + A_3  - A_6  \ne 0$ and has $h^2 a^2  = 2(\Lambda -3)/( 2A_1
- A_2 + A_3 - A_6)$. Thus such a solution can be expressed by
$\phi_n=(2/\sqrt{3})A\cos(2\pi n/3-\pi /6)$ with $A^2=[2(\Lambda-2)]/[
h^2(2A_1-A_2+A_3-A_6)]$.
The Goldstone mode is $U_0=U_2=k$, $U_1=-p$,
and the ratio $k/p$ can be found but it is not important for our
analysis. Calculations similar to that for the four-periodic
solutions give that the considered three-periodic solution possesses
the Goldstone mode or, in other words, is a TI solution under the
condition
\begin{eqnarray}\label{FinalCond33}
   \Big( - 4A_1^2 - A_3^2 + A_5^2 + 2A_6^2 - 2A_1A_6 + 2A_1A_2
   + 4A_1A_3 - 4A_1A_4 - A_2A_6 + A_3A_6 \nonumber \\
   + 2A_4A_6 - 3A_5A_6 - A_2A_3 + 2A_3A_4 + A_2A_5 - 2A_4A_5 \Big)
   (\Lambda^2 -3)^2 = 0\,.
\end{eqnarray}
It is easily checked that both Model 2 and Speight and Ward Model
7 support the three periodic solution and meet the condition Eq.
(\ref{FinalCond33}). Thus the three-periodic solution possesses the
Goldstone mode in both Model 2 and Model 7, but, as mentioned
above, one cannot claim that this solution is a TI one in these
models unless one demonstrates that the Goldstone mode exists for
any shift of the three-periodic solution along the lattice.

It may be noted here that the solution (...,a,0,-a,...) also
exists in case $\Lambda=2$ and $2A_1+A_3=A_2+A_6$, and in this case $a$ is
any real number. In this case the solution can be expressed in the form
$\phi_n=(2/\sqrt{3})a\cos[2\pi n/3-\pi /6]$.
Further, such a solution is {\em always} a
TI one since $\Lambda=3$.
Thus, such a solution exists at $\Lambda=2$ in Model 2
(in case $2\gamma+8\delta=1$),
Model 6, Model 8 (in case $4\alpha[\beta+\gamma]=3\alpha^2+\gamma^2
+\beta \gamma$), Model 10 (in case $2\alpha_1+\alpha_3=2\alpha_2$),
and in Models where only $A_2,A_3,A_5$ or only $A_2,A_3,A_4,A_5$ are
nonzero.

\begin{figure}
\includegraphics{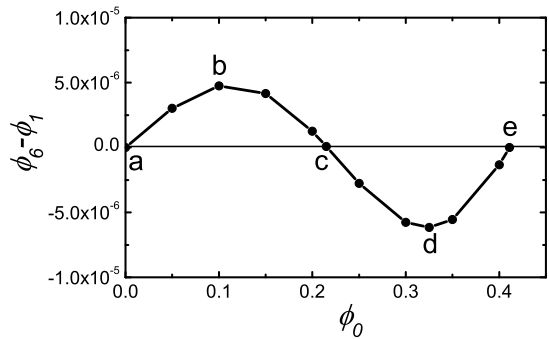}
\caption{Difference between $\phi_6$ and $\phi_1$ as function
of $\phi_0$ for the static solutions constructed for the Speight
and Ward Model 7 from the three-point map $\phi_{n+1}
=f(\phi_{n-1}, \phi_{n})$ for chosen $\phi_0$ and numerically
found $\phi_1$ such that $\phi_5=\phi_0$. If the three-point
problem is reducible to a two-point problem $\phi_{n+1}
=g(\phi_{n})$ then from having $\phi_5=\phi_0$ one must also have
$\phi_6=\phi_1$, which is not the case and thus, for this solution
the two-point reduction is impossible. Static solutions generated
by the three-point map for the points marked with a to e are shown
in the corresponding panels of Fig. \ref{Figure2}.}
\label{Figure1}
\end{figure}

\begin{figure}
\includegraphics{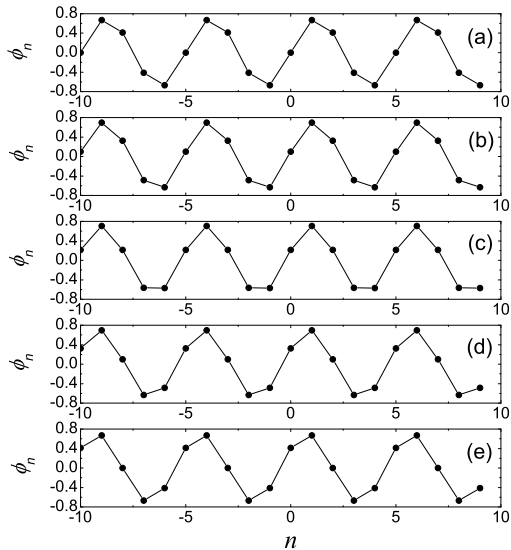}
\caption{Static solutions for the Speight and Ward Model 7
generated from the three-point map $\phi_{n+1} =f(\phi_{n-1},
\phi_{n})$ for chosen $\phi_0$ and numerically found $\phi_1$ such
that $\phi_5=\phi_0$. Panels (a) to (e) show the results for the
values of $\phi_0$ marked with a to e in Fig. \ref{Figure1}.
Solutions in (b) and (d) are {\em not} five-periodic solutions
because $\phi_6$ differs from $\phi_1$ by the amount shown in Fig.
\ref{Figure1}. Highly symmetric solutions in (a), (c), and (e) are
the five-periodic ones and for them $\phi_6=\phi_1$, as it can be
seen from Fig. \ref{Figure1}. Solution in (a) [and in (e)]
corresponds to a minimum of the Peierls-Nabarro potential while
the one in (c) to a maximum of this potential.} \label{Figure2}
\end{figure}

\section{Numerical results}
\label{NumericalResults}

\subsection{Five-periodic static solution in Speight and Ward Model 7}
\label{FivePeriodic}

Here we address the problem of integrability of the Speight and
Ward (SW) Model 7. As it was mentioned, this model has the two-point
map Eq. (\ref{SpeightMap}) to derive the kink solution but a general
two-point map, that includes the integration constant as a free
parameter, for this model is not known. We will give numerical
evidence that the Model 7 is not integrable and the two-point map
for obtaining a two-parameter set of static solutions cannot be
constructed. For this purpose we will make an attempt to construct
the static five-periodic solutions to this model and will observe
that it can be constructed only for highly symmetric positions of
the solution with respect to the lattice so that the solution is
not a TI one and it possesses the Peierls-Nabarro potential. We
have chosen the five-periodic solution for this study because it is
relatively simple and it corresponds to a non-factorized static
problem. Simple short-periodic solutions described in Sec.
\ref{ShortPeriodicSolutions} are not suitable for this study
because they are obtained from low-order algebraic equations,
i.e., from factorized static problems that do not represent the
model in its general formulation.

We set for the model parameters $\lambda=1$, $h=1.3$, and $A_k$
corresponding to the Model 7 as given in the Introduction.
The three-point static problem of Eq. (\ref{DModel}) is written in
the form of the map $\phi_{n+1} =f(\phi_{n-1}, \phi_{n})$ which,
for given $\phi_0$ and $\phi_1$, generates a static solution. For
chosen $\phi_0$ we numerically find $\phi_1$ such that in the
iteratively obtained solution $\phi_5=\phi_0$ and check whether in this
case we also have $\phi_6=\phi_1$. If the three-point map
$\phi_{n+1} =f(\phi_{n-1}, \phi_{n})$ is reducible to a two-point
map $\phi_{n+1} =g(\phi_{n})$ then having $\phi_5=\phi_0$ we must
also have $\phi_6=\phi_1$. However, as it can be seen from Fig.
\ref{Figure1}, $\phi_6-\phi_1$ is equal to zero only for a
discrete set of $\phi_0$. In Fig. \ref{Figure2} (a) to (e) we plot
the structures generated by the three-point map for different
$\phi_0$ indicated in Fig. \ref{Figure1} by, correspondingly,
letters a to e. Recall that for all structures presented in Fig.
\ref{Figure2} we have $\phi_5=\phi_0$ but the condition
$\phi_6=\phi_1$ is fulfilled only for the highly symmetric
structures shown in (a), (c), and (e) [structures in (a) and (e)
are equivalent]. Solutions in (b) and (d) are {\em not}
five-periodic solutions because $\phi_6$ differs from $\phi_1$ by
the amount shown in Fig. \ref{Figure1}. Solutions in (b) and (d)
are modulated five-periodic structures but this cannot be seen in
Fig. \ref{Figure2} because the period of the modulated structure
is very large.

Static solutions shown in Fig. \ref{Figure2} (a) and (c) are the
five-periodic equilibrium solutions for which the small-amplitude
vibrational spectrum can be calculated as described in Sec.
\ref{GoldstoneTI}. Doing so we find that these five-periodic
structures do not possess the zero-frequency Goldstone mode, but
they have a nearly translational mode with frequency $\omega
=0.0036$ for the structure in Fig. \ref{Figure2} (a) and purely
imaginary frequency $\omega =0.0036i$ for the structure in Fig.
\ref{Figure2} (c). Thus, the five-periodic structure in the
Speight and Ward model 7 is not a TI one and it experiences the
Peierls-Nabarro potential with a minimum energy corresponding to
the structure in (a) and a maximum energy corresponding to the
structure in (c).

We have studied some other periodic solutions, for example, seven-
and eight-periodic ones and have obtained the results
qualitatively similar to that for the five-periodic structure. In
all studied cases the structures had nonzero Peierls-Nabarro
potential.

We conclude that the static solutions supported by the Speight and
Ward Model 7, except for the kinks, anti-kinks, sine and staggered-sine
solutions, usually have the
Peierls-Nabarro potential. The corresponding three-point static
problem is non-integrable and cannot be reduced to a two-point
problem, again, except for the kink solutions.

This result is not surprising at all and it could be expected
taking into account that the derivation of the two-point map Eq.
(\ref{SpeightMap}), from which the kink solution can be derived,
was done for the integration constant $C=0$ in Eq. (\ref{FI}). The
resulting discrete model supports the TI solutions only for this
particular value of the integration constant, and those solutions
are kinks.

Hamiltonian TI discretization of the Klein-Gordon field that
generalizes the Speight and Ward model has been derived in Sec. II
C of \cite{DKYF_PRE2006}. The model includes the integration
constant and thus supports a two-parameter set of TI static
solutions, although it is rather complex even for the cubic
nonlinearity.

It is worth noting here that the Speight and Ward Model 7, apart
from the five-periodic solutions shown in Fig. \ref{Figure2} (a),
(c), and (e), supports the following non-TI five-periodic
solutions:
\begin{eqnarray}\label{FinalCond3}
   {\rm Solution\,\, (xiv)}&:&\,\, \phi=(1,1,1,-1,-1)\,\, {\rm if}\,\, \Lambda=9/2\,; \nonumber \\
{\rm Solution\,\, (xix)}&:&\,\, \phi=(1,1,0,-1,0)\,\, {\rm if}\,\, \Lambda=18/7\,; \nonumber \\
{\rm Solution\,\, (xvi)}&:&\,\, \phi=(a,-a,a,-a,0)\,\, {\rm if}\,\,  \Lambda=6\,\, {\rm and}\,\, a^2=3\,; \nonumber \\
{\rm Solution\,\, (xvi)}&:&\,\, \phi=(a,0,a,0-a)\,\, {\rm if}\,\, \Lambda=6\,\, {\rm and}\,\, a^2=3\,; \nonumber \\
{\rm Solution\,\, (xvi)}&:&\,\, \phi=(a,a,-a,-a,0)\,\, {\rm
if}\,\, \Lambda=12\,\, {\rm and}\,\, a^2=3/2\,.
\end{eqnarray}
The solutions shown in Fig. \ref{Figure2} exist for a continuously
varying $\Lambda$ (at least within a range of $\Lambda$ values) while the
above five solutions exist for a fixed $\Lambda$.

\begin{figure}
\includegraphics{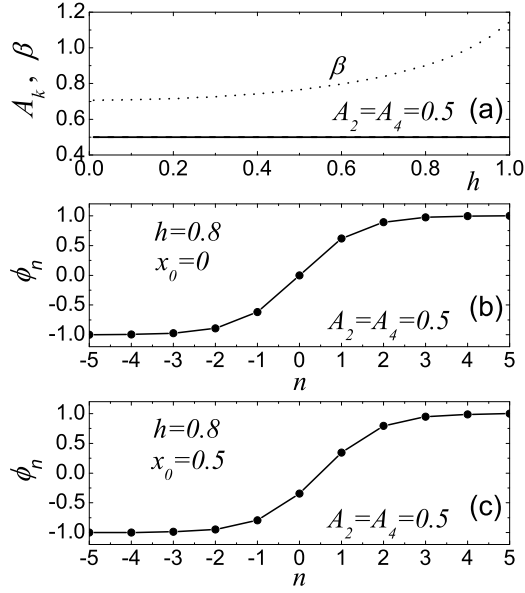}
\caption{Static kink in the case (iii) with only $A_2$ and $A_4$
nonzero: (a) model parameters $A_k$ and kink parameter $\beta$ as
functions of $h$; (b) on-site kink at $h=0.8$;  (c) inter-site
kink at $h=0.8$. This model admits TI static solutions, including
the kink solution, at constant ($h$-independent) model parameters
$A_k$, as can be seen in (a).} \label{Figure3}
\end{figure}

\begin{figure}
\includegraphics{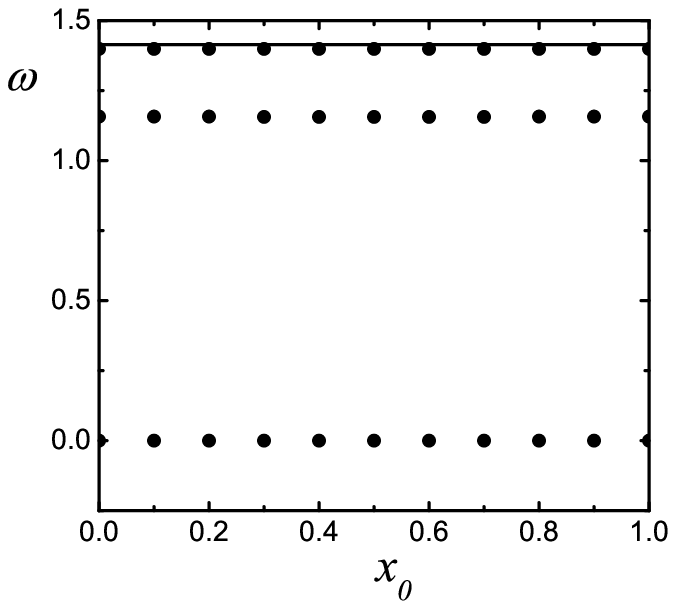}
\caption{Spectrum of the lattice with a kink in the case (iii)
with only $A_2$ and $A_4$ nonzero. Straight horizontal line at
$\omega=\sqrt{2}$ shows the lower bound of the phonon band while
dots show the kink's internal modes calculated for the kink at
various positions $x_0$ with respect to the lattice. At any
position $x_0$ the kink possesses the zero-frequency Goldstone
translational mode. The kink profiles at $x_0=0$ and $x_0=0.5$ are
shown in Fig. \ref{Figure3} (b) and (c), respectively. Model
parameters: $h=0.8$, $A_2=A_4=0.5$, $\lambda=1$.} \label{Figure4}
\end{figure}

\subsection{Static kinks}
\label{Kinks}

Here, after a brief discussion on the JEF solutions, we focus
on the analysis of the kink solutions because they are discussed
in applications more often than the periodic solutions.

As it was mentioned, the JEF solutions and their hyperbolic
function limit solutions like kink and pulse exist in the model
Eq. (\ref{DModel}) in the seven cases, of which the first three
cases with (i) only $A_2$ nonzero, (ii) only $A_4$ nonzero, and
(iii) only $A_2$ and $A_4$ nonzero are qualitatively different
from the other four cases discussed in Sec. \ref{JEFsolutions}.
The difference is in that for the first three cases one has two
conditions for finding the JEF solution parameters $A$ and $\beta$
while in the remaining four cases one has to satisfy one more
condition. This additional constraint couples the model parameters
$A_k$ to the lattice spacing $h$. As a result, in the last four
cases, for fixed $A_k$, one has TI solutions only at particular
$h$, while in the first three cases, even for fixed $A_k$, one has
TI solutions for any $h$.

Let us demonstrate this qualitative difference between two groups
of models by comparison of the properties of the static kinks. The
first group of models will be represented by the case (iii) with
only $A_2$ and $A_4$ nonzero while from the four models of the
second group we will choose the case (iv) with only $A_3$ and
$A_5$ nonzero, and the case (vi) with only $A_3$, $A_4$, and $A_5$
nonzero. For the kink solutions discussed below we will always
set $\lambda=1$.

{\em Kink in the case {\rm (iii)} with only $A_2$ and $A_4$
nonzero.} Parameters of the kink solution Eq. (\ref{tanh}) with
$S=1$ for this case are given by Eq. (\ref{5.11}). For given model
parameters $h$ and $A_4$ one can find the inverse kink width
$\beta$ solving the second equation in Eq. (\ref{5.11}). The model
parameter $A_2$ must satisfy the continuity constraint given by
the last expression in Eq. (\ref{5.11}). Particular feature of
this discrete $\phi_4$ model is that it admits the TI solutions at
constant $A_k$, see Fig. \ref{Figure3}. Contrary to that, as it
will be seen in the following examples, in the models (iv) to
(vii), model parameters $A_k$ are $h$-dependent.

Vibrational spectrum of the lattice containing a kink at different
positions with respect to the lattice $x_0$ is shown in Fig.
\ref{Figure4} for model parameters $h=0.8$, $A_2=A_4=0.5$,
$\lambda=1$. The corresponding kink profiles at $x_0=0$ and
$x_0=0.5$ are shown in Fig. \ref{Figure3} (b) and (c),
respectively. At any position $x_0$ the kink possesses the
zero-frequency Goldstone translational mode. Straight horizontal
line at $\omega=\sqrt{2}$ shows the lower bound of the phonon
band, see Eq. (\ref{VacuumSpectrum}).

\begin{figure}
\includegraphics{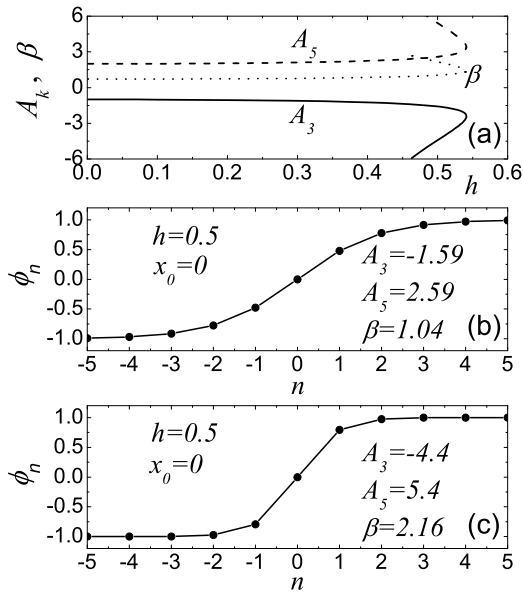}
\caption{Static kink in the case (iv) with only $A_3$ and $A_5$
nonzero. (a) Model parameters $A_k$ and kink inverse width $\beta$
as functions of $h$. In the region of lattice parameter around
$h=0.5$ it is possible to have two different static kink solutions
at the same $h$ which is illustrated in (b) and (c). In both cases
$h=0.5$, but model parameters $A_k$ and the inverse kink width
$\beta$ are different (shown in each panel). Both kinks are stable
and have zero-frequency translational Goldstone mode at any
position with respect to the lattice $x_0$.} \label{Figure5}
\end{figure}

{\em Kink in the case {\rm (iv)} with only $A_3$, and $A_5$
nonzero.} Model parameters and kink parameters in this case are
given by Eq. (\ref{3.11aa}). For chosen
$A_3$ (or $A_5$) one can find $A_5$ (or $A_3$) from the continuity
constraint [Eq. (\ref{3.11aa})] and then find
the inverse kink width $\beta$ solving the second equation in Eq.
(\ref{3.11aa}). Finally, one of the Eq. (\ref{3.11aa}) relates the model
parameters $A_k$ to the lattice spacing $h$. In Fig. \ref{Figure5}
we show (a) the model parameters $A_k$ and kink inverse width
$\beta$ as functions of $h$. In the region of lattice
parameter around $h=0.5$ it is possible to have two different
static kink solutions at the same $h$ which is illustrated in (b)
and (c). In both cases $h=0.5$, but model parameters $A_k$ and the
inverse kink width $\beta$ are different (shown in each panel).
Both kinks are stable and have a zero-frequency translational
Goldstone mode at any position with respect to the lattice $x_0$.

\begin{figure}
\includegraphics{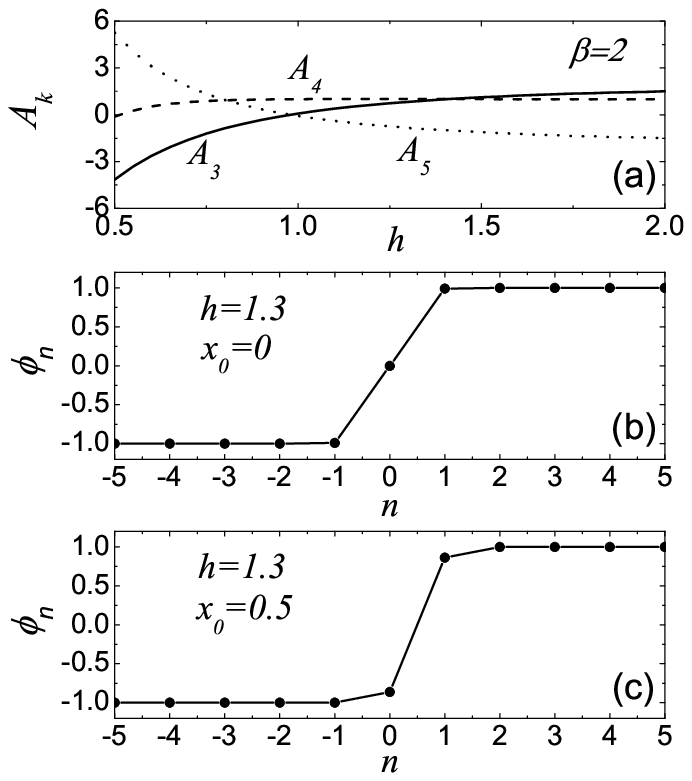}
\caption{Static kink in the case (vii) with only $A_3$, $A_4$, and
$A_5$ nonzero. (a) Model parameters $A_k$ as functions of $h$
at {\em fixed} inverse kink width $\beta=2$. (b) On-site kink at
$h=1.3$. (c) Inter-site kink at $h=1.3$. Other model parameters
for (b) and (c) are $A_3=0.8297$, $A_4=1.0092$, $A_5=-0.8389$, and
$\lambda=1$.} \label{Figure6}
\end{figure}

{\em Kink in the case {\rm (vii)} with only $A_3$, $A_4$, and
$A_5$ nonzero.} Kink and model parameters are related by Eq.
(\ref{3.11bb}). In Fig. \ref{Figure6} we plot (a) model parameters
$A_k$ as the functions of $h$ at {\em fixed} inverse kink width
$\beta=2$; (b) the on-site kink at $h=1.3$; and (c) the inter-site
kink at $h=1.3$. Other model parameters for (b) and (c) are
$A_3=0.8297$, $A_4=1.0092$, $A_5=-0.8389$, and $\lambda=1$.

Note that in Fig. \ref{Figure6} (a) all $A_k$ vary with $h$ but,
interestingly, the inverse kink width $\beta$ is constant ($=2$)
in a wide range of lattice spacing $h$. In the classical discrete
$\phi^4$ model and in the models with $h$-independent parameters
$A_k$ the kink width usually decreases with increase in $h$. On
the other hand, it is possible to get from Eq. (\ref{3.11bb}) one
constant model parameter with two other model parameters $A_k$ and
kink parameter $\beta$ being functions of $h$.

\begin{figure}
\includegraphics{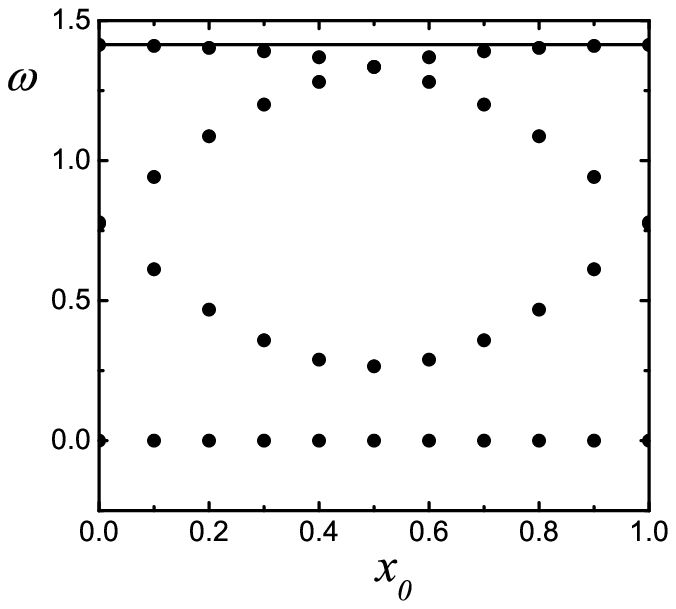}
\caption{Spectrum of the lattice with a kink in the case (vii)
with only $A_3$, $A_4$, and $A_5$ nonzero. Straight horizontal
line at $\omega=\sqrt{2}$ shows the lower bound of the phonon band
while dots show the kink's internal modes calculated for the kink
at various positions $x_0$ with respect to the lattice. At any
position $x_0$ the kink possesses the zero-frequency Goldstone
translational mode. The kink profiles at $x_0=0$ and $x_0=0.5$ are
shown in Fig. \ref{Figure6} (b) and (c), respectively. Model
parameters: $h=1.3$, $A_3=0.8297$, $A_4=1.0092$, $A_5=-0.8389$,
and $\lambda=1$.} \label{Figure7}
\end{figure}

In Fig. \ref{Figure7} one can see the spectrum of the lattice with
a kink. Straight horizontal line at $\omega=\sqrt{2}$ shows the
lower bound of the phonon band [see Eq. (\ref{VacuumSpectrum})]
while dots show the kink's internal modes calculated for the kink
at various positions $x_0$ with respect to the lattice. At any
position $x_0$ the kink possesses the zero-frequency Goldstone
translational mode. The kink profiles at $x_0=0$ and $x_0=0.5$ are
shown in Fig. \ref{Figure6} (b) and (c), respectively. Model
parameters are: $h=1.3$, $A_3=0.8297$, $A_4=1.0092$, $A_5=-0.8389$,
and $\lambda=1$.

\section{Discussion and conclusions}
\label{Discussion}

In this paper we have introduced a rather general discrete $\phi^4$ model
of which all known models in the literature are special cases.
We could find seven special cases when the model as given by Eq. (\ref{DModel})
supports the exact static JEF and hence hyperbolic kink and pulse solutions.
Two of those seven cases have been analyzed in \cite{DKYF_PRE2006}
and \cite{DKKS2007_BOP}, while for the remaining five cases, JEF
solutions were given in Sec. \ref{JEFsolutions}.

The exact solutions constructed for the considered discrete
$\phi^4$ model are important for the theory of the TI lattices.
Indeed, the JEF static solutions with an arbitrary shift along the
lattice $x_0$ are the TI solutions with the zero-frequency
Goldstone mode, i.e., solutions that are free of the
Peierls-Nabarro potential. The discrete $\phi^4$ model of Eq.
(\ref{DModel}) corresponding to the following four particular
cases
\begin{enumerate}

\item case (iv) only $A_3$ and $A_5$ nonzero;

\item case (v) only $A_2$, $A_3$, and $A_5$ nonzero;

\item case (vi) only $A_3$, $A_4$, and $A_5$ nonzero;

\item case (vii) only $A_2$, $A_3$, $A_4$, and $A_5$ nonzero;

\end{enumerate}
are the new TI models. Each of these models (like the other three)
supports a
two-dimensional set of TI static solutions that can be
parameterized by the points of the plane $(m,x_0)$. For {\em
fixed} model parameters $A_k$ these models support TI solutions
only for a particular lattice spacing $h$. Note, however, that the other
three models, i.e. cases (i) to (iii), for fixed model parameters $A_2$
and $A_4$,
support the TI solutions for any arbitrary value of $h$.
However, when $A_k$ are
considered to be functions of $h$, the TI solutions in (iv) to (vii) can
also be
constructed for continuously varying $h$ (see numerical examples
of Sec. \ref{Kinks}). In this context, it is worth noting that the TI
Model 5 given by Eq.
(\ref{Saxena}) and the TI Model 9 also have $h$-dependent nonlinearities.

In this paper,  we also showed that the general model, Eq. (\ref{DModel})
supports periodic $\sin$e and staggered
$\sin$e solutions. Remarkably, almost all the known models (even those not
supporting the JEF solutions), were found to support these solutions.
Besides, a large number of exact, short-periodic and aperiodic static solutions
admitted by Eq. (\ref{DModel}) were obtained in Sec.
\ref{ShortPeriodicSolutions}. While we do not have a rigorous proof, but the
few examples discussed in Sec. \ref{FactorizedMaps} suggest that
very likely, the short-periodic, aperiodic as well as trigonometric solutions,
in fact follow from low order algebraic equations. In this context, it is
worth pointing out that the $\sin$e solution does not follow from the map
for Model 2 as well as the map for case (iii) when only $A_2$ and $A_4$ are
nonzero. The factorization can also easily explain the appearance of the
aperiodic solutions that can be regarded as the solutions obtained
from different multipliers and linked together, as exemplified by
the discussion below Eq. (\ref{FactBT}).

It is worth pointing out that none of the  factorized problems
discussed in Sec. \ref{FactorizedMaps} contain an integration
constant and thus they generate only particular solutions. Some of
them are TI solutions, for example, the three-periodic solution to
the Speight and Ward Model 7 derivable from Eq.
(\ref{MapForThreePeriodic}) and Eq. (\ref{ForThreePeriodic}),
while others are not, for example, arbitrary sequence of $\pm 1$,
derivable from Eq. (\ref{FactBT}).

As it was shown in Sec. \ref{FactorizedMaps}, in some cases the
two-point map can be factorized and then the short-periodic
solutions can also be derived from a two-point problem. We also discussed
several examples in which the three-point problem can be reduced to a
set of two
lower-order finite-difference equations, and one of those
equations is a two-point one while another is a three-point one.
Based on these examples, we conjecture that all short-periodic, aperiodic
as well as $\sin$e and staggered $\sin$e solutions should follow from lower
order difference equations.

Note that the short-periodic solutions and, more generally, the
solutions derived from factorized problems very often do not
survive the continuum limit because factorized equations usually
have a different continuum limit than the original, non-factorized
one. In this context it is worth noting that $\sin$e is not a
solution of the continuum $\phi^4$ field equation. One exception
to this rule is the kink solution to the Speight and Ward Model
7 for which the reduced two-point problem Eq. (\ref{SpeightMap})
in the continuum limit obtains a form which is equivalent to the
first integral of the static $\phi^4$ field.

Coming back to the exact JEF solutions, we emphasize that they
are important because by using them one can construct the corresponding two-point
maps from which the corresponding solutions can be obtained iteratively.
Moreover, in
some cases, the map obtained for a particular JEF solution can be
transformed to the form of a general map from which majority of static
solutions including other JEF solutions admitted by the model can be
constructed. We conjecture that, except for
those obtainable from special cases when the three-point problem can
be substituted with a set of two equations, all other solutions can be obtained
recursively from this general map. Following this
way we could construct the map Eq. (\ref{Two_point_map_A2_A4})
from which any static solution of Eq. (\ref{DModel}) with only
$A_2$ and $A_4$ nonzero can be constructed (again, except for the
solutions that result from specially factorized three-point
problems). On the other hand, for cases (iv) to (vii), while one can obtain
a map from a JEF solution, so far we are unable to obtain a general map.

In Sec. \ref{FivePeriodic} we provided numerical evidence that the
Speight and Ward Model 7 does not support TI static solutions
other than those derivable from reduced lower-order algebraic
problems as discussed in Sec. \ref{FactorizedMaps}. In addition to
the well-known TI kink solution we have found the TI
$sin$  and staggered $\sin$ solutions to this model. We believe that in the
general (non-factorized) formulation, the static Speight and Ward
model is not integrable and a two-point map that includes
the integration constant as a parameter cannot be constructed for this
model.

Based on the results of the present study one can separate the TI
models into two classes. In the first class belong the models that
support a two-dimensional space of TI static solutions. These
solutions, if they are derivable from a two-point nonlinear map, can be
parameterized by the points of the plane $(C,\phi_0)$, where $C$
is the integration constant that can vary continuously within
certain range and $\phi_0$ is the initial value of the map that
can also vary continuously. Alternatively, if the JEF solutions are
known, then the solutions can be parameterized by the points of the
plane $(m,x_0)$ so that $m$ plays the role of the integration
constant $C$ while variation of $x_0$ plays the role of $\phi_0$, and
results in the shift of the solution along the lattice. The second class is
formed by the models that admit TI static solutions with an arbitrary
shift along the lattice (controlled by either $x_0$ or $\phi_0$)
but corresponding solutions do not include the integration
constant as a parameter.

The TI models in the first class have been investigated in
\cite{DKYF_PRE2006,DKYF_JPA_2007DNLSE,DKKS2007_BOP} and four more
TI cubic nonlinearities belonging to this class are found in the
present work. In particular, TI I models are cases (i) to (vii) and Model 2.
In this context it is worth noting that while a universal two-point map is
known for Model 2 (for arbitrary $\gamma$ and $\delta$), no JEF or any
other analytical solutions are known so far
which can be characterized by $C$ and $\phi_0$. On the other hand, no universal
two-point map is known for cases (iv) to (vii).

It is likely that the Speight and Ward model 7
\cite{SpeightKleinGordon} belongs to the second class of TI models
because it supports the well-known TI kink and the TI sine
solutions derived in the present work [see Eq. (\ref{zz0}) and Eq.
(\ref{zz3})] but these solutions are derived from reduced
equations, as shown in Sec. \ref{FactorizedMaps}. The reduced
equations do not contain the integration constant. On the other hand,
e.g., the five-periodic solution derived from the non-factorized
model possesses the Peierls-Nabarro potential, as shown in Sec.
\ref{FivePeriodic}.

At this stage, it may be worthwhile classifying the various known models.
As noted above, cases (i) to (vii) and Models 2 and 9 (and hence Models
3 to 6) are TI I models. On the other hand, Models
7, 8 and 10 are in general TI II models even though in the special cases of
$\alpha=\beta=0$ or $\alpha=0,\beta=-1/2$, Model 8 becomes a TI I model.

Note that while a TI I model can also be a TI II model,
the converse is obviously not true. For example, except for case (ii) when only
$A_4$ is nonzero, all other TI I models also admit $\sin$e and staggered $\sin$e
solutions. Besides, Model 2, Model 8 (in case $\gamma=0$) and Model 10
(in case $\alpha_1=\alpha_3$) also satisfy the four-periodic TI solution
$(...,a,b,-a,-b,...)$. At the special value of $\Lambda=2$, such a TI solution
is admitted by Model 9 (at $\beta=0$) as well as by Models where only
$A_3,A_4,A_5$ or only $A_2,A_3,A_4,A_5$ are nonzero.

It is thus clear that the general Model as given by Eq. (\ref{DModel}) is
only a TI II model even though in few special cases, it could be elevated to
a TI I model. There is one model, however, which is not a TI model with respect to
any known solution. We have in mind the model where only $A_1$ is nonzero,
i.e. the $\phi^4$ model with standard discretization. For example, while
it admits the short-period solutions $(...,a,0,-a,...),(...,a,0,-a,0,...),
(...,a,a,-a,-a,...)$, neither of them possesses the Goldstone mode.

Before closing, we spell out some of the open problems.
\begin{enumerate}

\item For the cases (iv) to (vii), can one obtain a unified general
two-point map from which all solutions, including the JEF solutions can
be derived? Note that, at the moment by starting from $\sn,\cn,\dn$ solutions,
one can obtain three different maps from which only the respective
solution can be obtained.

\item For Model 2, while a general two-point map is known, to date no
analytic solution is known which is characterized by the two parameters
$C$ and $\phi_0$. Can one find few such analytic solutions?

\item Can one rigorously show that all short period, $\sin$e and staggered
$\sin$e solutions for any TI I or TI II model, follow from the lower
order equations?

\item  Can one rigorously prove that Model 7 of Speight and Ward is only
a TI II and not a TI I model?

\item There is a belief that all TI models (at least TI I models) must
have some conserved quantity. Unfortunately, for cases (iii) to (vii)
(which are all TI I models) no such conserved quantity is known at
present.
Can one find such a quantity or disprove the conjecture?

\item While it has been demonstrated that no discrete model can simultaneously
have conservation of $P_1$ and energy $E$, it is not known whether one
can have a model where both $P_2$ and $E$ can be simultaneously
conserved. The obvious guess would be {\it no}. It would be
nice to prove or disprove this conjecture.

\item Can one find particular TI solutions for the discrete models that
do not belong to the TI I class, i.e., finding the isolated TI
solutions to the discrete models that are not considered by many
researchers as the TI models. In the present study we have given
several examples of such solutions, for instance, four-periodic TI
solution $(...,a,b,-a,-b,...)$ and the TI trigonometric solutions given
in Sec. \ref{TrigonometricSolutions}. It can be so that isolated
TI solutions exist for many discrete models. Many of the isolated TI
solutions result from factorized static problems and thus,
finding various factorizations of the original static problem can
be a method for their derivation.

\end{enumerate}

Finally, the results obtained in this paper are easily extended to the
case of the general nonlinear Schr\"odinger equation \cite{NewJPA}. We hope
to address these issues in a forthcoming publication.

\section*{Acknowledgements}
SVD gratefully acknowledges the financial support provided by the
Russian Foundation for Basic Research, grant 07-08-12152.

\section{Appendix}

In this Appendix, using the various short-period solutions given in Sec. IV,
we spell out the solutions which are admitted by some of the Models discussed
in the paper. We shall only mention the allowed solutions with the lowest
period. Needless to say that the corresponding solutions with arbitrarily
large period or aperiodic solutions (if they exist) will also be valid
in that case. For example, in case only $A_2$ is nonzero, as shown below,
the solution $\phi_n=(...,a,a,-a,...)$ is allowed provided $\Lambda=2$ and
$a^2=1$. It is then clear from the discussion in Sec. IV that in that case
a solution of the form $\phi_n =(...,a,-a, (a,-a~p~ times),-a,...)$
with period $2p+1~(p \ge 1)$ as well as an {\it aperiodic solution} with any
number of "a" and "-a" kept at random but with the constraint that at most two
"a" or two "-a" are always together, is also an exact solution.

\subsection{Model 1: Only $A_1$ nonzero, $A_1 h^2=\Lambda$}

This is the case of the classical discretization of $\phi^4$ term. In this case
the admitted solutions are

(i) $\phi_n = (...,a,-a,...)$ in case $a^2=\frac{\Lambda-4}{\lambda}$.
Thus such a solution is valid if $\Lambda>4$ or if $\Lambda<0$.

(ii) $\phi_n = (...,a,-a,0,...)$ in case $a^2=\frac{\Lambda-3}{\lambda}$.
Thus such a solution is valid if $\Lambda>3$ or if $\Lambda<0$.

(iii) $\phi_n = (...,a,a,-a,-a,...)$ and $\phi_n =(...,a,0,-a,0,...)$ in case
$a^2=\frac{\Lambda-2}{\lambda}$.
Thus such a solution is valid if $\Lambda>2$ or if $\Lambda<0$.

(iv) $\phi_n = (...,a,a,0,-a,-a,0,...)$ in case $a^2=\frac{\Lambda-1}{\lambda}$.
Thus such a solution is valid if $\Lambda>1$ or if $\Lambda<0$.

\subsection{Model 3: Only $A_2$ nonzero, $A_2 h^2=\Lambda$}

In this case the admitted solutions are

(i) $\phi_n = (...,a,-a,...)$ in case $a^2=\frac{4-\Lambda}{\lambda}$.
Thus such a solution is valid if $0<\Lambda<4$.

(ii) $\phi_n = (...,a,-a,0,...)$ in case $a^2=\frac{3-\Lambda}{\lambda}$.
Thus such a solution is valid if $0<\Lambda<3$.

(iii) $\phi_n = (...,a,a,-a,-a,...)$ and $\phi_n =(...,a,0,-a,0,...)$ in case
$\Lambda=2$ while $a$ is {\it any} real number.

(iv) $\phi_n = (...,a,b,-a,-b,...)$ in case
$\Lambda=2$ while $a,b$ both are {\it any} arbitrary real numbers.

(v) $\phi_n = (...,a,a,0,-a,-a,0,...)$ in case $a^2=\frac{\Lambda-1}{\lambda}$.
Thus such a solution is valid if $\Lambda>1$ or if $\Lambda<0$.

(vi) In addition, there are a large number of solutions which are admitted in
case $\Lambda=2$ and $a^2=1$. These allowed solutions are
\bea
&&\phi_n=(...,a,a,-a,...)\,,~(...,a,a,a,-a,...)\,,~(...,a,0,-a,a,...)\,,~
(...,a,a,a,-a,-a,...)\,, \nonumber \\
&&(...,a,-a,a,-a,0,...)\,,~(...,a,a,-a,-a,0,...)\,,~(...,a,a,0,-a,0,...)\,,~
(...,a,0,-a,a,a,...)\,, \nonumber \\
&&(...,a,a,a,0,-a,-a,...)\,,~(...,a,a,0,-a,0,a,...)\,,
~(...,a,-a,a,a,-a,0,...)\,,~ (...,a,a,0,-a,-a,0,a,...)\,, \nonumber \\
&&(...,a,0,-a,a,0,-a,0,...)\,,~(...,a,-a,0,a,a,0-,a,...)\,,
~(...,a,a,-a,a,a,-a,0,...)\,,~ (...,a,0,-a,a,a,a,-a,...)\,, \nonumber \\
&&(...,a,-a,0,a,a,a,0,-a,...)\,,~(...,a,a,0,-a,-a,0,a,-a,0,...)\,, \nonumber \\
&&~(...,a,a,0,-a,-a,0,a,-a,0,a,...)\,,~ (...,a,-a,0,a,a,a,0,-a,a,a,-a,...)\,.
\eea

\subsection{Model 6: Only $A_4$ nonzero, $A_4 h^2=\Lambda$}

In this case the admitted solutions are

(i) $\phi_n = (...,a,-a,...)$ in case $a^2=\frac{\Lambda-4}{\lambda}$.
Thus such a solution is valid if either $\Lambda>4$ or $\Lambda <0$.

(ii) $\phi_n = (...,a,a,-a,-a,...)$ in case $a^2=\frac{2-\Lambda}{\lambda}$.
Thus such a solution is valid if $0<\Lambda<2$.

(iii) $\phi_n = (...,a,1/a,a,1/a,...)$ and $\phi_n =(...,a,0,-a,0,...)$ in case
$\Lambda=2$ while $a$ is {\it any} real number.

(iv) $\phi_n = (...,a,-a,0,...)$ in case
$\Lambda=3$ while $a$ is {\it any} real number.

In addition, there are a large number of solutions which are admitted in
case $\Lambda=1$ and $a^2=1$. These allowed solutions are
\bea
&&\phi_n=(...,a,a,a,-a,-a,...)\,,~(...,a,a,-a,-a,0,...)\,,
~(...,a,a,0,-a,-a,0,...)\,,~(...,a,a,a,0,-a,-a,...)\,, \nonumber \\
&&(...,a,a,0,-a,-a,0,a,...)\,.
\eea

\subsection{Case (iii): Only $A_2,A_4$ nonzero, $(A_2+A_4) h^2=\Lambda$}

In this case the admitted solutions are

(i) $\phi_n = (...,a,-a,...)$ in case $h^2a^2(A_4-A_2)=\Lambda-4$.

(ii) $\phi_n = (...,a,1/a,a,1/a,...)$ in case $h^2A_4=2\,,~h^2A_2
=\Lambda-2$ while $a$ is any real number.

(iii) $\phi_n = (...,a,-a,0,...)$ in case $a^2h^2 A_2=2(3-\Lambda)$.

(iv) $\phi_n = (...,a,a,-a,-a,...)$ in case $a^2h^2 A_4=2-\Lambda$.

(v) $\phi_n = (...,a,0,-a,0,...)$ in case $\Lambda=2$ while $A_2,A_4$
as well as $a$ are arbitrary real numbers.

(vi) $\phi_n = (...,a,a,-a,-a,0,...)$ in case $a^2h^2A_2=2\Lambda-2$.

(vii) $\phi_n=(...,a,a,0,-a,0,...)\,,~(...,a,0,-a,a,0,-a,0,...)\,, \\
~(...,a,a,0,-a,-a,0,a,-a,0,...)$\,, in case $\Lambda=2, h^2a^2A_2=2$.

(viii) $\phi_n=(...,a,a,a,-a,-a,...)\,,~(...,a,a,-a,-a,0,...)\,,
~(...,a,a,0,-a,-a,0,a,...)$\,, in case $a^2=1, h^2A_2=2\Lambda-2,
h^2a^2A_4=2-\Lambda$.

(ix) $\phi_n=(...,a,a,-a,...)\,,~(...,a,-a,a,-a,0,...)\,,
~(...,a,-a,a,a,-a,0,...)$\,, in case
\be
a^2=\frac{8-3\Lambda}{\Lambda}\,,~~h^2A_2=\frac{2\Lambda(3-\Lambda)}
{8-3\Lambda}\,,~~h^2A_4=\frac{\Lambda(2-\Lambda)}{8-3\Lambda}\,.
\ee
Thus this solution is valid provided $0<\Lambda<8/3$.

\subsection{Case (iv): Only $A_3,A_5$ nonzero, $(A_3+A_5) h^2=\Lambda$}

In this case the admitted solutions are

(i) $\phi_n = (...,a,-a,...)$ in case $h^2a^2(A_3-A_5)=\Lambda-4$.

(ii) $\phi_n = (...,a,0,...)$ in case $a^2 h^2A_5=2\,,\Lambda=2$.

(iii) $\phi_n = (...,a,-a,0,...)$ in case $a^2h^2 A_3=2(\Lambda-3)$.

(iv) $\phi_n=(...,a,a,0,...)$ in case $a^2=1, h^2 A_3=\Lambda-1, h^2 A_5=1$.
(
v) $\phi_n = (...,a,0,-a,0,...)$ in case $\Lambda=2$ and $a$ is any real
number.

(vi) $\phi_n = (...,a,a,-a,-a,...)$ in case $a^2 h^2 A_3=\Lambda-2$.

(vii) $\phi_n= (...,a,a,0,-a,0,...)$ in case $\Lambda=2, a^2 h^2 A_3= 2$.

(viii) $\phi_n = (...,a,a,0,-a,-a,0,...)$ in case $a^2h^2A_3=2\Lambda-2$.

(ix) $\phi_n=(...,a,a,a,0,-a,-a,0,...)$ in case $a^2=1, h^2 A_3=2(\Lambda-1),
h^2 A_5=2-\Lambda$.

(x) $\phi_n=(...,a,0,-a,a,0,-a,0,...)$ in case $\Lambda=2, a^2 h^2 A_3=-2$.

(xi) $\phi_n=(...,a,-a,0,-a,a,0,...)$ in case
\be
a^2 =\frac{2(\Lambda-2)}{\Lambda}\,,~h^2 A_3=\frac{\Lambda(\Lambda-3)}
{\Lambda-2}\,,~~h^2 A_5=\frac{\Lambda}{\Lambda-2}\,.
\ee
Thus this solution is valid provided either $\Lambda>2$ or $\Lambda<0$.

(xii) $\phi_n=(...,a,-a,a,-a,0,...)$ in case
\be
a^2 =\frac{3\Lambda-8}{\Lambda}\,,~h^2 A_3=\frac{2\Lambda(\Lambda-3)}
{3\Lambda-8}\,,~~h^2 A_5=\frac{\Lambda(\Lambda-2)}{3\Lambda-8}\,.
\ee
Thus this solution is valid provided either $\Lambda>8/3$ or $\Lambda<0$.

(xiii) $\phi_n=(...,a,a,-a,...)\,,~(...,a,a,a,-a,...)\,,
~(...,a,a,a,-a,-a,...)$\,, in case $a^2=1, h^2A_3=\Lambda-2,
h^2a^2A_5=2$.

(xiv) In addition, there are a large number of solutions which are admitted in
case $A_3=A_5, \Lambda=4,a^2=1$. These allowed solutions are
\bea
&&\phi_n=(...,a,-a,0,a,...)\,,~(...,a,a,-a,0,-a,...)\,,
~(...,a,-a,a,a,-a,0,...)\,,~(...,a,a,-a,0,-a,a,...)\,, \nonumber \\
&&(...,a,0,-a,a,a,a,-a,...)\,,~(...,a,a,-a,a,a,-a,0,a,...)\,.
\eea

\subsection{SW Model 7: $A_1=\frac{2\lambda}{9},A_2=A_3=\frac{\lambda}{3},
A_4=A_5=0,A_6=\frac{\Lambda}{9}$}

In this case the admitted solutions are

(i) $\phi_n = (...,a,-a,...)$ in case $a^2=\frac{9(\Lambda-4)}{\lambda}$.
Thus such a solution is valid if either $\Lambda>4$ or $\Lambda <0$.

(ii) $\phi_n = (...,a,-a,0,...)$ in case $a^2=\frac{6(\Lambda-3)}{\lambda}$.
Thus such a solution is valid if either $\Lambda>3$ or $\Lambda <0$.

(iii) $\phi_n = (...,a,a,-a,-a,...)$ in case $a^2=\frac{9(\Lambda-2)}
{5\lambda}$.
Thus such a solution is valid if either $\Lambda>2$ or $\Lambda <0$.

(iv) $\phi_n = (...,a,0,-a,0,...)$ in case $a^2=\frac{9(\Lambda-2)}
{2\lambda}$.
Thus such a solution is valid if either $\Lambda>2$ or $\Lambda <0$.

(v) $\phi_n = (...,a,a,0,-a,-a,0,...)$ in case $a^2=\frac{18(\Lambda-1)}
{11\lambda}$.
Thus such a solution is valid if either $\Lambda>1$ or $\Lambda <0$.

(vi) $\phi_n = (...,a,a,0,-a,0,...),(...,a,a,0,-a,0,a,...)$ in case
$a^2=1,\Lambda=18/7$.

(vii) $\phi_n = (...,a,a,0,...),(...,a,a,-a,-a,0,...)$ in case
$a^2=3/2,\Lambda=12$.

(viii) $\phi_n = (...,a,a,-a,...),(...,a,a,a,-a,-a,...)$ in case
$a^2=1,\Lambda=9/2$.

(ix) $\phi_n = (...,a,a,0,-a,-a,0,a,-a,0,...)$ in case
$a^2=6/5,\Lambda=15/4$.

(x) In addition, there are a large number of solutions which are admitted in
case $\Lambda=6$ and $a^2=3$. These allowed solutions are
\bea
&&\phi_n=(...,a,0,...)\,,~(...,a,-a,a,-a,0,...)\,,
~(...,a,0,a,0,-a,...)\,,~(...,a,-a,0,-a,a,0,...)\,, \nonumber \\
&&(...,a,0,a,0,a,-a,...)\,,(...,a,0,-a,a,0,-a,0,...)\,.
\eea

\subsection{Model 2 with $\delta=0,\gamma=1/4$: Only $A_5,A_6$ nonzero with
 $A_5 h^2=A_6h^2=\Lambda /2$}

This is same as the Model 4 due to Kevrekidis. In this case the admitted
solutions are

(i) $\phi_n = (...,a,-a,...)$ in case $a^2=\frac{4-\Lambda}{\lambda}$.
Thus such a solution is valid if $0<\Lambda<4$.

(ii) $\phi_n = (...,a,-a,0,...)$ in case $a^2=\frac{4(3-\Lambda)}{\lambda}$.
Thus such a solution is valid if $0<\Lambda<3$.

(iii) $\phi_n = (...,a,a,-a,-a,...)$ and $\phi_n =(...,a,0,-a,0,...)$ in case
$\Lambda=2$ while $a$ is {\it any} real number.

(iv) $\phi_n = (...,a,b,-a,-b,...)$ in case
$\Lambda=2$ while $a,b$ both are {\it any} arbitrary real numbers.

(v) $\phi_n = (...,a,a,0,-a,-a,0,...)$ in case $a^2=\frac{4(\Lambda-1)}
{\lambda}$.
Thus such a solution is valid if $\Lambda>1$ or if $\Lambda<0$.

(vi) In addition, there are a large number of solutions which are admitted in
case $\Lambda=2$ and $a^2=1$. These allowed solutions are
\bea
&&\phi_n=(...,a,a,-a,...)\,,~(...,a,a,a,-a,...)\,,~(...,a,0,-a,a,...)\,,~
(...,a,a,a,-a,-a,...)\,, \nonumber \\
&&(...,a,-a,a,-a,0,...)\,,~(...,a,a,-a,-a,0,...)\,,~(...,a,a,0,-a,0,...)\,,~
(...,a,0,-a,a,a,...)\,, \nonumber \\
&&(...,a,a,a,0,-a,-a,...)\,,~(...,a,a,0,-a,0,a,...)\,,
~(...,a,-a,a,a,-a,0,...)\,,~ (...,a,a,0,-a,-a,0,a,...)\,, \nonumber \\
&&(...,a,0,-a,a,0,-a,0,...)\,,~(...,a,-a,0,a,a,0-,a,...)\,,
~(...,a,a,-a,a,a,-a,0,...)\,,~ (...,a,0,-a,a,a,a,-a,...)\,, \nonumber \\
&&(...,a,-a,0,a,a,a,0,-a,...)\,,~(...,a,a,0,-a,-a,0,a,-a,0,...)\,, \nonumber \\
&&~(...,a,a,0,-a,-a,0,a,-a,0,a,...)\,,~ (...,a,-a,0,a,a,a,0,-a,a,a,-a,...)\,,
\eea

\subsection{Model 2 with $\delta=1/4,\gamma=0$: Only $A_1,A_3,A_4$ nonzero with
 $A_1 h^2=A_4 h^2=A_3 h^2 /2=\Lambda /4$}

In this case the admitted solutions are

(i) $\phi_n = (...,a,-a,...)$ in case $a^2=\frac{\Lambda-4}{\lambda}$.
Thus such a solution is valid if either $\Lambda>4$ or $\lambda<0$.

(ii) $\phi_n = (...,a,-a,0,...)$ in case $a^2=\frac{2(\Lambda-3)}{\lambda}$.
Thus such a solution is valid if either $\Lambda>3$ or $\Lambda<0$.

(iii) $\phi_n = (...,a,b,-a,-b,...)$ in case
$(a^2+b^2)=\frac{2(\Lambda-2)}{\Lambda}=2$.

(iv) $\phi_n = (...,a,a,a,-a,...)$ in case
$a^2=1, \Lambda=4$.

(v) $\phi_n = (...,a,a,0,-a,0,...)$ in case
$a^2=4/3, \Lambda=3$.

(vi) $\phi_n = (...,a,a,0,-a,-a,0,a,...)$ in case
$a^2=1, \Lambda=2$.

(vii) In addition, there are a large number of solutions which are admitted in
case $a^2=\frac{2(\Lambda-1)}{\Lambda}$. These allowed solutions are
\be
\phi_n=(...,a,a,-a,...)\,,~(...,a,a,-a,-a,...)\,,~(...,a,0,-a,0,...)\,,~
(...,a,a,0,-a,-a,0,...)\,.
\ee

\subsection{Model 2 with arbitrary $\delta,\gamma$: $A_1=A_4=A_3 /2
=\delta \lambda, A_5=A_6=2\gamma \lambda, A_2=(1-4\delta-4\gamma)\lambda$}

This is the full Model 2. Note that we have already obtained solutions in three
special cases. In this general case, many more solutions are admitted. The
admitted solutions are

(i) $\phi_n = (...,a,-a,...)$ in case $a^2
=\frac{\Lambda-4}{(8\delta-1)\Lambda}$.
Thus such a solution is valid if either $\Lambda>4$ or $\lambda<0$.

(ii) $\phi_n = (...,a,0,...)$ in case
\be
a^2=\frac{\Lambda-2}{\delta \Lambda}\,,~~2\gamma \Lambda a^2=1\,.
\ee

(iii) $\phi_n = (...,a,-a,0,...)$ in case
\be
a^2=\frac{2(\Lambda-3)}{(8\delta+2\gamma-1) \Lambda}\,.
\ee

(iv) $\phi_n = (...,a,a,-a,...)$ in case
\be
a^2=\frac{3\Lambda-4}{\Lambda}\,,~~\delta=\frac{\Lambda-2}{2(3\Lambda-4)}\,.
\ee

(v) $\phi_n = (...,a,a,0,...)$ in case
\be
a^2=\frac{1}{2\gamma \Lambda}=\frac{2}{1+2\delta}\,.
\ee

(vi) $\phi_n = (...,a,a,a,0,...)$ in case
\be
a^2=1\,,~~\delta=\frac{1}{2}\,,\gamma=\frac{1}{2\Lambda}\,.
\ee

(vii) $\phi_n = (...,a,a,a,-a,...)$ in case
\be
a^2=1\,,~~\delta=0\,,\Lambda=2\,.
\ee

(viii) $\phi_n = (...,a,b,-a,-b,...)$ in case
\be
(a^2+b^2)=\frac{\Lambda-2}{\delta \Lambda}\,.
\ee

(ix) $\phi_n = (...,a,a,-a,-a,...)$ in case
\be
a^2=\frac{\Lambda-2}{2\delta \Lambda}\,.
\ee

(x) $\phi_n = (...,a,0,-a,0,...)$ in case
\be
a^2=\frac{\Lambda-2}{\delta \Lambda}\,.
\ee

(xi) $\phi_n = (...,a,-a,0,a,...)$ in case
\be
a^2=\frac{1}{6\gamma}\,,~~\Lambda=3\,,~~8\delta+2\gamma=1\,.
\ee

(xii) $\phi_n = (...,a,a,-a,-a,...)$ in case
\be
a^2=1\,,~~\delta=\frac{\Lambda-2}{2 \Lambda}\,.
\ee

(xiii) $\phi_n = (...,a,-a,0,a,...)$ in case
\be
a^2=\frac{1}{6\gamma}\,,~~\Lambda=3\,,~~8\delta+2\gamma=1\,.
\ee

(xiv) $\phi_n = (...,0,a,0,a,a,...)$ in case
\be
a^2=\frac{\Lambda+2}{\Lambda}\,,
~~\delta=\frac{\Lambda-2}{2 (\Lambda+2)}\,,
~~\gamma=\frac{1}{2 (\Lambda+2)}\,.
\ee

(xv) $\phi_n = (...,a,-a,a,-a,0,...)$ in case
\be
a^2=\frac{\Lambda-2}{2\gamma \Lambda}\,,
~~(8\delta-1)(\Lambda-2)=2\gamma (\Lambda-4)\,.
\ee

(xvi) $\phi_n = (...,a,0,a,0,-a,...)$ in case \be
a^2=\frac{3(2\Lambda-3)}{\Lambda}\,, ~~\delta=\frac{\Lambda-2}{3
(2\Lambda-3)}\,, ~~\gamma=\frac{1}{6 (2\Lambda-3)}\,. \ee

(xvii) $\phi_n = (...,a,a,-a,-a,0,...)$ in case
\be
a^2=\frac{(\Lambda-2)}{2\delta \Lambda}=\frac{1}{1-4\gamma}\,.
\ee

(xviii) $\phi_n = (...,a,a,0,-a,0,...)$ in case \be
a^2=\frac{(\Lambda-2)}{2\delta
\Lambda}=\frac{2}{\Lambda(1-4\gamma)}\,. \ee

(xix) $\phi_n = (...,a,-a,0,-a,0,...)$ in case \be
a^2=\frac{(2\Lambda-1)}{\Lambda}\,,
~~\delta=\frac{\Lambda-2}{2(2\Lambda-1)}\,,
~~\gamma=\frac{1}{2(2\Lambda-1)}\,. \ee

(xx) $\phi_n = (...,a,a,0,a,-a,...)$ in case
\be
a^2=\frac{5}{3}\,,~~\Lambda=3\,,~~\delta=\gamma=\frac{1}{10}\,.
\ee

(xxi) $\phi_n = (...,a,a,0,-a,-a,0,...)$ in case
\be
a^2=\frac{2(\Lambda-1)}{(1+2\delta-4\gamma) \Lambda}\,.
\ee

(xxii) $\phi_n = (...,a,-a,0,-a,a,0,...)$ in case
\be
a^2=\frac{1}{2\gamma \Lambda}\,,
~~(8\delta-1)=2\gamma (2\Lambda-7)\,.
\ee

(xxiii) $\phi_n = (...,a,a,a,0,-a,-a,...)$ in case
\be
a^2=1\,,~~\gamma=0\,,~~\delta=\frac{\Lambda-2}{\Lambda}\,.
\ee

(xxiv) $\phi_n = (...,a,a,0,-a,0,a,...)$ in case
\be
a^2=1\,,~~\gamma=\frac{\Lambda-2}{4\Lambda}\,,~~
\delta=\frac{\Lambda-2}{\Lambda}\,.
\ee

(xxv) $\phi_n = (...,a,a,0,a,-a,0,...)$ in case \be
a^2=\frac{(6\Lambda+7)}{5\Lambda}\,,
~~\delta=\frac{4\Lambda-7}{2(6\Lambda+7)}\,,
~~\gamma=\frac{5}{2(6\Lambda+7)}\,. \ee

(xxvi) $\phi_n = (...,a,a,-a,0,-a,a,...)$ in case
\be
a^2=1\,,~~\Lambda=1\,,~~\gamma=\frac{1}{2}\,,~~\delta=\frac{-1}{2}\,.
\ee

(xxvii) $\phi_n = (...,a,0,a,0,a,-a,...)$ in case
\be
a^2=3\,,~~\Lambda=3\,,~~\gamma=\frac{1}{18}\,,~~\delta=\frac{1}{9}\,.
\ee

(xxviii) $\phi_n = (...,a,0,a,0,a,a,...)$ in case
\be
a^2=1\,,~~\Lambda=2\,,~~\gamma=\frac{1}{4}\,,~~\delta=0\,.
\ee

(xxix) $\phi_n = (...,a,-a,a,a,0,a,...)$ and $(...,a,a,0,a,a,0,-a,...)$
in case
\be
a^2=\frac{5}{3}\,,~~\Lambda=3\,,~~\gamma=\delta=\frac{1}{10}\,.
\ee

(xxx) $\phi_n = (...,a,a,0,-a,-a,0,a,...)$ in case
\be
a^2=1\,,~\delta-2\gamma=\frac{\Lambda-2}{2\Lambda}\,.
\ee

(xxxi) $\phi_n = (...,a,0,-a,a,0,-a,0,...)$ in case
\be
a^2=\frac{\Lambda-2}{\delta \Lambda}\,,
~~2\delta(5-3\Lambda)=(2\gamma-1) (\Lambda-2)\,.
\ee

(xxxii) $\phi_n = (...,a,a,0,-a,-a,a,-a,...)$ in case \be
a^2=\frac{(7\Lambda-12)}{\Lambda}\,,
~~\delta=\frac{\Lambda-2}{2(7\Lambda-12)}\,,
~~\gamma=\frac{7(\Lambda-2)}{4(7\Lambda-12)}\,. \ee

(xxxiii) In case
\be
a^2=1\,,~\delta=\gamma=0\,,~~\Lambda=2\,,
\ee
then the following solutions are allowed
\bea
&&\phi_n = (...,a,0,-a,a,a,...),(...,a,0,-a,a,a,a,-a,...),
(...,a,-a,0,a,a,a,0,-a,...), (...,a,0,a,0,a,a,a,-a,...)\,, \nonumber \\
&&(...,a,-a,0,a,0,a,a,0,-a,...), (...,a,-a,0,a,0,a,a,a,0,-a,...),
(...,a,-a,0,a,a,a,0,-a,a,a,-a,...)\,.
\eea

(xxxiv) $\phi_n = (...,a,a,0,a,0,-a,-a,...)$ in case
\be
a^2=2\,,~\delta=0\,,~~\gamma=\frac{1}{4\Lambda}\,.
\ee

(xxxv) $\phi_n = (...,a,a,-a,0,-a,0,-a,...)$ in case
\be
a^2=\frac{5-2\Lambda}{\Lambda}\,,~\delta=0\,,~~\gamma=\frac{1}{2(5-2\Lambda)}\,.
\ee

(xxxvi) $\phi_n = (...,a,-a,a,a,a,0,-a,...)$ in case
\be
a^2=1\,,~\delta=0\,,~~\gamma=\frac{1}{4}\,,~~\Lambda=2\,.
\ee

(xxxvii) $\phi_n = (...,a,a,0,-a,-a,0,a,-a,0,...)$ in case
\be
a^2=\frac{2}{\Lambda(1-3\delta-3\gamma)}\,,~~6(1-3\gamma)
=(3\Lambda-1)(1-3\delta-3\gamma)\,.
\ee

(xxxviii) $\phi_n = (...,a,a,0,-a,-a,0,a,-a,0,a,......)$ in case
\be
a^2=1\,,~\delta=\frac{7(\Lambda-2)}{18\Lambda}\,,~~\gamma=\frac{1}{9\Lambda}\,.
\ee

(xxxiv) In case
\be
a^2=\frac{3\Lambda-4}{\Lambda}\,,~\delta=\gamma
=\frac{\Lambda-2}{2(3\Lambda-4)}\,,
\ee
then the following solutions are allowed
\be
\phi_n = (...,a,0,-a,a,...), (...,a,-a,a,a,-a,0,...),
(...,a,-a,0,a,a,0,-a,...), (...,a,a,-a,a,a,-a,0,...)\,.
\ee

\subsection{Case (v): Only $A_3,A_4,A_5$ nonzero, $(A_3+A_4+A_5) h^2=\Lambda$}

In this case the admitted solutions are

(i) $\phi_n = (...,a,-a,...)$ in case $h^2a^2(A_3+A_4-A_5)=\Lambda-4$.

(ii) $\phi_n = (...,a,0,...)$ in case $a^2 h^2A_5=2\,,\Lambda=2$.

(iii) $\phi_n = (...,a,-a,0,...)$ in case $a^2h^2 A_3=2(\Lambda-3)$.

(iv) $\phi_n = (...,a,a,-a,...)$ in case $a^2h^2 (A_3-A_4)=(\Lambda-2),
a^2h^2(H+F-J)=\Lambda-4$.

(v) $\phi_n = (...,a,a,0,...)$ in case $a^2h^2 A_3=2(\Lambda-1),
a^2 h^2 A_5=2$.

(vi) $\phi_n = (...,a,0,-a,0,...)$ in case $\Lambda=2$ and $a$ is any real
number.

(vii) $\phi_n = (...,a,a,-a,-a,...)$ in case $a^2 h^2 A_3=\Lambda-2$.

(viii) $\phi_n = (...,a,a,a,0,...)$ in case $a^2=1, h^2 A_3=2(\Lambda-1)\,,
h^2 A_5=2, h^2 A_4=-\Lambda$.

(ix) $\phi_n = (...,a,-a,0,a,...)$ in case $a^2=1, h^2 A_3=2(\Lambda-3),
h^2 A_4=4-\Lambda, h^2 A_5=2$.

(x) $\phi_n = (...,a,b,-a,-b,...)$ in case $A_3=A_4, \Lambda=2$ and $a,b$
are arbitrary real numbers.

(xi) $\phi_n=(...,a,a,0,-a,0,...)$ in case $\Lambda=2, h^2a^2A_3=2$.

(xii) $\phi_n=(...,a,-a,a,-a,0,...)$ in case
$h^2 a^2(A_3+A_4-A_5) =\Lambda-4, h^2a^2 A_3=2(\Lambda-3)$.

(xiii) $\phi_n=(...,a,a,a,-a,-a,...)$ in case $a^2=1, h^2(A_3-A_4)=\Lambda-2$.

(xiv) $\phi_n=(...,0,a,0,a,a,...)$ in case $A_3=A_5, \Lambda=2, h^2 a^2 A_5 =2$.

(xv) $\phi_n=(...,a,0,a,0,-a,...)$ in case
$\Lambda=2, A_3=-A_5, h^2A_4=2, h^2 a^2 A_5 =2$.

(xvi) $\phi_n=(...,a,a,-a,-a,0,...)$ in case $h^2a^2A_4=\Lambda,
h^2a^2A_3=2(\Lambda-1)$.

(xvii) $\phi_n=(...,a,-a,0,-a,a,...)$ in case
\be
a^2 =\frac{3\Lambda-8}{\Lambda}\,,~h^2 A_3=\frac{2\Lambda(\Lambda-3)}
{3\Lambda-8}\,,
~~h^2 A_4=\frac{\Lambda(\Lambda-4)}{3\Lambda-8}\,,
~~h^2 A_5=\frac{2\Lambda}{3\Lambda-8}\,.
\ee

(xviii) $\phi_n = (...,a,a,0,-a,-a,0,...)$ in case $a^2h^2A_3=2(\Lambda-1)$.

(xix) $\phi_n = (...,a,-a,0,-a,a,0,...)$ in case $a^2h^2A_3=2(\Lambda-3),
h^2 a^2 A_5=2$.

(xx) $\phi_n = (...,a,a,a,0,-a,-a,...)$ in case $a^2=1, A_3=-A_5,
h^2A_3=2(\Lambda-1), h^2 A_4=\Lambda$.

(xxi) $\phi_n = (...,a,a,0,-a,0,a,...)$ in case $a^2=1, \Lambda=2,
h^2A_3=2$.

(xxii) $\phi_n = (...,a,-a,a,a,-a,0,...)$ in case
\be
a^2 =\frac{5\Lambda-16}{\Lambda}\,,~~A_3=A_5\,,~~
h^2 A_3=\frac{2\Lambda(\Lambda-3)}
{5\Lambda-16}\,,
~~h^2 A_4=\frac{\Lambda(\Lambda-4)}{5\Lambda-16}\,.
\ee

(xxiii) $\phi_n=(...,a,a,0,-a,-a,0,a,...)$ in case $a^2=1,
h^2 A_3=2(\Lambda-1)$.

(xxiv) $\phi_n=(...,a,0,-a,a,0,-a,0,...)$ in case $\Lambda=2, a^2 h^2 A_3=-2$.

(xxv) $\phi_n=(...,a,a,0,-a,-a,a,-a,...)$ in case $\Lambda=2, h^2a^2A_3=2\,,
h^2a^2(A_4-A_5)=-4$.

(xxvi) $\phi_n=(...,a,a,0,a,0,-a,-a,...)$ in case $\Lambda=2, a^2=3,
h^2A_3=h^2A_4=h^2A_5=2/3$.

(xxvii) $\phi_n=(...,a,-a,0,a,0,a,a,a,0,-a,...)$ in case $\Lambda=2,
h^2a^2A_3=2, A_5=2A_3+A_4$.

\end{document}